\documentclass[a4paper,11pt]{article}
\pdfoutput=1 

\usepackage{jheppub} 

\usepackage[T1]{fontenc}
\usepackage[italian,english]{babel}
\usepackage{hyperref}
\usepackage{ifpdf}
\usepackage{subfigure}
\usepackage{amssymb}
\usepackage{amsfonts}
\usepackage{epsf}
\usepackage{rotating}
\usepackage{graphicx}
\usepackage{amsmath}
\usepackage{fancyhdr}
\usepackage{lineno}
\usepackage{babel}
\usepackage{graphics}
\usepackage{pstricks}
\usepackage{color}

\makeatletter
\def\cleardoublepage{\clearpage\if@twoside
\ifodd\c@page
\else\hbox{}\thispagestyle{empty}\newpage
\if@twocolumn\hbox{}\newpage\fi\fi\fi}
\makeatother
\makeatletter
\def\cleardoublepage{\clearpage\if@twoside
\ifodd\c@page
\else\hbox{}\thispagestyle{empty}\newpage
\if@twocolumn\hbox{}\newpage\fi\fi\fi}
\makeatother

      \let\g=\gamma

\def\sq{\,\raise.5pt\hbox{$\nbox{.09}{.09}$}}
\def\sqb{\,\raise.5pt\hbox{$\overline{\nbox{.09}{.09}}$}}

\def\slashed{\ds}

\def\g{\gamma}

\def\ds#1{#1\kern-1ex\hbox{/}}
\def\dsh{h\kern-1.2ex /}

\newcommand{\bea}{\begin{eqnarray}}
\newcommand{\eea}{\end{eqnarray}}

\def\nn{\nonumber}

\def\beq{\begin{equation}}
\def\eeq{\end{equation}}

\def\ba{\begin{eqnarray}}
\def\ea{\end{eqnarray}}

\newcommand{\beqa}{\begin{eqnarray}}
\newcommand{\eeqa}{\end{eqnarray}}

\newcommand{\ksl}{k \! \! \!  /}

\newcommand{\pd}{\partial}

\def\th{\theta}

\newcommand{\bes}{\begin{subequations}}
\newcommand{\ees}{\end{subequations}}

\title{\boldmath Neutrino and Photon Lensing by Black Holes: Radiative Lens Equations and Post-Newtonian Contributions}

\author[a,b]{Claudio Corian\`o}
\author[a]{Antonio Costantini}
\author[a]{Marta Dell'Atti}
\author[a]{Luigi Delle Rose}

\affiliation[a]{Dipartimento di Matematica e Fisica "Ennio De Giorgi", \\ Universit\`a del Salento and INFN-Lecce, \\ Via Arnesano, 73100 Lecce, Italy}

\affiliation[b]{STAG Research Centre and Mathematical Sciences,\\ University of Southampton, Southampton SO17 1BJ, UK}

\emailAdd{claudio.coriano@le.infn.it}
\emailAdd{antonio.costantini@le.infn.it}
\emailAdd{marta.dellatti@le.infn.it}
\emailAdd{luigi.dellerose@le.infn.it}

\abstract{We extend a previous phenomenological analysis of photon lensing in an external gravitational background to the case of a massless neutrino, 
and propose a method to incorporate radiative effects in the classical lens equations of neutrinos and photons.
The study is performed for a Schwarzschild metric, generated by a point-like source, and expanded in the Newtonian potential at first order.  We use a semiclassical approach, where the perturbative corrections to neutrino scattering, evaluated at one-loop in the Standard Model, are compared with the Einstein formula for the deflection using an impact parameter formulation. For this purpose, we use the renormalized expression of the graviton/fermion/fermion vertex presented in previous studies. We show the agreement between the classical and the semiclassical formulations, for values of the impact parameter $b_h$ of the neutrinos of the order of $b_h\sim 20$, measured in units of the Schwarzschild radius. The analysis is then extended with the inclusion of the post Newtonian corrections in the external gravity field, showing that this extension finds application in the case of the scattering of a neutrino/photon off a primordial black hole. The energy dependence of the deflection, generated by the quantum corrections, is then combined with the standard formulation of the classical lens equations.  We illustrate our approach by detailed numerical studies, using as a reference both the thin lens and the Virbhadra-Ellis lens.}

\begin{document}
\maketitle
\flushbottom

\section{Introduction}
According to classical general relativity (GR) massless particles follow null spacetime geodesics which bend significantly in the presence of very massive sources. The gravitational lensing enforced on their spatial trajectories provides important information on the underlying distributions of matter and, possibly, of dark matter, which act as sources of the gravitational field.\\
Several newly planned weak lensing experiments such as the Dark Energy Survey (DES) \cite{DES}, the Large Synoptic Survey Telescope (LSST)\cite{LSST}, both ground based, or from space with the Wide-Field Infrared Survey Telescope (WFIRST) \cite{WFIRST} and Euclid \cite{EUCLID}, are expected to push forward, in the near future, the boundaries of our knowledge in cosmology.\\
 In the analysis of the deflection by a single compact and spherically symmetric  source, one significant variable, beside the mass of the source, is the impact parameter of the incoming particle beam, measured respect to the center of the source, which determines the size of the deflection. It is very convenient to measure the impact parameter $(b)$, which is typical of a given collision, in units of the Schwarzschild radius $r_s\equiv 2 G M $, denoted as $b_h\equiv b/r_s$. In the Newtonian approximation for the external background, this allows to scale out the entire mass dependence of the lensing event. \\
For an impact parameter of the beam of the order of $10^5-10^6$, the corresponding deflection is rather weak, of the order of 1-2 arcseconds, as in the case of a photon skimming the sun. Stronger lensing effects are predicted as the particle beam nears a black hole, with deflections which may reach 30 arcseconds or more. These are obtained for impact parameters $b_h$ of the order of $2\times 10^4$. Even larger deflections, of 1 to 2 degrees or a significant fraction of them, are generated in scatterings which proceed closer to the event horizon \cite{Coriano:2014gia}. In fact,  as we are going to show, for closer encounters, with the beam located between 20 and 100 $b_h$, such angular deflections are around 
$10^{-2}$ radians in size, as predicted by classical GR. A high energy cosmic ray of 10-100 GeV  will then interact with the field of the source by exchanging momenta far above the MeV region, and will necessarily be sensitive to radiative effects, such as those due to the electroweak corrections.\\
Interactions with such momentum exchanges cannot be handled by an effective Newtonian potential, as derived, for instance, from the (loop corrected) scattering amplitude. We recall that, in general, in the derivation of such a potential, one has to take into account only non-analytic terms in the momentum transfer $q$. These are obtained from a given amplitude and/or gravitational form factor of the incoming particle after an expansion at small momentum. The analytic terms in the expansion correspond to contact interactions which are omitted from the final form of the potential, being them proportional to Dirac delta functions. \\
 As one can easily check by a direct analysis, non-analytic contributions originate from massless exchanges in the loops, 
which approximate the full momentum dependence of the radiative corrections only for momentum transfers far below the MeV region. Therefore, the validity of the method requires that the typical impact parameter of the beam, for a particle with the energy of few GeV's, be of the order of $10^6$ Schwarzschild radii and not less. For such a reason, if we intend to study a lensing event characterized by a close encounter between a cosmic ray and a black hole, we need to resort to an alternative approach, which does not suffer from these limitations. \\
Finally, with the photon sphere located at $b_h\sim 2.5$ for a Schwarzschild metric, one expects that very strong deflections are experienced by a beam for scattering events running close to such a value of the impact parameter. This is also the radial distance from the black hole center at which the scattering angle diverges. A simple expansion of the Einstein formula for the deflection shows that this singularity is logarithmic \cite{Coriano:2014gia}. In such extreme cases the beam circulates around the source one or more times before escaping to infinity, generating a set of relativistic images \cite{Bozza:2001xd}. This is also the region where the simple Newtonian approach, discussed in 
\cite{Coriano:2014gia},  fails to reproduce the classical GR prediction, as expected. \\

\subsection{Comparing classical and semiclassical effects}
 The analysis of possible extensions of the classical GR prediction for lensing, with the inclusion also of quantum effects 
in the interaction between the particle source and the deflector (lens), has not drawn much attention in the past, except for a couple of very original proposals \cite{Delbourgo:1973xe, Berends:1974gk}. While these effects are expected to be small, even for huge gravitational sources such as massive/supermassive black holes, they could provide, in principle, a way to test the impact of quantum gravity and of other radiative corrections to the propagation of cosmic rays. Close encounters of a beam with a localized source, which could be a large black hole or a neutron star, are expected to be quite common in our universe, although the probability of identifying a lensing event characterized by a close alignment between the source, the lens and an earth based detector, especially for neutrinos, is exceedingly rare \cite{Mena:2006ym}. The situation might be more promising for photons in close encounters with primordial black holes, revealed by resorting to spaceborne detectors. \\ 
Such is the FERMI satellite \cite{FERMI}, with source beams given by Gamma Ray Bursts (GRBs) \cite{Gould1992}, which could detect fringes between primary and secondary paths of the GRBs on its ultra sensitive camera, generated by a gravitational time delay. This approach was termed in 
\cite{Gould1992} "femtolensing", due to the size of the Einstein radius characteristic of these events, which was estimated to be of the order of a femtoarcsecond. As shown in \cite{Gould1992}, a classical GR analysis based on the thin lens equation can be applied quite straightforwardly also to this extreme situation.\\
An important point which needs to be addressed, in this case, concerns the quantum features of these types of lensing events, since the Schwarzschild radius of a primordial black hole, for a gamma ray photon, is comparable to its wavelength. Our analysis draws a path in this direction.

The classical deflections of photons, as pointed out in the past and in a recent work \cite{Coriano:2014gia}, can be compared at classical and quantum
 levels by equating the classical gravitational cross section, written in terms of the impact parameter of the incoming photon beam, to the perturbative cross section. The latter is expanded in ordinary perturbation theory with the inclusion of the corresponding radiative corrections. 
 The result is a differential equation for the impact parameter of the beam, whose solution provides the link between the two descriptions. In particular, the energy dependence, naturally present in the cross section starting at one-loop order, allows to derive a new formula which 
 relates $b_h$ to the energy $E$ of the beam and to the angle of deflection $\alpha$, $b_h(E, \alpha)$. This dependence, which is absent in Einstein's formula, propagates into all the equations for the usual observables of any lensing process: magnifications, cosmic shears, the light curve of microlensing events and Shapiro time delays. Clearly, such a dependence implies, as noted in \cite{Accioly1}, that radiative corrections induce a violation of the classical equivalence principle in General Relativity.  
The violation of the equivalence principle, viewed from a quantum perspective, is not surprising, since this principle is inherently classical and requires the localization of the point particle trajectory on a geodesic. It can be summarized in the statement that an experiment will not be able to determine the nature of the point particle which is subjected to gravity, except for its mass. The notion of a point particle clearly clashes with the quantum description, which is, on the other hand, inherently tight to Heisenberg's indetermination principle. For this reason, one expects that the inclusion of radiative corrections will cause a violation of such principle.
 
 Gravity, in this approach, is treated as an external background and the transition amplitude involves on the quantum side, in the photon case, the $TVV$ vertex, where $T$ denotes the energy momentum tensor (EMT) of the Standard Model and $V$ the electromagnetic current. In the fermion case (f), the corresponding vertex is the $Tff$, with $f$ denoting a neutrino. The comparison between the classical and the semiclassical formula for the deflection derived by this method can then be performed at numerical level, as shown in \cite{Coriano:2014gia} for the photons. The energy dependence of the bending angle, for a given impact parameter of the photon beam, though small, is found to become more pronounced at higher energies, due to the logarithmic growth of the electroweak corrections with the energy.    

The goal of our present work is to propose a procedure which allows to include these effects in the ordinary lens equations, illustrating in some detail how this approach can be implemented in a complete numerical study. We mention that our semiclassical analysis is quite general, and applies both to macroscopic and to microscopic black holes. In the case of macroscopic black holes the procedure has to stop at Newtonian level in the external field. In fact, post-Newtonian corrections, though calculable, render the perturbative expansion in the external (classical) gravitational potential divergent, due to the macroscopic value of the Schwarzschild radius. On the other hand, in the case of primordial black holes, the very same corrections play a significant role in the deflection of a cosmic ray, and bring to a substantial modification of the classical formulas.

 \subsection{Organization of this work}
In the first part of our work we will extend a previous analysis of photon lensing \cite{Coriano:2014gia}, developed along similar lines, to the neutrino case, presenting a numerical study of the complete one-loop corrections derived from the electroweak theory. 
The formalism uses a retarded graviton propagator with the effects of back reaction of the scattered beam on the source not included, as in a typical scattering problem by a static external potential.  
In this case, however, because of the presence of a horizon, we search for a lower bound on the size of the impact parameter of the collision where the classical GR prediction and the quantum one overlap. Indeed, above the bound the two descriptions are in complete agreement.
As already mentioned above, both in the fermion as in the photon case \cite{Coriano:2014gia}, this bound can be reasonably taken to lay around 20 ${b_h}$, which is quite close to the horizon of the classical source.\\
 For smaller values of $b_h\, (4< b_h <20)$, the two approaches are in disagreement, since the logarithmic singularity in the angle of deflection, once the beam gets close to the photon sphere, starts playing a significant role. This is expected, given the assumption of weak field for the gravitational coupling, which corresponds to the Newtonian approximation in the metric.\\
 The second part of our work deals with the implementation of the semiclassical deflection within the formalism of the classical lens equations. We use the energy dependence of the angular deflection to derive new lens equations, which are investigated numerically. We quantify the impact of these effects both in the thin lens approximation, where the trigonometric relations in the lens geometry are expanded to first order, and for a lens with deflection terms of 
 higher order included.
As an example, in this second case, we have chosen the 
Virbhadra-Ellis \cite{Virbhadra:2002ju} lens equation. The observables that we discuss are limited to solutions of these equations and to their magnifications, although time delays, shears and the light curves of a typical microlensing event can be easily included in this framework.  We anticipate that the effects that we quantify are small and cover the milliarcsecond region, remaining quite challenging to detect at experimental level. We hope though,  that the framework that we propose can draw further interest on this topic in the future, both at theoretical and at phenomenological level.

In the third part of our study we discuss the post Newtonian formulation of the impact parameter formalism, and apply it to the case of a compact source with a microscopic Schwarzschild radius. This is the only case in which the gravitational corrections to the Newtonian cross section can be consistently included in our approach in a meaningful way. We then summarize our analysis and discuss in the conclusions some possible future directions of possible extensions of our work.

 \section{Gravitational interaction of neutrinos}
\label{Sec.TheorFram}
We start our analysis with a brief discussion of the structure of the gravitational interaction of neutrinos, building on the results of \cite{Coriano:2012cr, Coriano:2013iba}, to which we refer for additional details, and that we are going to specialize to the case of a massless neutrino. An analysis of gravity with the fermion sector is contained in \cite{Degrassi:2008mw}.
We simply recall that the dynamics of the Standard Model in external gravity is described by the Lagrangian  
\beq S = S_G + S_{SM} + S_{I}= -\frac{1}{\kappa^2}\int d^4 x \sqrt{-{g}}\, R+ \int d^4 x
\sqrt{-{g}}\mathcal{L}_{SM} + \chi \int d^4 x \sqrt{-{g}}\, R \, H^\dag H.      \, 
\label{thelagrangian}
\eeq
This includes the Einstein term $\mathcal{S}_G$, the $\mathcal{S}_{SM}$ action and a term $\mathcal{S}_I$ involving the Higgs doublet $H$ \cite{Callan:1970ze}, called the term of improvement. $\mathcal{S}_{SM}$, instead, is obtained by extending the ordinary Lagrangian of the Standard Model to a curved metric background. The term $\chi$ is a parameter which, at this stage, is arbitrary and that at a special value $(\chi\equiv\chi_c=1/6)$ guarantees the renormalizability of the model at leading order in the expansion in $\mathcal{\kappa}$. \\ 
Deviations from the flat metric $\eta_{\mu\nu}=(+,-,-,-)$ will be parametrized in terms of the gravitational coupling $\kappa$, with $\kappa^2= 16 \pi G$ and with $G$ being the gravitational Newton's constant. At this order the metric is given as $g_{\mu\nu}=\eta_{\mu\nu} + \kappa h_{\mu\nu}$, with $h_{\mu\nu}$ describing its fluctuations. We will consider two spherically symmetric and static cases, corresponding to the Schwarzschild and Reissner-Nordstrom metrics. The first, in the weak field limit and in the isotropic form is given by

\beq
ds^2\approx\left(1- \frac{2 G M}{|\vec{x}|}\right)dt^2 -\left(1 + \frac{2 G M}{|\vec{x}|}\right)d\vec{x}\cdot d\vec{x}
\label{SCH3}.
\eeq
In this case the fluctuation tensor takes the form 
\bea
h_{\mu\nu}(x)
&=& \frac{2 G M}{\kappa |\vec{x}|}\bar{S}_{\mu\nu}, \qquad  \bar{S}_{\mu\nu}\equiv \eta_{\mu\nu}-2 \delta^0_{\mu}\delta^0_{\nu}.
\label{hh}
\eea
The inclusion of higher order terms in the weak field expansion will be discussed in the following sections. \\
The coupling of the gravitational fluctuations to the fields of the Standard Model involves the 
 EMT, which is defined as  
\beq
T_{\mu\nu}=\frac{2}{\sqrt{-g}}\frac{\delta \left(S_{SM}+S_{I}\right)}{\delta g^{\mu\nu}} \bigg|_{g=\eta} \,
\label{stmn}
\eeq 
with a tree-level coupling summarized by the action 
\beq
\mathcal{S}_{int}=-\frac{\kappa}{2}\int d^4 x \, T_{\mu\nu} h^{\mu\nu} \,,
\label{inter}
\eeq
where $T_{\mu \nu}$ is symmetric and covariantly conserved. The complete expression of the EMT of the Standard Model, including ghost and gauge-fixing contributions can be found in \cite{Coriano:2011zk}. \\
The Higgs field is parameterized in the form 
\beq
H = \left(\begin{array}{c} -i \phi^{+} \\ \frac{1}{\sqrt{2}}(v + h + i \phi) \end{array}\right)
\eeq
 in terms of $h$, $\phi$ and $\phi^{\pm}$, which denote the physical Higgs and the Goldstone bosons of the $Z$ and $W'$ s respectively. $v$ is the Higgs vacuum expectation value. The terms of the Lagrangian $\mathcal{S}_I$, generate an extra contribution to the EMT which is given by
\bea
\label{Timpr}
T^{\mu\nu}_I = - 2 \chi (\partial^\mu \partial^\nu - \eta^{\mu \nu} \Box) H^\dag H = - 2 \chi (\partial^\mu \partial^\nu - \eta^{\mu \nu} \Box) \left( \frac{h^2}{2} + \frac{\phi^2}{2} + \phi^+ \phi^- + v \, h\right) \, ,
\eea
the term of improvement, which can be multiplied by an arbitrary  constant ($\chi$). As mentioned above, it is mandatory to choose 
the value $\chi=1/6$ for any insertion of the EMT on the correlators of the Standard Model. These are found to be ultraviolet finite only if $T^{\mu\nu}_I$ is included \cite{Callan:1970ze,Coriano:2011zk,Freedman:1974gs}.\\
We will be dealing with the $T f \bar{f}$ vertex, where $T$ denotes the EMT and $f\equiv \nu_f$ a neutrino of flavour $f$, and work in the limit of zero mass of the neutrinos. The vertex, to lowest order, is obtained from the EMT of the neutrino. For instance, the explicit expression of the EMT for the (left-handed, $\nu\equiv \nu_L$) electron neutrino is given by
\begin{equation}
\begin{split}
T^{\nu^e}_{\mu\nu}
=& \frac{i}{4}\bigg\{\bar\nu^e\g_\mu\stackrel{\rightarrow}{\pd}_\nu\nu^e - \bar\nu^e\g_\mu\stackrel{\leftarrow}{\pd}_\nu\nu^e + \frac{2e}{\sin2\th_W}\bar\nu^e \g_\mu\frac{1-\g^5}{2}\nu^e Z_\nu \\
&- 2i\frac{e}{\sqrt{2}\sin\th_W}\bigg(\bar\nu^e \g_\mu \frac{1-\g^5}{2}e\,W^+_\nu
+ \bar e\g_\mu\frac{1-\g^5}{2}\nu^e\,W^-_\nu\bigg) \\
& + (\mu \leftrightarrow \nu) \biggr\} - \eta_{\mu\nu} \mathcal{L}_{\nu^e} \,,
\end{split}
\end{equation}
with 
\begin{equation}
\begin{split} 
\mathcal{L}_{\nu_e}
&= i\bar\nu^e\gamma^\mu\partial_\mu\nu^e + \frac{e}{\sin2\vartheta_{\text{W}}}\bar\nu^e \gamma^\mu\frac{1-\gamma^5}{2}\nu^e Z_\mu \\
&+ \frac{e}{\sqrt{2}\sin\vartheta_{\text{W}}}\bigg(\bar\nu^e \gamma^\mu \frac{1-\gamma^5}{2}e\, W^+_\mu
    + \bar e\gamma^\mu\frac{1-\gamma^5}{2}\nu^e\, W^-_\mu\bigg).\\
\end{split} 
\end{equation}
In momentum space, in the case of a massless fermion, the vertex takes the form
\beq
V^{(0) \mu\nu}=\frac{i}{4}\left( \gamma^\mu(p_1 +p_2)^\nu + \gamma^\nu(p_1 +p_2)^\mu -
2 \eta^{\mu\nu}(\slashed{p}_1+\slashed{p}_2)\right).
\eeq
while in the case of neutrinos we have
\beq
V_\nu^{(0)\mu\nu}=V^{(0)\mu\nu}\,P_L
\eeq
with $P_L=(1-\gamma_5)/2$ being the chiral projector.
We refer to appendix \ref{rules} for a list of the relevant Feynman rules necessary for the computation. \\
We will denote with
\beq
\hat{T}^{(0) \mu\nu}=\bar u(p_2)V^{(0) \mu\nu}u(p_1),
\eeq
the corresponding invariant amplitude, a notation that we will use also at one-loop level in the electroweak expansion.  
We introduce the two linear combinations of momenta  $p = p_1 + p_2$ and $q = p_1 - p_2$ to express our results. %
It has been shown that the general $Tf\bar{f}$ vertex, for any fermion $f$ of the Standard Model, decomposes into six different contributions  \cite{Coriano:2012cr}, but in the case of a massless neutrino only three amplitudes at one-loop level are left, denoted as 
\bea
\label{hatT}
\hat T^{\mu\nu} =\hat T^{\mu\nu}_{Z} + \hat T^{\mu\nu}_{W} + \hat T^{\mu\nu}_{CT}. 
\eea
In the expression above, the subscripts indicate the contributions mediated by virtual $Z$ and $W$ gauge bosons, while $CT$ indicates  the contribution from the counterterm.

\begin{figure}[t]
\centering
\subfigure[]{\includegraphics[scale=0.7]{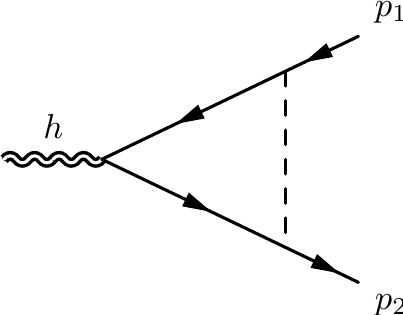}} \hspace{.5cm}
\subfigure[]{\includegraphics[scale=0.7]{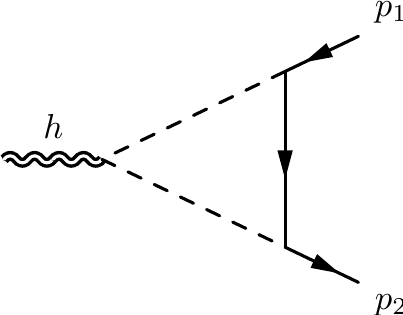}} \hspace{.5cm}
\subfigure[]{\includegraphics[scale=0.7]{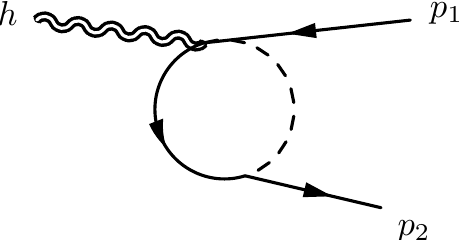}} \hspace{.5cm}
\subfigure[]{\includegraphics[scale=0.7]{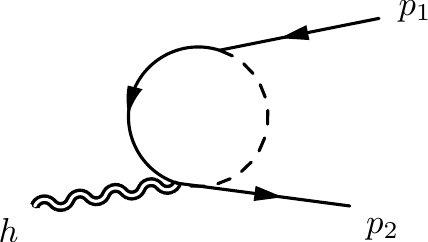}}
\caption{The one-loop Feynman diagrams of the neutrino vertex in a gravitational background. The dashed lines can be $Z$ and $W$.
\label{diagrams}}
\end{figure}

We show in Fig. \ref{diagrams} some of the typical topologies appearing in their perturbative expansion.\\
Two of them are characterized by a typical triangle topology, while the others denote terms where the insertion of the EMT and of the fermion field occur on the same point. The computation of these diagrams is rather involved and has been performed in dimensional regularization using the on-shell renormalization scheme. %
Neutrinos interactions, in the limit of massless neutrinos, involve only few of the structures of the $Tf\bar{f}$ tensor decomposition presented in \cite{Coriano:2012cr}. In this case we are left with only one tensor structure and hence only one form factor for each sector
\bea
\hat T^{\mu\nu}_Z &=& i \, \frac{G_F}{16 \pi^2 \sqrt{2}}   f^{Z}_1(q^2, m_Z) \, \bar u(p_2) \, O^{\mu\nu}_{C  1} \, u(p_1) \,, \nn \\ 
\hat T^{\mu\nu}_W &=& i \, \frac{G_F}{16 \pi^2 \sqrt{2}}  f^{W}_1(q^2, m_f, m_W) \, \bar u(p_2) \, O^{\mu\nu}_{C  1} \, u(p_1) \,, 
\label{expans}
\eea
where we have defined the vertex
\bea
\label{chiralbasis}
O^{\mu\nu}_{C  1} &=& \left( \gamma^\mu \, p^\nu + \gamma^\nu \, p^\mu \right) P_L.
\eea
The counterterms needed for the renormalization of the vertex can be obtained by promoting the counterterm Lagrangian of the Standard Model from a flat spacetime to the curved background, and then extracting the corresponding Feynman rules, as for the bare one. We obtain  
\bea
\label{TCT}
\hat T^{\mu \nu}_{CT}  = - \frac{i}{4}  \Sigma^L(0)\,\bar u(p_2)O^{\mu\nu}_{C 1}u(p_1),
\eea
where we have denoted with $\Sigma^L$ the neutrino self-energy
\bea
\Sigma^L(p^2) = \frac{G_F}{16 \pi^2 \sqrt{2}} \bigg[ \Sigma^L_Z (p^2) +  \Sigma^L_W(p^2) \bigg],
\eea
which is a combination of the self-energy contributions 
\bea
&&\Sigma^L_W (p^2) = - 4\bigg[ \left( m_f^2 + 2 m_W^2 \right) \mathcal B_1 \left( p^2, m_f^2, m_W^2 \right) + m_W^2 \bigg]\\
&& \Sigma^L_Z (p^2) = -2 m_Z^2\bigg[ 2 \, \mathcal B_1 \left( p^2, 0, m_Z^2 \right)  +1 \bigg]  \,,
\eea
with
\bea
\mathcal B_1 \left( p^2, m_0^2, m_1^2 \right) = \frac{m_1^2 -m_0^2}{2 p^2} \bigg[ \mathcal B_0(p^2, m_0^2, m_1^2) -  \mathcal B_0(0, m_0^2, m_1^2) \bigg] -\frac{1}{2} \mathcal B_0(p^2, m_0^2, m_1^2),
\eea
expressed in terms of the scalar form factor $\mathcal{B}_0$, given in appendix \ref{scalarint} together with all the other relevant scalar integrals. We have denoted with 
$m_Z $ and $m_W$ the masses of the $Z$ and $W$ gauge bosons; with $q^2$ the virtuality of the incoming momentum of the EMT and $m_f$ is the mass of the fermion of flavor $f$ running in the loops. \\
The explicit expressions of the form factors appearing in (\ref{expans}) is given by 
\bea
f^Z_1&=&-2\,m_Z^2-\frac{4\,m_Z^4}{3\,q^2}+\left(2+\frac{7\,m_Z^2}{3\,q^2}\right)\,\mathcal A_0(m_Z^2)-\left(\frac{17\,m_Z^2}{6}+\frac{7\,m_Z^4}{q^2}+\frac{4\,m_Z^6}{q^4}\right)\,\mathcal B_0(q^2, 0, 0)\nn\\
&&+\frac{2}{3\,q^4}\,m_Z^2(2\,m_Z^2+q^2)\,(3m_Z^2+2q^2)\,\mathcal B_0(q^2, m_Z^2, m_Z^2)\nn\\
&&-\frac{4}{q^4}\,m_Z^6\,(m_Z^2+q^2)\,\mathcal C_0(0, m_Z^2, m_Z^2)-\frac{1}{q^4}m_Z^2\,(m_Z^2+q^2)^2(4\,m_Z^2+q^2)\,\mathcal C_0(m_Z^2, 0, 0),
\eea
with $\mathcal C_0$ denoting the scalar 3-point function, and with the form factor $f^W_1$ related to the exchange of the $W$' s given by 
\begin{align}
f^W_1&=\frac{m_f^2}{2}-4\,m_W^2+\frac{4}{3\,q^2}\,(m_f^4+m_f^2\,m_W^2-2\,m_W^4)-\frac{1}{3\,q^2}(m_f^2+2\,m_W^2)\left(\mathcal A_0(m_f^2)-\mathcal A_0(m_W^2)\right)\nn\\
&-\frac{2}{q^2}\Big(m_f^4+m_f^2\,m_W^2-2m_W^2\,(m_W^2+q^2)\Big)\,\mathcal B_0(0, m_f^2, m_W^2)+\frac{1}{6\,q^4}\Big(-24\,m_f^6-10\,m_f^4\,q^2\nn\\
&+m_f^2\,(72\,m_W^4+46\,m_W^2\,q^2+q^4)-2\,m_W^2\,(24\,m_W^4+42\,m_W^2\,q^2+17\,q^4)\Big)\,\mathcal B_0(q^2, m_f^2, m_f^2)\nn\\
&+\frac{1}{3\,q^4}\Big(12\,m_f^6+12\,m_f^4\,q^2+4\,m_W^2\,(2\,m_W^2+q^2)(3\,m_W^2+2\,q^2)\nn\\
&+m_f^2\,(-36\,m_W^4-16\,m_W^2\,q^2+q^4)\Big)\,\mathcal B_0(q^2, m_W^2, m_W^2)+2\Big(m_f^4+\frac{2}{q^4}\,(m_f^2-m_W^2)^3\,(m_f^2+2\,m_W^2)\nn\\
&+\frac{1}{q^2}\left(3\,m_f^6-4\,m_f^4\,m_W^2+5\,m_f^2\,m_W^4+4\,m_W^6\right)\Big)\,\mathcal C_0(m_f^2, m_W^2, m_W^2)\nn\\
&+\frac{1}{q^2}\Big(4\,m_f^8+m_f^6\,(q^2-4\,m_W^2)-2\,m_W^2\,(m_W^2+q^2)^2\,(4\,m_W^2+q^2)\nn\\
&-m_f^4\,(2\,m_W^2+q^2)\,(6\,m_W^2+q^2)\Big)\,\mathcal C_0(m_W^2, m_f^2, m_f^2)\nn\\
&+\frac{m_f^2}{q^2}\,(20\,m_W^6+25\,m_W^4\,q^2+6\,m_W^2\,q^4)\,\mathcal C_0(m_W^2, m_f^2, m_f^2).
\end{align}
Being the computations rather involved, the correctness of the results above has been secured by appropriate Ward identities, whose general structure has been discussed in \cite{Coriano:2011zk}. 
As an example, by requiring the invariance of the generating functional of the theory under a diffeomorphic change of the spacetime metric, one derives the following Ward identity
\bea
\label{WI}
q_{\mu} \, \hat T^{\mu\nu}& =& \bar u(p_2) \bigg\{ p_2^{\nu} \,  \Gamma_{\bar f f}(p_1) -  p_1^{\nu} \, \Gamma_{\bar f f}(p_2)
 + \frac{q_\mu}{2} \left( \Gamma_{\bar f f}(p_2) \, \sigma^{\mu\nu} - \sigma^{\mu\nu} \, \Gamma_{\bar f f}(p_1) \right) \bigg\} u(p_1) \,,
\eea
where $ \Gamma_{\bar f f}(p)$ is the fermion two-point function, diagonal in flavor space \cite{Coriano:2012cr}.
From this equation one obtains 
\bea
0 &=&  f^Z_1 - \frac{1}{4} \Sigma^L_Z(0) \nn \\
0 &=&  f^W_1 - \frac{1}{4} \Sigma^L_W(0),
\eea
which, as one can check, are identically satisfied by the explicit expressions of $f_Z$ and $f_W$ given above. \\
In the case of MeV neutrinos, the expressions of the two form factors simplify considerably, since the typical momentum transfer $q^2=-4 E^2 \sin^2(\theta/2)$ may be small. These expansions, in fact, are useful in the case of scattering and lensing of neutrinos far from the region of the event horizon, of the order of $10^3-10^6$ horizon units.  As we are going to see, an expansion in $q^2$ provides approximate analytical expressions of the $b_h(\alpha)$ relation, connecting the impact parameter to the angle of deflection $\alpha$, 
valid at momentum transfers which are smaller compared to the electroweak scale, i.e.  $q^2/m_W^2 \ll1$. 
We will come back to illustrate this point more closely in the following sections. \\
In these cases the expression of the renormalized $f_Z$ form factor takes the form
\bea
f^{Z\,(ren)}_{low\,q}=-\frac{11}{18} q^2,
\eea
while the $W$ form factor is slightly lengthier
\bea
f^{W\,(ren)}_{low\,q}=& -  \dfrac{q^2}{36\,(m_f^2-m_W^2)^4}\Biggl[ 5\,m_f^8-98\,m_f^6m_W^2+243\,m_f^4m_W^4-194\,m_f^2m_W^6+44\,m_W^8  \nn\\ 
&+6\left(10\,m_f^6m_W^2-15\,m_f^4m_W^4+2\,m_f^2m_W^6 \right) \ln \left( m_f^2 / m_W^2\right) \Biggr]  .
\eea

\section{Cross Sections for photons, massive fermions and scalars}

\begin{figure}[t]
\centering
\includegraphics[width=0.65\textwidth]{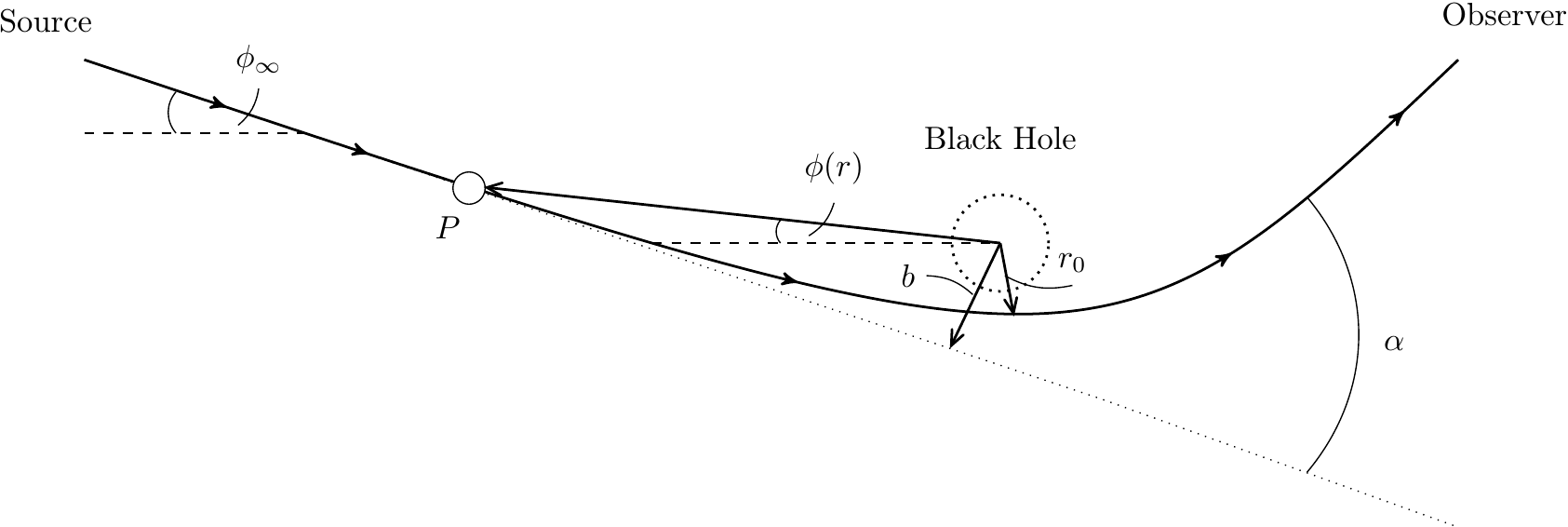}
\caption{The deflection of the trajectory of a massless particle $P$ approaching a black hole. }
\label{picx}
\end{figure}

\begin{figure}[t]
\centering
\subfigure[]{\includegraphics[scale=.7]{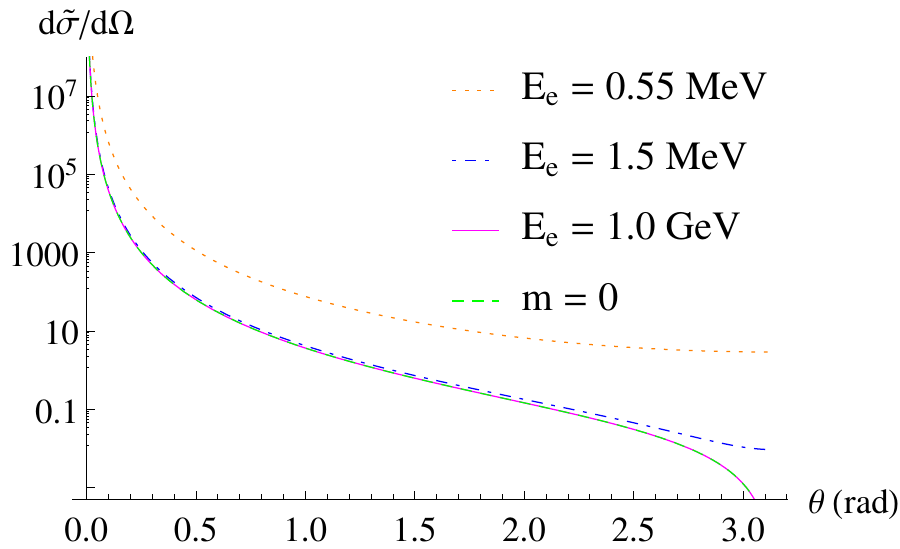}} \hspace{.5cm}
\subfigure[]{\includegraphics[scale=.7]{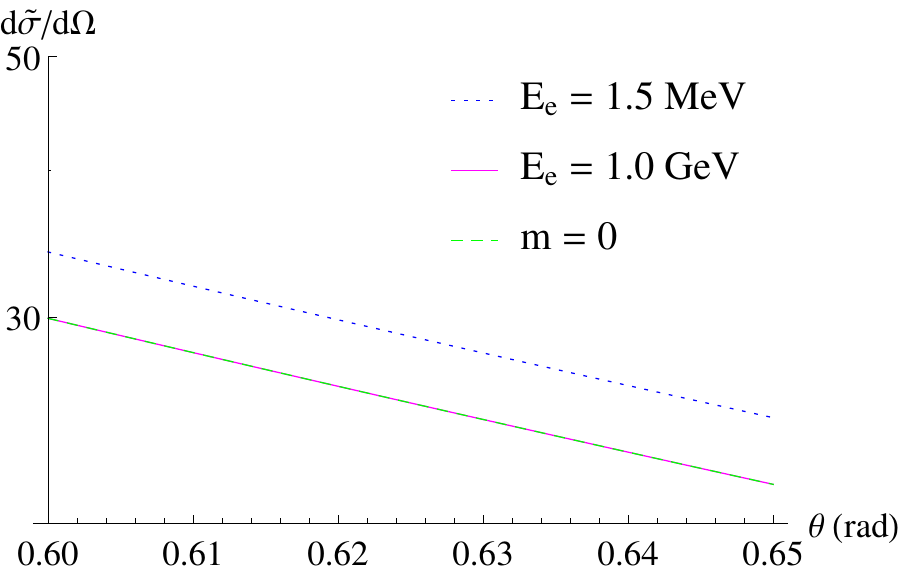}}
\subfigure[]{\includegraphics[scale=.7]{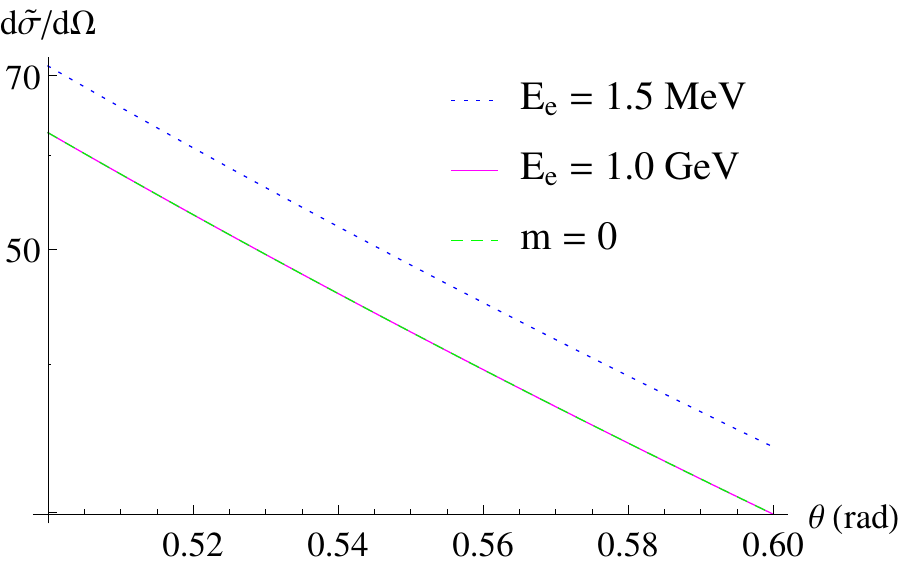}}
\subfigure[]{\includegraphics[scale=.5]{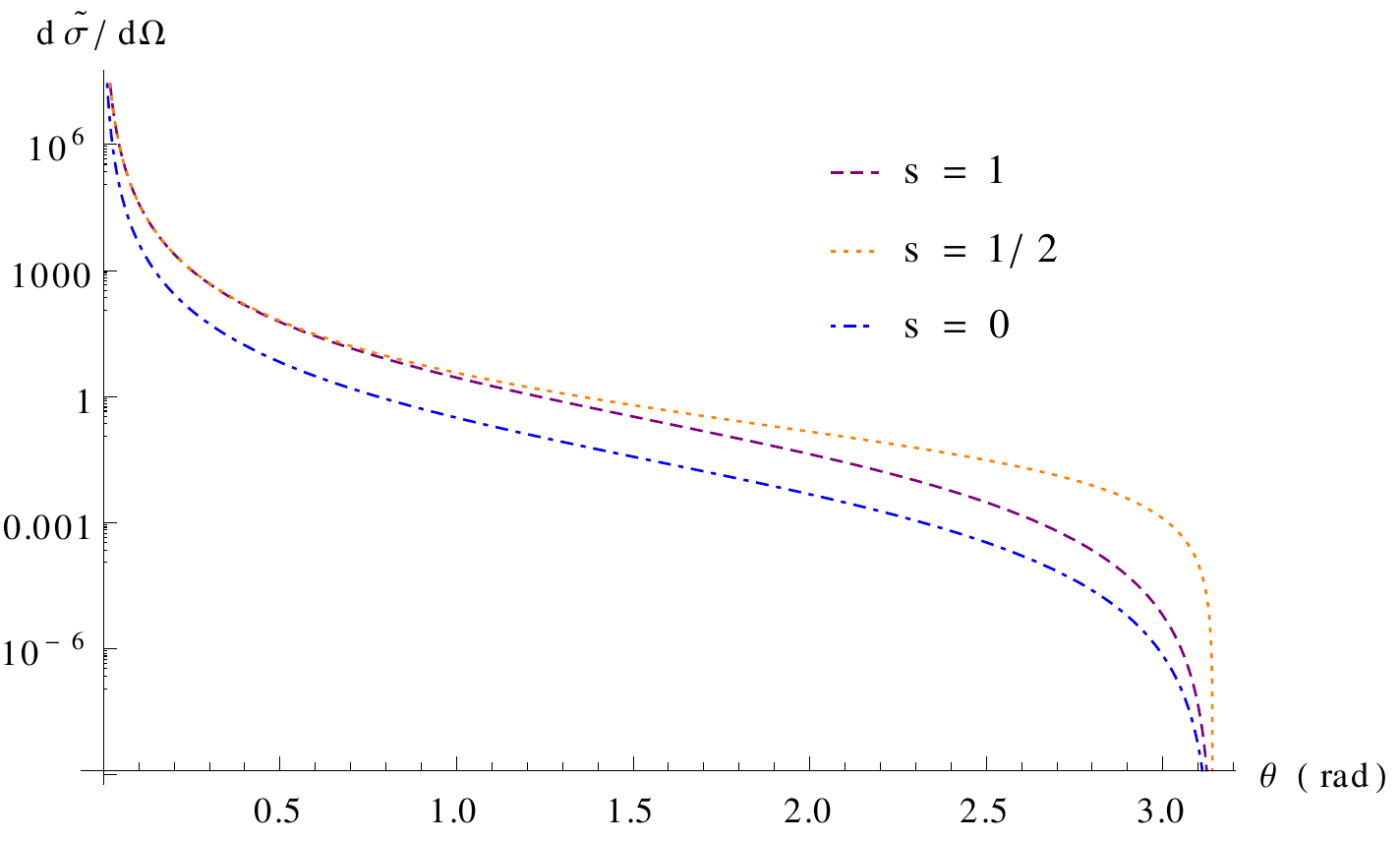}}
\caption{Normalized ($\tilde{\sigma}=\sigma/(2 G M)^2$) cross sections for massive and massless fermions. In the massive case $m$ is the electron mass (a). Two enlargements of (a) are in (b) and (c). Panel (d) 
shows the cross sections for photons ($s=1$), massless neutrinos ($s=1/2$) and conformally coupled scalars ($s=0$).  }
\label{fig:TreeLevel}
\end{figure}

Before coming to a discussion of the 1-loop effects in the scattering of neutrinos, we briefly summarize the result for the 
leading order cross sections for fermions, photons and scalars using in an external static background \cite{Accioly1} \cite{Coriano:2012cr,Coriano:2013iba}. We just recall that the scattering matrix element is written as 
\beq
i\mathcal{S}_{if}=-\frac{\kappa}{2}\int_ V d^4 x \langle p_2 | h_{\mu\nu}(x) T^{\mu\nu}(x)| p_1 \rangle 
\label{volume},
\eeq
where ${ V}$ is the integration volume where the scattering occurs, which gives
\bea
 \langle p_2 |h_{\mu\nu}(x) T^{\mu\nu}(x)| p_1 \rangle&=& h_{\mu\nu}(x) \bar{\psi}(p_2)V^{\mu\nu}\psi(p_1) e^{i q\cdot x}.
  \eea
 Denoting with $i$ and $f$ the initial and final neutrino, we have introduced plane waves normalized as 
\beq
\psi_{i}(p_{1})={\mathcal{N}_{i} }u(p_{1}), \qquad \mathcal{N}_{i}=\sqrt{\frac{1}{E_{1} V}}, \qquad \bar{u}(p_1)u(p_1)=1,
\eeq
and similarly for $\psi_f$,  while $V$ denotes a finite volume. The $E_1$ ($E_2$) are the energy of the incoming (outgoing) particle respectively.\\
In momentum space the matrix element is given by
 \bea
i \mathcal{S}_{fi}= -\frac{\kappa}{2} h_{\mu\nu}(q) \bar{\psi}(p_2)V^{\mu\nu}\psi(p_1)
=  -\frac{\kappa}{2}  h_{\mu\nu}(q)  \mathcal{N}_i\mathcal{N}_f \hat{T}^{\mu\nu}
\label{sfi}
\eea
in terms of the gravitational fluctuations in momentum space $h_{\mu\nu}(q)$. For a static external field the energies of the incoming/outgoing fermions are conserved ($E_1=E_2\equiv E$).  \\
The Fourier transform of $h_{\mu\nu} $ in momentum space is given by 
\bea
 h_{\mu\nu}(q_0,\vec{q})&=&\int d^4 x e^{i q\cdot x} h_{\mu\nu}(x),
 \eea
 which for a static field can be expressed as 
  \beq
 h_{\mu\nu}(q_0,\vec{q})=2 \pi \delta(q_0) h_{\mu\nu}(\vec{q} ),
 \label{h1}
 \eeq
 in terms of a single form factor $h_0(\vec{q})$
 \beq
 h_{\mu\nu}(\vec{q})\equiv h_0(\vec{q}) \bar{S}_{\mu\nu} \qquad \textrm{with}\qquad  h_0(\vec{q})\equiv \left(\frac{\kappa M}{2 \vec{q}^2}\right).
 \label{h2}
 \eeq
The squared matrix element in each case takes the general form
\begin{equation}
\label{eq:sfi}
\left|iS_{fi}\right|^2=\frac{\kappa^2}{16 V^2 E_1 E_2}\, 2\pi \delta(q_0)\, \,\mathcal{T} \,\frac{1}{2} \, \mathcal{J}^{\mu\nu\rho\sigma}(p_1,p_2)  \,h_{\mu\nu}(\vec{q}\,) \,h_{\rho\sigma}(\vec{q}\,), \,
  \end{equation} 
where $\mathcal{T}$ is the transition time. Specifically, in the case of a massive (Dirac) fermion one obtains 
\begin{equation}
\mathcal{J}^{\mu\nu\rho\sigma}_f(p_1,p_2)= \text{tr} \left[ (\slashed{p}_2+m) V_m^{\mu\nu}(p_1,p_2)( \slashed{p}_1+m) V_m^{\rho\sigma}(p_1,p_2) \right] \,,
\end{equation}
where the $V_m^{\mu\nu}$ vertex is in this case given by
\begin{equation}
\label{eq:hff}
V_m^{\mu\nu}(p_1,p_2)=\frac{i}{4} \Bigl( \gamma^{\mu} (p_1+p_2)^{\nu} + \gamma^{\nu}(p_1+p_2)^{\mu} - 2\eta^{\mu\nu} (\slashed{p}_1+\slashed{p}_2 - 2 m )\Bigr) \,
\end{equation}
which gives a cross section 
\begin{equation}
\label{eq:crosssecFerm}
\left.\frac{d \sigma}{d \Omega}\right|^{(0)}_f= \Biggl( \frac{G M}{\sin^2 (\theta/2)} \Biggr)^{\!2} \left( \cos^2\vartheta/2 + \frac{1}{4} \frac{m^2}{|\vec{p}_1|^2} + \frac{1}{4} \frac{m^4}{|\vec{p}_1|^4} + \frac{3}{4} \frac{m^2}{|\vec{p}_1|^2} \cos^2\vartheta/2      \right)\,.
\end{equation}
In the case of a neutrino, the corresponding cross section is obtained by sending the fermion mass $m$ of the related Dirac cross section to zero, giving  
\beq
\left.\frac{d \sigma}{d \Omega}\right|_\nu^{(0)} =\left(\frac{G M }{\sin^2\frac{\theta}{2}}\right)^2\cos^2\frac{\theta}{2}
\label{leading},
\eeq
which is energy independent. Notice that the inclusion of the chiral projector $P_L$ in the expression of the neutrino amplitude, which carries a factor $1/2$, makes the neutrino and Dirac cross sections coincide. The same $1/2$ factor, in the Dirac case, appears in the average over the two states of helicity, while the axial-vector terms induced by $P_L$ are trivially zero (see \cite{ChangCorianoGordon} for typical studies of polarized processes).\\
In the photon case one obtains
\begin{equation}
  \mathcal{J}^{\alpha\beta\rho\sigma}_{\gamma}(k_1,k_2)= \sum_{\lambda_1,\lambda_2} V^{\alpha\beta\kappa\lambda} (k_1,k_2) e_{\kappa}(k_1,\lambda_1) e_{\lambda}^{*}(k_2,\lambda_2) V^{\rho\sigma\mu\nu}(k_1,k_2)  e_{\mu} (k_2,\lambda_2) e^{*}_{\nu}(k_1,\lambda_1)\,,
  \end{equation}
  where $e_{\mu}$ denotes the polarization vector of the photon, with an interaction vertex which is given by
\begin{equation}
\label{eq:hAA}  
V^{\mu\nu\alpha\beta}(k_1,k_2)=i \bigg\{ \left( k_1 \cdot k_2 \right) C^{\mu\nu\alpha\beta}
+ D^{\mu\nu\alpha\beta}(k_1,k_2) \bigg\}\,,
\end{equation}
where
\begin{gather*}
 C_{\mu\nu\rho\sigma} = \eta_{\mu\rho}\, \eta_{\nu\sigma} +\eta_{\mu\sigma} \, \eta_{\nu\rho} -\eta_{\mu\nu} \, \eta_{\rho\sigma} \,,   \\
 D_{\mu\nu\rho\sigma} (k_1, k_2) = \eta_{\mu\nu} \, k_{1 \, \sigma}\, k_{2 \, \rho} - \biggl[\eta^{\mu\sigma} k_1^{\nu} k_2^{\rho} + \eta_{\mu\rho} \, k_{1 \, \sigma} \, k_{2 \, \nu}
  - \eta_{\rho\sigma} \, k_{1 \, \mu} \, k_{2 \, \nu}  + (\mu\leftrightarrow\nu)\biggr] \,.
\end{gather*} 
The cross section for a photon is then given by 
\begin{equation}
\label{eq:crsechVV}
\left.\frac{d \sigma}{d \Omega}\right|^{(0)}_\gamma= (G M)^2\cot^4 (\theta/2) \,.
\end{equation}
Finally, in the case of a scalar the relative expression is given by
\begin{equation}
  \mathcal{J}^{\alpha\beta\rho\sigma}_{s}(p_1,p_2)=V_s^{\alpha\beta}(p_1,p_2)V_s^{\rho\sigma}(p_1,p_2)\,,
  \end{equation} 
  with 
\begin{equation}
V^{\mu\nu}_s=-i\left\{ p_{1\,\rho} p_{2\,\sigma}  C^{\mu\nu\rho\sigma} - 2\chi \left[ \left( p_1+p_2 \right)^{\mu} \left( p_1+p_2 \right)^{\nu} - \eta^{\mu\nu} (p_1+p_2)^2 \right]  \right\} \,,
\end{equation}
where we have included the minimal and the term of improvement \cite{Coriano:2011zk}. For a conformally coupled scalar $\chi=1/6$.
The cross sections, in this case, are given by
\begin{equation}
\label{eq:crsechSS}
\left.\frac{d \sigma}{d \Omega}\right|^{(0)}_s= \left\{
\begin{array}{l}
(GM)^2\csc^4 (\theta/2)\qquad\chi=0\\
\left(\frac{ G M}{3}\right)^2\cot^4 (\theta/2)\qquad\chi=1/6
\end{array}
\right. \\
\end{equation}
We show in Fig. \ref{fig:TreeLevel} the expressions of these three cross sections at different energies, normalized by $1/(2 GM)^2$ and denoted as $\tilde{\sigma}$. In panel (a) we consider the scattering of a massive fermion, together with the massless limit, which applies in the neutrino case. We have included in (b) and (c) 
two enlargements of (a) which show how the massive and the massless cross sections tend to overlap for energies of the order of 1 GeV. In panel (d)  we show the cross sections for the photon ($s=1$), for the neutrino ($s=1/2$) and for the conformally coupled scalar
($s=0$). 

\subsection{The neutrino cross section at 1-loop}
\begin{figure}[t]
\centering
\subfigure[]{\includegraphics[scale=.73]{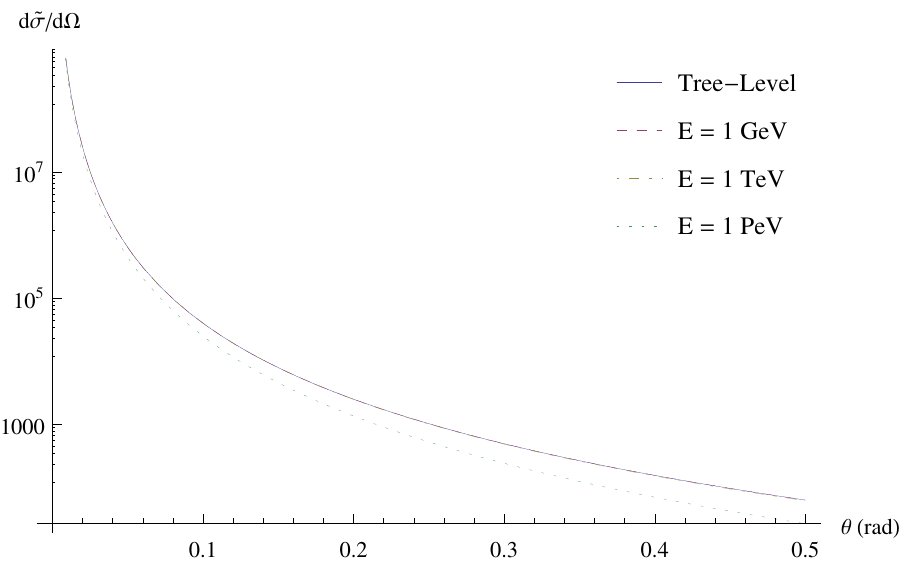}} \hspace{.5cm}
\subfigure[]{\includegraphics[scale=.5]{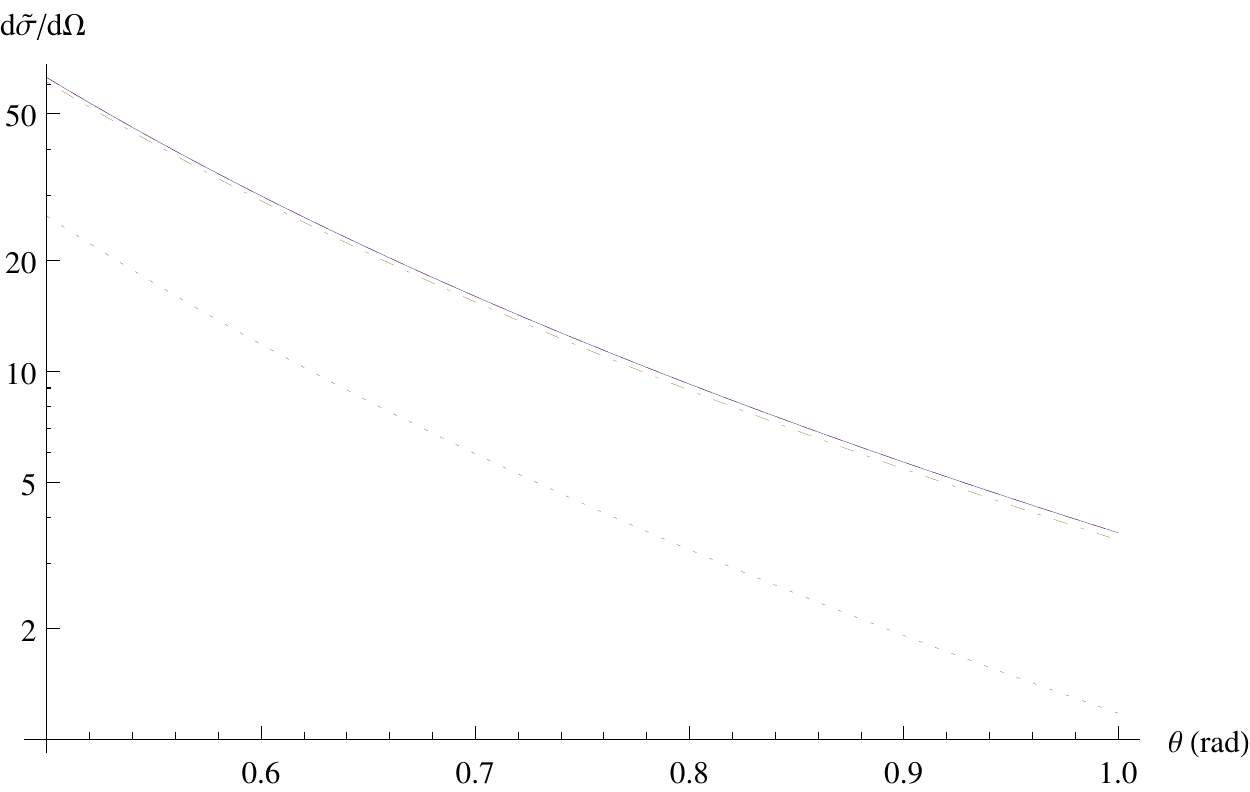}}
\caption{Differential cross section: tree level and one-loop contribution for a wide range of energies.\label{energies}}
\label{diff2}
\end{figure}

In the neutrino case, at 1-loop level, Eq. (\ref{leading}) is modified in the form 
\beq
\label{sigmaOL}
\frac{d \sigma}{d \Omega}=G^2M^2\frac{\cos^2\theta/2}{\sin^4\theta/2}\left\{1+\frac{4\,G_F}{16\,\pi^2 \sqrt{2}}  \left[ \, f_W^1(E,\theta) + f_Z^1(E,\theta) - \frac{1}{4} \Sigma_Z^L - \frac{1}{4} \Sigma_W^L\right] \right\},
\eeq
whose explicit expression has been given in appendix \ref{fcs}. In the massless approximation for the neutrino masses, loop corrections do not induce flavor transition vertices, such as those computed in  \cite{Coriano:2013iba}.\\
\begin{figure}[t]
\centering
\subfigure[]{\includegraphics[scale=.7]{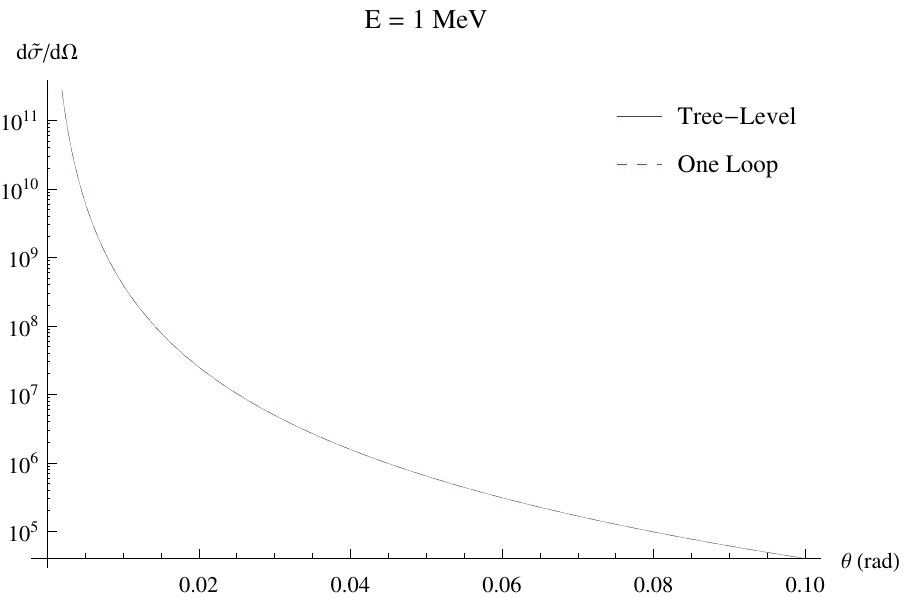}} \hspace{.5cm}
\subfigure[]{\includegraphics[scale=.5]{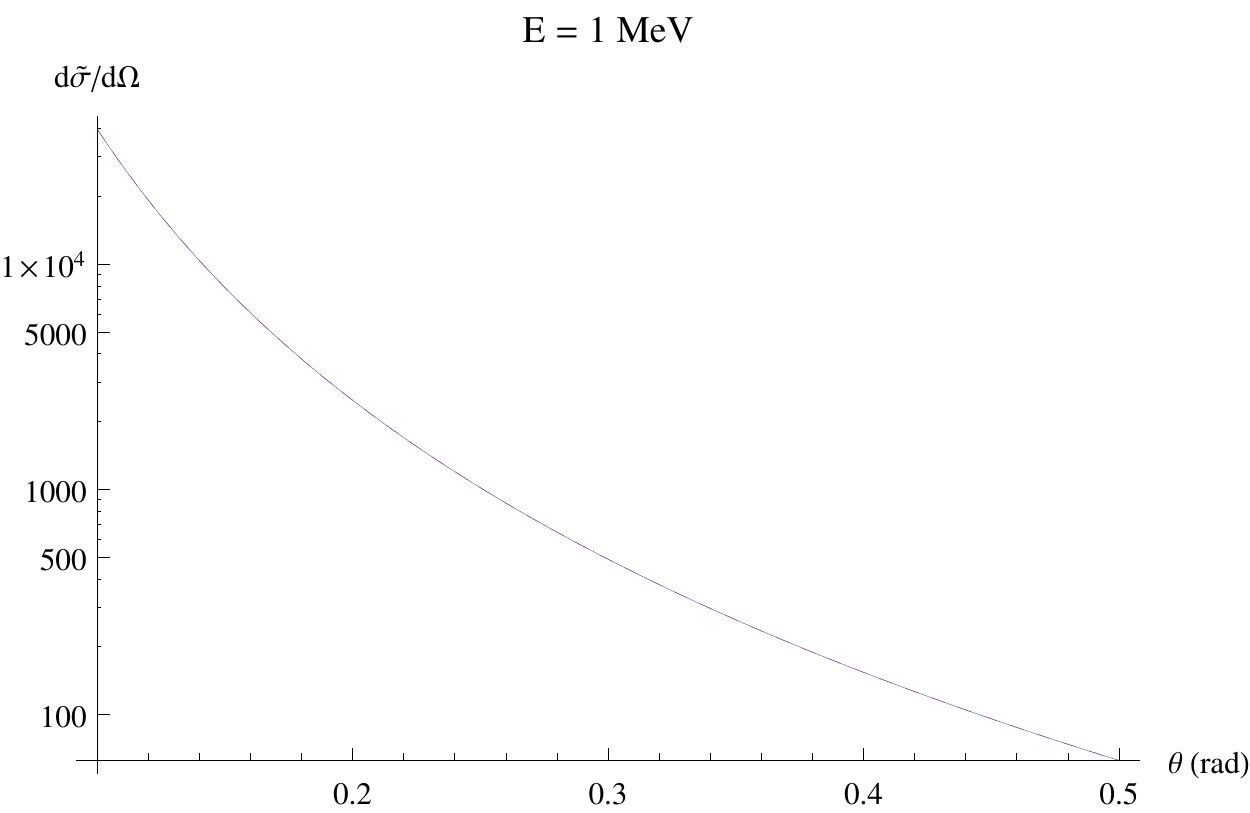}} \hspace{.5cm}
\subfigure[]{\includegraphics[scale=.5]{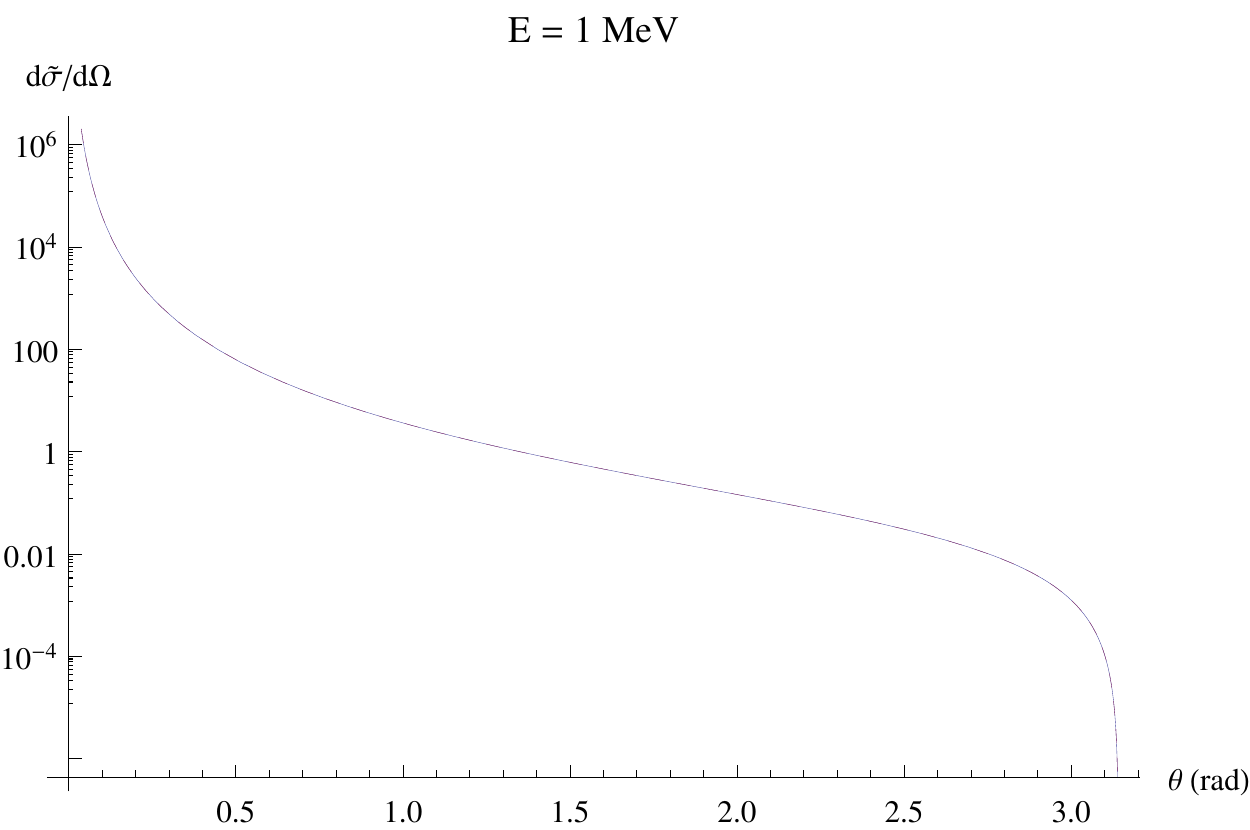}}
\caption{Differential cross section for MeV neutrinos in units of $r_s^2$, with $r_s$ the Schwarzschild radius.\label{OLMeV}}
\label{diff1}
\end{figure}
In the case of neutrinos of an energy $E$ in the MeV range, the expression above simplifies considerably and takes the form
\begin{align}
\frac{d\sigma}{d\Omega}=&\,G^2 M^2 \frac{\cos^2\theta/2}{\sin^4\theta/2} \Biggl\{ 1 + \frac{G_F}{\pi^2 \sqrt{2}} \Biggl[  \frac{11}{18}  +  \frac{1}{36\,(m_f^2-m_W^2)^4} \Biggl(  5\,m_f^8-98\,m_f^6m_W^2+243\,m_f^4m_W^4 \nonumber\\ 
 &  -194\,m_f^2m_W^6  +44\,m_W^8  +\,\, 6\, \Bigl( 10\,m_f^6m_W^2-15\,m_f^4m_W^4+2\,m_f^2m_W^6 \Bigr) \ln \dfrac{m_f^2}{m_W^2} \Biggr)  \Biggr] E^2 \sin^2\frac{\theta}{2}  \Biggr\} .
\end{align}
We show in Fig. \ref{diff1} three plots of the tree level and one-loop cross sections for an energy of the incoming neutrino beam of 1 MeV, for 2 different angular regions (plots $(a)$ and $(b)$), together with a global plot of the entire cross section (plot ($c$)) for the rescaled differential cross section $d\tilde{\sigma}/d\Omega\equiv 1/r_s^2\,\, d\sigma/d\Omega$. Notice that the tree-level and one-loop results are superimposed. We can resolve the differences between the two by zooming-in in some specific angular regions of the two results, varying the energy of the incoming beam. The result of this analysis is shown in Fig. \ref{diff2}, where in plots $(a)$ and $(b)$ we show the rescaled cross section $d\tilde{\sigma}/d\Omega$ as a function of the scattering angle 
$\theta$, for three values of the incoming neutrino beam equal to $1$ GeV,  $1$ TeV and $1$ PeV.  PeV neutrinos events are 
rare, due to the almost structureless cosmic ray spectrum, which falls dramatically with energy. They could be 
produced, though, as secondaries from the decays of primary protons of energy around the GZK
\cite{Greisen:1966jv, Zatsepin:1966jv} cutoff, and as such they are part of our analysis, which we try to keep as general as possible. \\ 
It is clear from these two plots that the tree-level and the one-loop result are superimposed at low energies, with a difference which becomes slightly more remarked at higher energies.
A similar behaviour is noticed in the cross section for scatterings at larger angles. Also in this case the radiative corrections tend to raise as the energy of the incoming beam increases. This  behaviour is expected to affect the size of 
the angle of deflection $\alpha$ as we approach the singular region of a black hole. In fact, $\alpha$ is obtained by integrating the semiclassical equation (\ref{semic}), introduced below, and large deviations are expected as the impact parameter $b_h$ reaches the photon sphere. As we are going to illustrate in the next sections, the $b_h(E, \alpha)$ relation is significantly affected by the behaviour of the cross section at large $\theta$ as $b_h\to 3/2 r_s$. This is the closest radial distance allowed to a particle approaching the black hole from infinite distance without being trapped. Therefore, these differences in $\tilde{\sigma}$ for large $\theta$ are going to render $b_h$ sensitive on the changes in energy of the neutrino beam for such close encounters of the neutrinos with a black hole.

\section{Impact parameter formulation of the semiclassical scattering} 
As pointed out in previous studies \cite{Coriano:2014gia,Delbourgo:1973xe,Coriano:2013iba, Berends:1975ah}, the computation of the angle of deflection for a fermion or a photon involves a simple semiclassical analysis, in which one introduces the 
impact parameter representation of the specific classical cross section and equates it to the quantum one. The classical/semiclassical scattering process is illustrated in Fig.~\ref{picx}, with $\alpha$ denoting the angle of deflection.
By assuming that the incoming particle is moving along the $z$ direction, with the source localized at the origin, and denoting with $\theta$ the azimuthal scattering angle present in the quantum cross section, we have the relation
\beq
\frac{b}{\sin\theta}\vline\frac{d b}{d\theta}\vline=\frac{d \sigma}{d\Omega}
\label{semic}
\eeq
between the impact parameter $b$ and $\theta$, as measured from the $z$-direction. This semiclassical equation \cite{Delbourgo:1973xe, Berends:1975ah} allows to relate the quantum and the classical features of the interaction between the particle beam and the gravitational source. The explicit expression of $b(\alpha)$, at least for small deflection angles, which correspond to large values of the impact parameter, can be found either analytically, such as at Born level and, for small momentum transfers also at one-loop, but it has to be obtained numerically otherwise.     
The solution of (\ref{semic}) takes the general form 
\beq
b_h^2({\alpha})=b_h^2(\bar{\theta}) +2\int_{\alpha}^{\bar{\theta}} d\theta' \sin\theta' \frac{d \tilde\sigma}{d\Omega'}, 
\label{intg}
\eeq
with $b_h^2(\bar{\theta})$ denoting the constant of integration. The semiclassical scattering angle $\alpha$ is obtained from (\ref{intg}) as a boundary value of the integral in $\theta$ of the quantum cross section. As discussed in \cite{Coriano:2014gia}, the integration constant derived from (\ref{intg}) has to be set to zero (for $\bar\theta=\pi$) in order for the solution of (\ref{semic}) to match the classical GR result for a very large $b_h$.

In the case of a point-like gravitational source and of neutrino deflection, one obtains from (\ref{intg}) the differential equation
\bea
\frac{d b^2}{d\theta}&=&- 2  \left(\frac{G M }{\sin^2\frac{\theta}{2}}\right)^2\cos^2\frac{\theta}{2}\, \sin\theta.
\label{semic1}
\eea

\begin{figure}[t]
\centering
\subfigure[]{\includegraphics[scale=0.9]{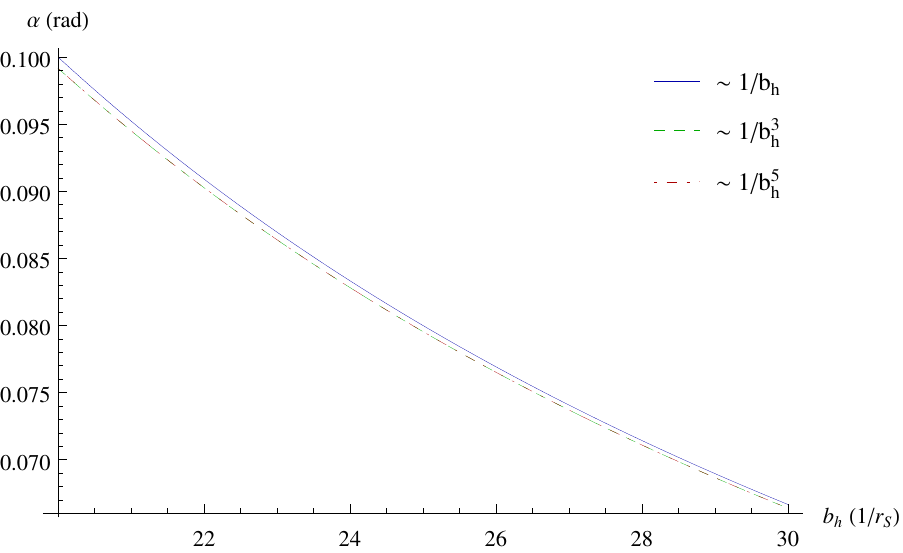}}
\caption{Plot of $\alpha$, the angle of deflection, versus the impact parameter $b_h$ for $20<b_h<100$, having inverted the $b(\alpha)$ solution of (\ref{semic}) at various $1/b^n$ orders. \label{alphathetainv}}
\end{figure}

\begin{figure}[t]
\centering
\subfigure[]{\includegraphics[scale=.8]{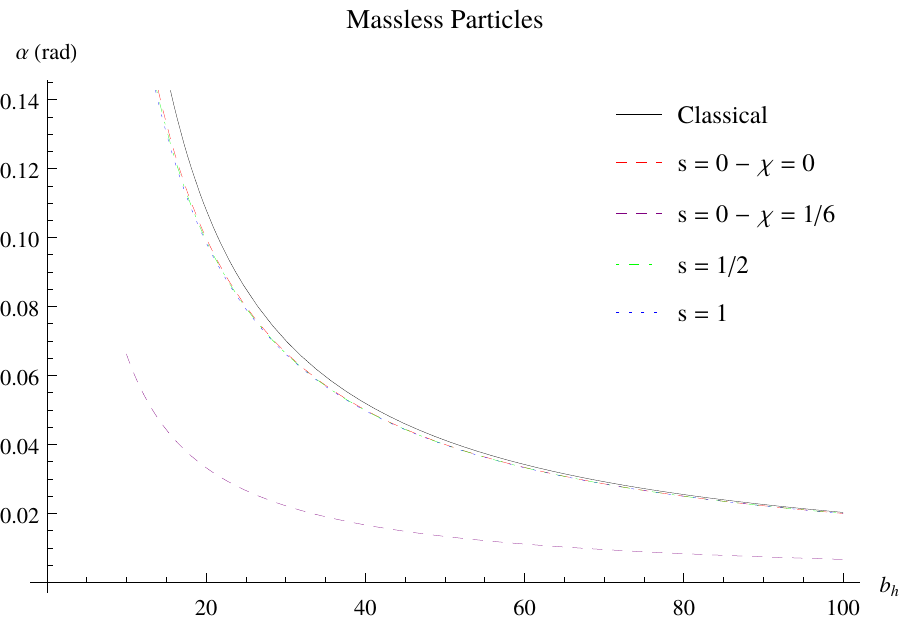}} \hspace{.5cm}
\subfigure[]{\includegraphics[scale=.8]{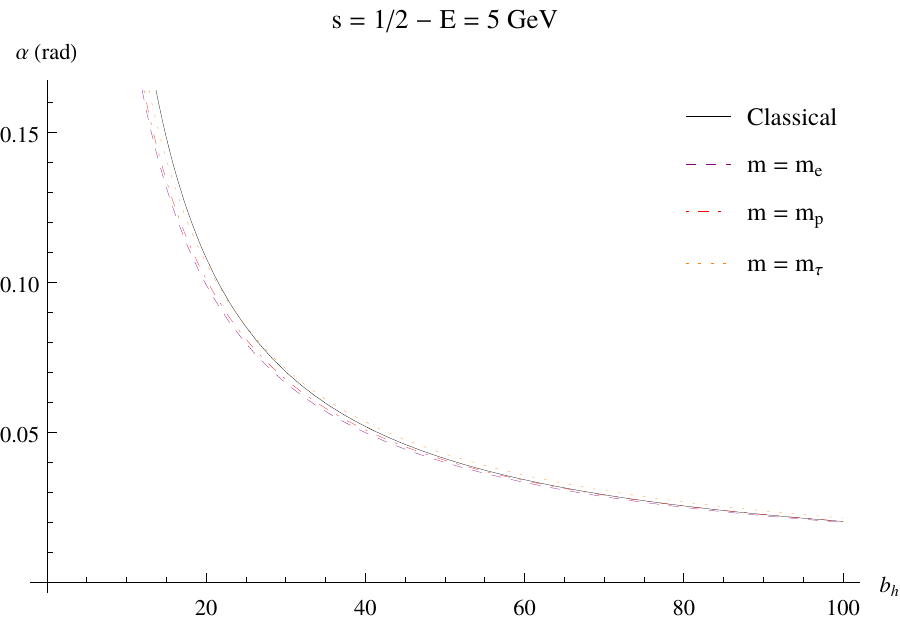}}
\caption{Angle of deflection as a function of the impact parameter: (a) spin dependence; (b) mass dependence.  }
\label{spinmass}
\end{figure}

Notice that the variation of $b$ with the scattering angle $\theta$ is negative, since the impact parameter decreases as $\theta$ grows, as we approach the center of the massive source. A comparison of this expression with the analogous relation in the photon case $(\gamma)$ shows that the two equations differ by a simple prefactor 
\bea
\frac{d b^2}{d\theta} =\frac{1}{\cos^2\frac{\theta}{2}}\frac{d b^2}{d\theta}\Big|_\gamma \qquad  \textrm{with} \qquad
\frac{d b^2}{d\theta}\Big|_\gamma = - 2 \,G^2 M^2\,\cot^4\frac{\theta}{2}\, \sin\theta.
\eea
The solution of (\ref{semic1}) takes the form 
\beq
\label{classb}
b^2(\alpha)=4\,G^2M^2\left(-1+\csc^2\frac{\alpha}{2}+2\ln\left(\sin\frac{\alpha}{2}\right)\right),
\eeq
and in the small $\alpha$ (i.e. large $b$) limit takes the asymptotic form  
\beq
b = G M\left(\frac{4}{\alpha} +\frac{\alpha}{3}(1+\ln\,8-3\ln\alpha)\right) + {\cal O}(\alpha^2)
\label{blocal}
\eeq
which allows us to identify the deflection angle as 
\beq
\alpha\sim 4 \frac{G M}{b}
\label{impact}
\eeq
in agreement with Einstein's prediction for the angular deflection.
This is the result expected from the classical (GR) analysis. The inversion of the asymptotic expansion (\ref{blocal}) generates the asymptotic behaviour
\bea
\alpha=\frac{2}{b_h} -\frac{2}{b_h^3}(\ln b_h +\frac{1}{3})+\frac{3}{b_h^5}(\ln^2 b_h-\frac{1}{5})+ \mathcal{O}(1/b_h^7)
\label{inv1}
\eea
which corresponds to the general functional form 
\beq
\label{genex}
\alpha= \frac{2}{b_{h}} + \sum_{k\geq1} \frac{a_{2k}}{b_{h}^{2k}} + \sum_{k\geq1} \frac{1}{b_h^{2k+1}} \left( a_{2k+1} + d_1 \ln b_h + d_2 \ln^2 b_h + \cdots +d_k\ln^k b_h \right) \,.       
\eeq
As shown in Fig.~\ref{alphathetainv}, the analytic inversion of (\ref{blocal}), given by (\ref{inv1}), is very stable under an increase of the order of the asymptotic expansion over a pretty large interval of $b_h$, from low to very high values. Solutions (\ref{inv1}) and (\ref{genex}) can be obtained by an iterative (fixed point) procedure, 
which generates a sequence of approximations $\alpha_0\to\alpha_1\to\ldots\to\alpha_n$ to $\alpha(b_h$) implemented after a Laurent expansion of (\ref{blocal}) and the use of the initial condition $\alpha_0=2/b_h$. The approach can be implemented also at one-loop and with the inclusion of the post-Newtonian corrections, if necessary.

The logarithmic corrections present in (\ref{genex}) are a genuine result of the quantum approach and, as we are going to discuss below, are not present in the classical formula for the deflection. Radiative and post-Newtonian effects, not included in (\ref{inv1}), give an expression for 
$\alpha(b_h)$ which coincides with the form (\ref{genex}), with specific coefficients $(a_n, d_n)$ which are energy dependent. This is at the origin of the phenomenon of light dispersion (gravitational rainbow) induced by the quantum corrections, which is absent at classical level \cite{Accioly1}. 

Eq. (\ref{genex}) will play a key role in our proposal for the inclusion of the radiative corrections in the classical lens equation. Such equation will relate the angular position of the source in the absence of lensing, $\beta$, to $\alpha(b)$.\\
We give, for completeness, the analogous expressions in the case of the scalar and for a massive fermion. For a massless scalar we have the relation 
\bea
\alpha = \frac{2}{3\,b_h}-\frac{1}{b^3_h}\left(\frac{12\,\ln 3 - 1}{243}+\frac{4}{81}\ln b_h\right)+\mathcal O (1/b_h^5),
\eea
while for a massive fermion the corresponding expression becomes more involved and takes the form
\begin{align}\label{massf}
\alpha &= \frac{8\,E^4}{4\,E^4-2\,E^2m_f^2+m_f^4}\frac{1}{b_h}-\frac{1}{b_h^3}\left[\frac{8\,E^4}{3(2\,E^2-m_f^2)(4\,E^4-2\,E^2m_f^2+m_f^4)^2}\times \right.\nn\\
&\left.\times \left(m_f^6+8\,E^6(1+\ln 8)+E^4m_f^2\ln 64 -6E^4(4\,E^2+m^2)\ln \frac{2}{1-\frac{m_f^2}{2\,E^2}}\right)\right.\nn\\
&\left.+\frac{4\,E^4(4\,E^2+m_f^2)}{8\,E^6-8\,E^4m_f^2+4\,E^2m_f^4-m_f^6}\ln b_h\right]
+\mathcal O(1/b_h^5),
\end{align}
where $E$ and $m_f$ are the energy and the mass of the fermion respectively.  One can easily check that in the limit $E\gg m_f$ Eq.~(\ref{massf}) reproduce the formula for the massless fermion (neutrino). We have plotted the behaviour of 
the formulas for the deflection in Fig. \ref{spinmass}. As one can immediately notice from the plots presented, the angular deflection is much less enhanced in the scalar case compared to the remaining cases, showing a systematic difference respect to the classical prediction form Einstein's deflection integral (\ref{alpha1}). The angular deflection in the scalar case is significantly affected by the choice of $\chi$ the free coupling factor of a scalar field to the external curvature $R$. We have chosen in panel (a) the two cases of $\chi=0$ (minimal coupling) and of conformal coupling $(\chi=1/6)$ as typical examples.

\subsection{Bending at 1-loop}
Moving to the one-loop expression given in (\ref{sigmaOL}), we can derive an analytic solution of the corresponding semiclassical equation (\ref{semic}) for $b=b(E,\alpha)$, in the limit of small momentum transfers. For this reason we perform an expansion of (\ref{sigmaOL}) in $q^2/m_W^2$ up to $\mathcal{O}((q^2/m_W^2)^2)$ and solve (\ref{semic}) in this approximation for $b_h^2(E, \alpha)$, obtaining
\bea
\label{OLb}
b_h^2(E, \alpha)&=&\left[-1+\csc^2\frac{\alpha}{2}+2\ln\left(\sin\frac{\alpha}{2}\right)\right]+C_1(E)\left[1+\cos\alpha+4\ln\left(\sin
\frac{\alpha}{2}\right)\right]+C_2(E)\cos^4 \frac{\alpha}{2}\nn\\
&&+20\,D_2(E)\ln\left(\sin\frac{\alpha}{2}\right)-4\,F_2(E)\cos \alpha-8\,D_2(E)\cos \alpha\ln\left(\sin^2\frac{\alpha}{2}\right)-G_2(E)\cos 2\alpha\nn\\
&&-2\,D_2(E)\cos 2\alpha\ln\left(\sin^2\frac{\alpha}{2}\right)-E_2(E), 
\eea
with the coefficients $C,D$, $F$ and $G$ are functions of the energy and of the masses of the weak gauge bosons. Their explicit expressions can be found in appendix \ref{expansion}.
The impact parameter $b_h(\alpha)$, as shown in the same appendix, has a dependence on the angular deflection $\alpha$ which can be summarized by an expression of the form 
\bea
\label{btheta}
b_h(E, \alpha) &=& \frac{2}{\alpha}+c(E)\,\alpha+d(E)\,\alpha\,\ln(\alpha)+f(E)\,\alpha^3+g(E)\,\alpha^3\ln \alpha+h(E)\,\alpha^3\ln^2 \alpha + \mathcal{O}(\alpha^5)\nn\\
\eea
that we can invert in order to get $\alpha(E, b_h)$. This is given by
\beqa
\alpha(E, b_h)&=&\frac{2}{b_h}-\frac{1}{b_h^3}\Big[\big(2+4\,C_1(E)\big)\log b_h + \mathcal{A}(E)\Big] + \mathcal{O}(1/b_h^5)\nonumber \\
\mathcal{A}(E) &=& -2\,C_1(E)-C_2(E)+E_2(E)+4F_2(E)+G_2(E)+\frac{2}{3} .
\eea

\begin{figure}[t]
\centering
\subfigure[]{\includegraphics[scale=.6]{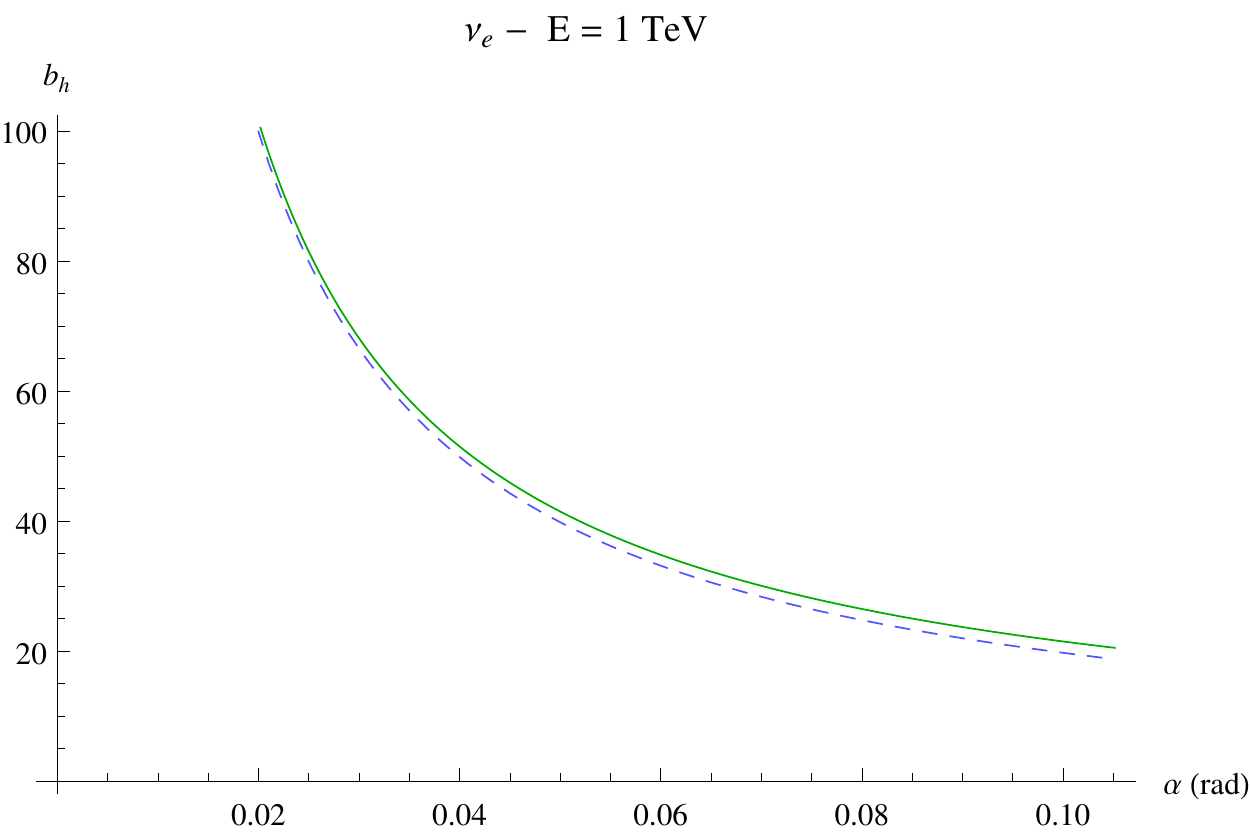}} \hspace{.5cm}
\subfigure[]{\includegraphics[scale=.82]{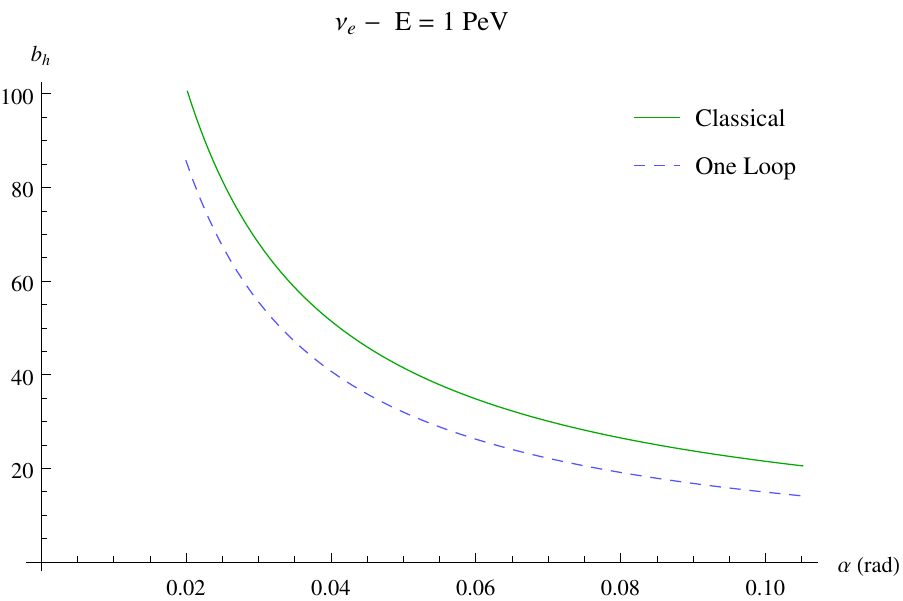}}
\caption{Plots of  the impact parameter $b_h$ versus $\alpha$, the angle of deflection, for $20<b_h<100$ for the classical and quantum solution. \label{bthetamev1}}
\end{figure}

We show in Fig.~\ref{bthetamev1} some plots of the impact parameter $b_h$ as a function of the deflection angle in a range closer to the horizon of a black hole, computed using the Newtonian approximation derived from the metric (\ref{SCH3}). The region involved covers the interval 
between 20 and 100 horizons. The numerical results refer to the GR solution and to the full one-loop prediction respectively.
 The classical expression and the quantum one start differing as we approach the value of $b_h\sim 20$, and are characterized by a certain dependence on the energy of the incoming beam. Shown are the plots corresponding to neutrinos of energies in the TeV and the PeV range respectively. In these regions the lensing is very strong, corresponding to $10^3$ arcseconds and larger. 
  As the neutrino (or the photon) beam gets closer to the photon sphere ($x_0=3/2 r_s$), which is the point of maximum approach, the angular deflection diverges. This is the impact parameter region where one expects the formation of relativistic images. The divergence can be parameterized by an integer $n$, with $\alpha_n=2 \pi n$, and $n$ tending to infinity. The integer is the winding number of the beam path around the photon sphere. In the external neighborhood of the point of closest approach the beam still escapes to infinity, forming an infinite set of images which are parameterized by the same integer $n$ \cite{Bozza:2001xd}. 
\section{$1/b^n$ contributions to the deflection }
It is interesting to compare the classical GR prediction for the deflection with the result of (\ref{genex}), by resorting to a similar expansion for the deflection integral. This has been studied quite carefully in the literature, especially in the limit of strong lensing \cite{Amore:2006pi,Keeton:2005jd}. The $1/b_h^n$ expansion 
has been shown to appear quite naturally in the post-Newtonian approach applied to the Einstein integral 
for light deflection. \\
We recall that Einstein's expression in GR is given by the integral 
\beq
\alpha(r_0)=\int_{r_0}^\infty dr \frac{2}{r^2}\left[ 
\frac{1}{r_0^2}\left(1 -\frac{2 M}{r_0}\right) -
\frac{1}{r^2}\left(1 -\frac{2 M}{r}\right)\right]^{-1/2} -\pi
\label{exactT}
\eeq
and can be re-expressed in the form 

\beq
\alpha=2\int_0^1 \frac{dy}{\sqrt{1 - 2 s - y^2 + 2 s y^3}} -\pi, 
\eeq
with the variable $s\equiv  r_s/ (2 r_0)$ being related to the ratio between the Schwarzschild radius and the distance of closest approach between the particle and the source, $r_0$. The exact computation of this integral is discussed in appendix \ref{deflection}, and involves elliptic functions. Additional information on 
$\alpha(r_0)$ is obtained via an expansion of the integrand in powers of $s$ and a subsequent integration. This method shows that the result can be cast in the form 
\beq
\alpha(b_h)=\frac{a_1}{b_h} +\frac{a_2}{b_h^2} +\frac{a_3}{b_h^3} + \frac{a_4}{b_h^4} + \frac{a_5}{b_h^5}\ldots
\label{exp}
\eeq
with 
\beq
a_1=2, \qquad a_2=\frac{15}{16}\pi, \qquad a_3=\frac{16}{3}, \qquad a_4=\frac{3465}{1024}\pi, \qquad a_5=\frac{112}{5}.
\eeq
The coefficients $a_i$ differ from those given in \cite{Keeton:2005jd} (up to $a_7$) just by a normalization. They are obtained by re-expressing 
$s=s(r_0)$ in terms of the impact parameter $b_h$ using the relation 
\beq
\label{change}
b_h=x_0 \left( 1 -\frac{1}{x_0}\right)^{-1/2}
\eeq
between the impact parameter and the radial distance of closest approach, having redefined $x_0 \equiv r_0/(2 \, G M)$. This can also be brought into the form 
\bea
\label{form1}
x_0=\frac{2\,b_h}{\sqrt 3}\cos\left[\frac{1}{3}\cos^{-1}\left(-\frac{3^{3/2} }{2\,b_h}\right)\right].
\eea
An expression equivalent to (\ref{form1}) can be found in \cite{Coriano:2014gia}. Eq. (\ref{form1}) can be given in a $1/b_h$ expansion
\beq
x_0=b_h-\frac{3}{8 \, b_h} -\frac{1}{2\,
   b_h^2} -\frac{105}{128\, b_h^3} -\frac{3}{2\, b_h^4} +\mathcal{O}(1/b_h^5),
\label{inversion}
\eeq
which will turn useful below.

We can invert (\ref{exp}) obtaining the relation
\begin{align}
b_h(\alpha)=&\frac{2}{\alpha}+\frac{a_2}{2}+\frac{\alpha}{8}\big(2\,a_3-a_2^2\big)+\frac{\alpha^2}{16}\big(a_2^3-3a_2\,a_3+2\,a_4\big)\nn\\
&+\frac{\alpha^3}{128}\big(8\,a_5-16\,a_2\,a_4-8\,a_3^2+20\,a_2^2a_3-5a_2^4\big)+\mathcal{O}(\alpha^4),
\end{align}
which differs from (\ref{genex}) by the absence of logarithmic terms in the impact parameter $b_h$ and by the energy independence of the coefficients. The inclusion of the extra contributions mentioned above, 
in the classical GR expression, becomes relevant in the case of strong lensing.   
We show in appendix \ref{continuous} how the inclusion of the additional $1/b_h^n $ terms in the expansion of the angular deflection can be extended to the case of a continuous distribution of sources/deflectors. This provides a simple generalization of the standard approach to classical lensing for such distributions.  

\section{Lens equations and $1/b^n$ corrections}
The standard approach to gravitational lensing in GR is based on an equation, derived from a geometrical construction, which relates the angular position of the image ($\theta_I$) to that of the source ($\beta$), with an intermediate angular deflection ($\alpha$) generated on the lens plane. In this section we are going to briefly review this construction, which is based on the asymptotic expression for the angular deflection ($\alpha\sim 2/b_h$), and discuss its extension when one takes into account more general expansions of $\alpha(b_h)$ of the form given by Eq. (\ref{exp}). The extension that we consider covers the case of a thin lens and concerns only the extra $1/b_h^n$ terms derived from classical GR. The discussion is preliminary to the analysis  of the next section, where we will consider the inclusion of the radiative effects, parameterized by 
(\ref{genex}), into the classical lens equation. 

\begin{figure} 
\centering 
\includegraphics[width=0.4\textwidth]{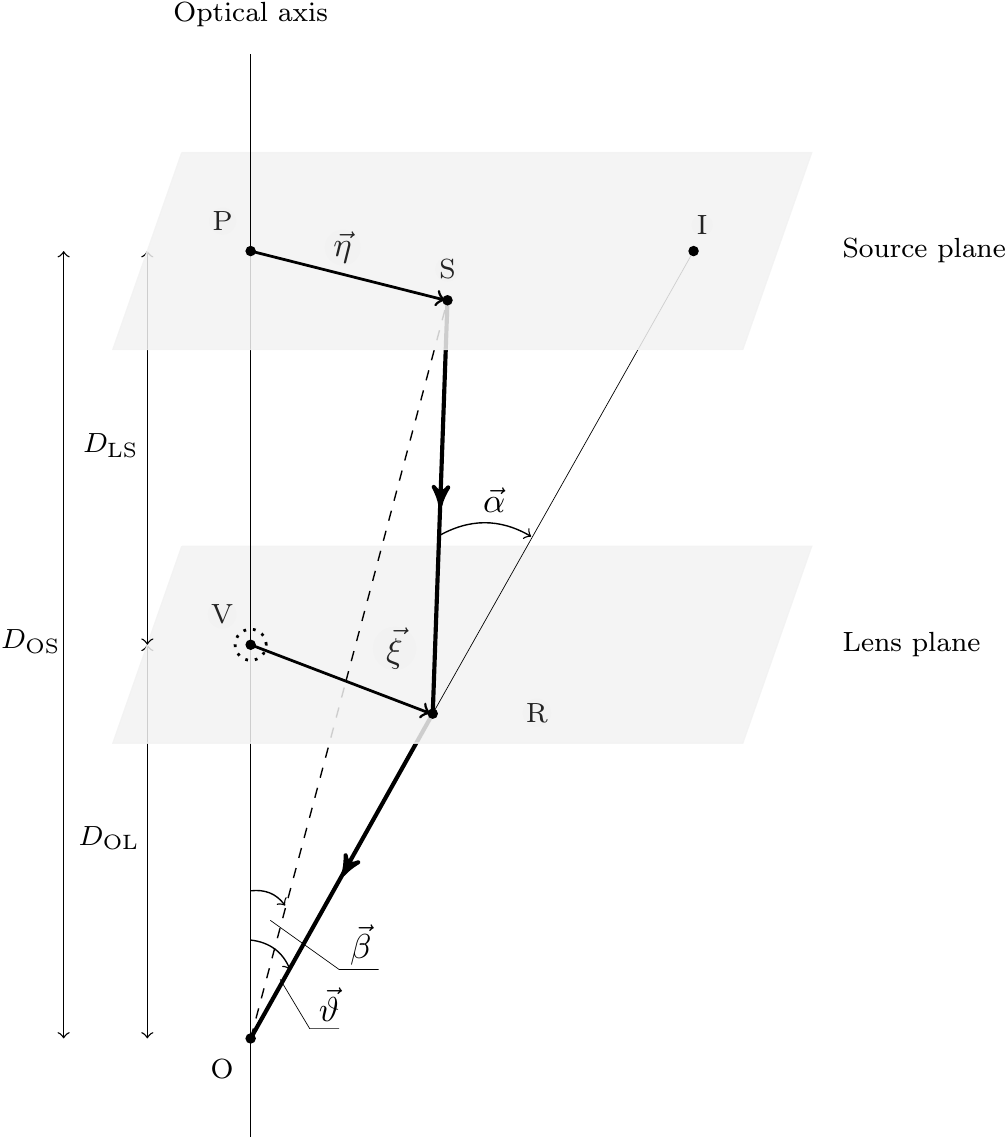}
\caption{Geometric construction of the lens for a continuous distribution of sources. Shown are the plane $S$ of the source distribution and the 
plane of the lens $L$. The line $OI$ identifies the direction at which the observer sees the image after the angular deflection $\alpha$.}
\label{lenspicture}
\end{figure}

\begin{figure}[t]
\centering
\includegraphics[width=0.35\textwidth]{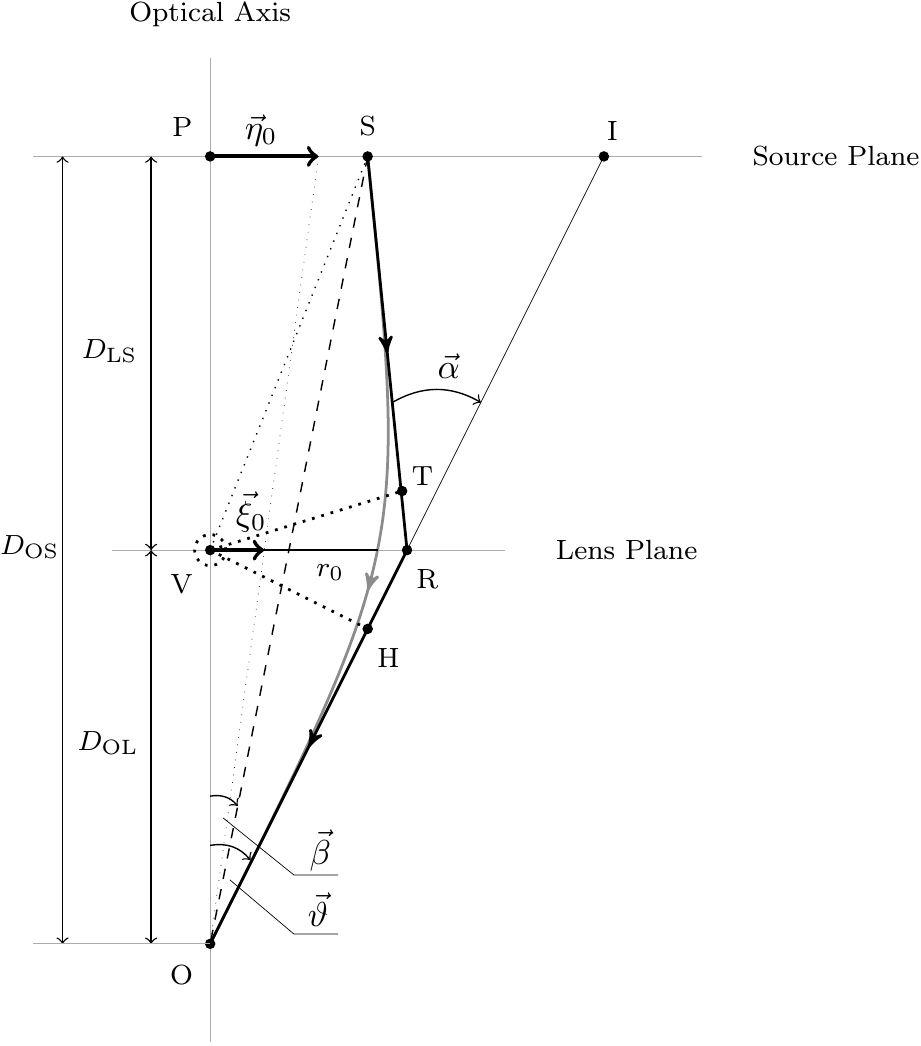}
\caption{The thin lens geometric construction where the source S, the lens V and the observer O lie on the same plane. Notice that figure is not to scale, since $D_{OL}$ and $D_{LS}$ are far larger than the length of $VR$.}
\label{geo}
\end{figure}

\begin{figure}
\centering 
\includegraphics[width=0.4\textwidth]{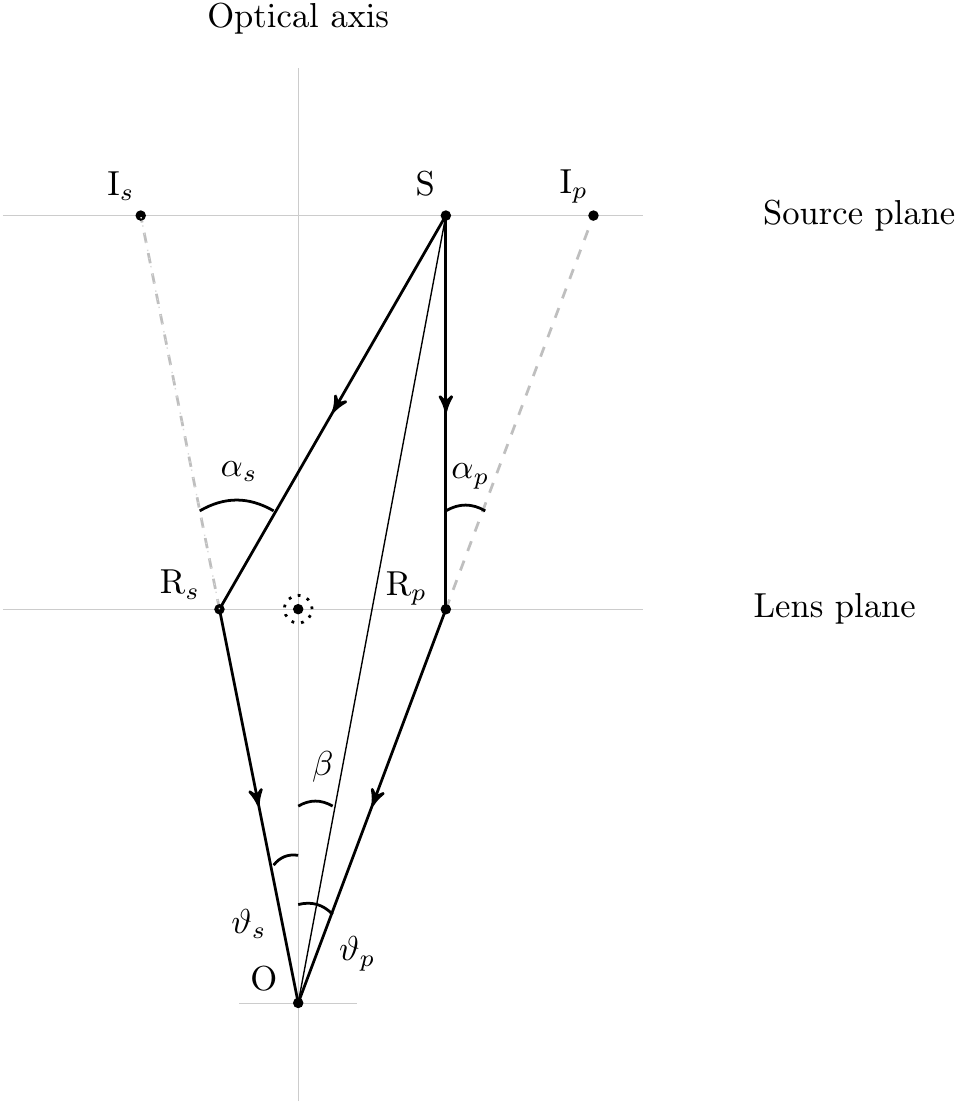} 
\caption{Geometrical construction for the primary I$_p$ and secondary I$_s$ images generated by the two geodesics of the isotropic emission. 
Shown are the source S, the lens, represented by the dotted circle, the observer O and the primary $I_p$ and secondary $I_s$ angular positions involved in the discussion.}
\label{prim}
\end{figure}

\subsection{The lens geometry}
We show in Fig.~\ref{lenspicture} the lens geometry in the case of a continuous distributions of sources and deflectors. A simplified picture of the geometry, with pointlike source and deflector is shown in Fig.~\ref{geo}. 
We indicate with $\vec{\beta}$ the oriented angle between the optical axis $(OP)$ (taken as the $z$ axis) and the unlensed direction of the source $(OS)$. $\vec{\theta}_I$ denotes the angle formed by the visual line of the image $(OI)$ with the optical axis. We also denote with $D_{OL}$  the distance between the observer and the lens plane; with $D_{LS}$ the distance between the lens plane and the source plane and with $D_{OS}$ the distance of the source plane from the observer. $\hat{\alpha}$ is the (oriented) angle of deflection, measured clockwise as all the other angles appearing in the geometrical construction. We also introduce the relations, valid for $D_{LS}, D_{OL}$ much larger than the size of the lens, typical of a linear lens,
\beq
\vec{\eta}\equiv \vec{PS}=\vec{\beta} D_{OS} \qquad \qquad \vec{SI}=\hat{\vec{\alpha}} D_{LS} \qquad \qquad \vec{PI}=\vec{\theta}_I D_{OS}.
\eeq
 The thin lens equation follows from the approximate geometrical relation
 \beq
\label{lin}
\vec{PI}=\vec{PS} +\vec{SI} \qquad \textrm{ i.e.} \qquad \vec{\beta}=\vec{\theta}_I -\hat{\vec{\alpha}}\frac{D_{LS}}{D_{OS}}.
\eeq
Denoting with $\vec{\xi}$ a 2-D vector in the lens plane, it is convenient to introduce two scales $\eta_0$ and $\xi_0$ defined as
\bea
\vec{\eta}=\eta_0\,\vec y\qquad \qquad\vec \xi\equiv\vec{VR}=\xi_0\vec x\qquad \qquad \frac{\eta_0}{\xi_0}=\frac{D_{OS}}{D_{LS}}.
\eea
Using the lens equation in the geometric relation
\bea
\frac{|\vec{PI}|}{|\vec{VR}|}=\frac{D_{OS}}{D_{OL}},
\eea
we find the relation
\bea
\label{thin}
\vec y=\vec x - \hat{\vec \alpha} \frac{D_{LS}\,D_{OL}}{D_{OS}\,\xi_0}\equiv \vec x -\vec \alpha
\qquad \qquad \textrm{with} \qquad\qquad\vec \alpha=\hat{\vec \alpha} \frac{D_{LS}\,D_{OL}}{D_{OS}\,\xi_0},
\eea
which defines the thin lens equation. It is possible to give a simpler expression to the equation above if we go back to (\ref{lin}) and perform simple 
manipulations on the angular dependence. On the lens plane (Fig.~\ref{geo}) the equation takes the scalar form 
\bea
\label{thin1}
\beta=\theta_I- \alpha \frac{D_{LS}}{D_{OS}}, 
\eea
which can be extended to the case of stronger lensing by the inclusion of the contributions of the $1/b^n$ corrections in $\alpha(b)$. Use of the Einstein relation $\alpha=4 G M/b$ and of the relation $b\sim\theta_I D_{OL}$ 
brings (\ref{thin1}) into the typical form
\beq
\beta=\theta_I -\frac{\theta_E^2}{\theta_I} \qquad   \theta_E^2 =\frac{D_{LS}}{D_{OS}}\frac{4 G M}{D_{OL}},
\eeq
which defines the thin lens approximation, with $\theta_E$ being the Einstein radius. For a source $S$ aligned on the optical axis together with the deflector and the observer $O$ (see Fig.~\ref{prim}) - which is defined by the segment connecting the observer, the lens and the plane of the source (with $\beta=0$) - the images will form radially at an opening $\theta_I=\theta_E$ and appear as a circle perpendicular to the lens plane. For a generic $\beta$, instead, the primary and secondary image solutions are given by the well-known expressions 
\beq
\label{img}
\theta_{I {\pm}}=\frac{\beta}{2} \pm \frac{1}{2}\left(\beta^2 + 4 \theta_E^2\right)^{1/2}.
\eeq
It is quite straightforward to extend this derivation with the inclusion of the $1/b^n$ corrections in the $\alpha(b)$ relation and test their effect numerically \cite{Keeton:2005jd}. This is part of a possible improvement of the ordinary (quadratic) thin lens equation which can be investigated more generally in conditions of strong lensing. In that case one can also adopt an equation which includes deflections of higher orders, as we will discuss in the following sections. For the moment we just mention that the inclusion of the higher order $1/b^n$ contributions given by (\ref{exp}) modifies (\ref{thin1}) into the form 
\beq
\label{thin2}
\beta=\theta_I - \frac{\theta_E^2}{\theta_I}-\sum_{n\geq 2} \frac{\theta_E^{(n)}}{\theta_I^n}, 
\eeq
with 
\beq
\theta_E^{(n)}\equiv r_s^n a_n \frac{D_{LS}}{D_{OS} D_{OL}^n}.
\eeq
Another observable that we will investigate numerically is going to be the lens magnification. For this purpose we recall that light beams are subject to deflections both as a whole but also locally, due to their bundle structure. Rays which travel closer to the deflector are subject to a stronger deflection compared to those that travel further away. This generates a difference in the solid angles under which the source is viewed by the observer in the unlensed and in the lensed cases. In the simple case of an axi-symmetric lens the ratio between the two solid angles can be defined in the scalar form 
\bea
\mu=\left|\left(\frac{\partial\beta}{\partial\theta_I}\frac{\sin\beta}{\sin\theta_I}\right)\right|^{-1}.
\eea
In the case of a thin lens (\ref{thin}), the analogous expression is given by 
\bea
\mu^{(0)}_\pm\equiv\left(\frac{\partial\beta}{\partial\theta_I}\frac{\beta}{\theta_I}\right)^{-1}.
\eea
For this lens the analysis simplifies quite drastically. Using the expression of the two images $\theta_{I\pm}$ given in (\ref{img}) one obtains  the simple expression for the primary and secondary images
\beq
\mu_{\pm}=\pm\left(1 -\left(\frac{\theta_E}{\theta_{I\pm}}\right)^4\right)^{-1},
\label{magni}
\eeq
where the Einstein angle is defined as usual
\bea
\theta_E=\sqrt{4\,G M\frac{x}{D_{OL}}}\qquad\qquad \textrm{with}\qquad \qquad x=\frac{D_{LS}}{D_{OS}}.
\eea
It is convenient to measure the angular variables in terms of the Einstein angle $\theta_E$, as $\bar{\beta}\equiv \beta/\theta_E$, 
$\bar{\theta}\equiv \theta_I/\theta_E$, with 
\beq
\bar{\theta}_{I \pm}=\frac{\bar{\beta}}{2} \pm \sqrt{1 + \frac{\bar{\beta}^2}{4}},
\eeq
then the total magnification takes a rather simple form 
\beq
\mu\equiv \mu_+  + \mu_-=\frac{2 + \bar{\beta}^2}{\bar\beta\sqrt{4 + \bar\beta^2}}.
\eeq
This equation is commonly used to calculate the light curve in the microlensing case. We refer to \cite{Mao:2008kp} for a short review on this point.
\subsection{Nonlinear effects in strong deflections}
In conditions of strong lensing, the linear approximations in the trigonometric expressions are not accurate enough and one has to turn to a fully nonlinear description of the geometry, expressed in terms of the angular variables which are involved. 
  We illustrate this point by taking as an example a typical lens equation, which in our case is given by the Virbhadra-Ellis construction (VE) \cite{Virbhadra:1999nm}. \\
 Following Fig.~\ref{geo}, we recall that the VE lens equation is based on the geometrical relation \cite{Virbhadra:1999nm,Bozza:2008ev} 

\begin{figure}[t]
\centering
\subfigure[]{\includegraphics[scale=0.9]{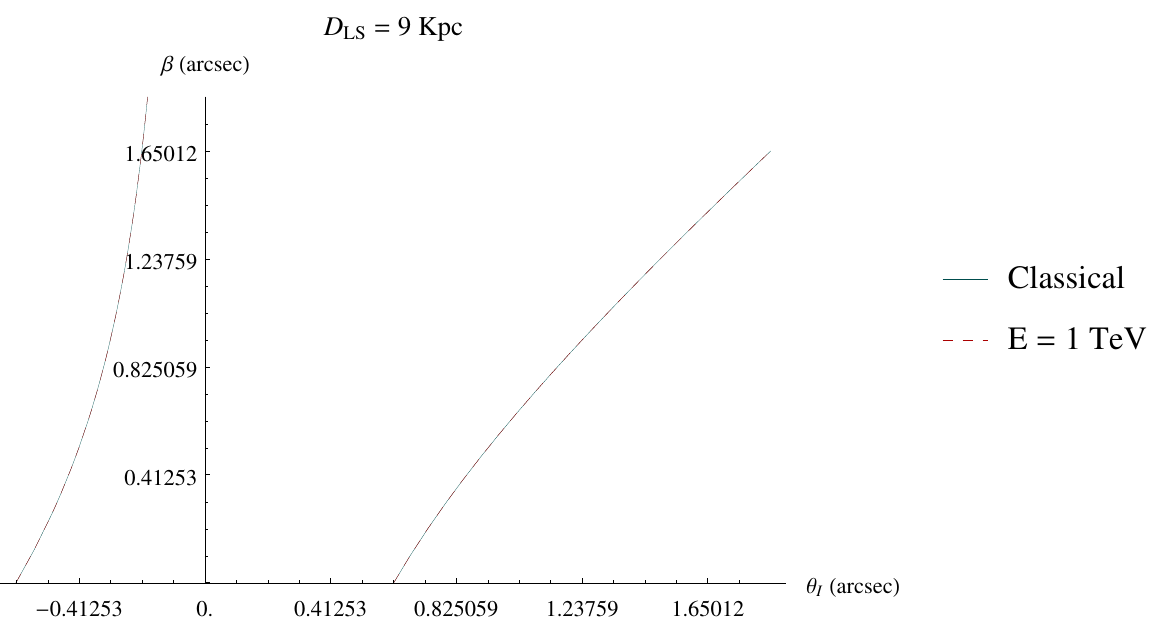}}\hspace{.5cm}
\subfigure[]{\includegraphics[scale=0.9]{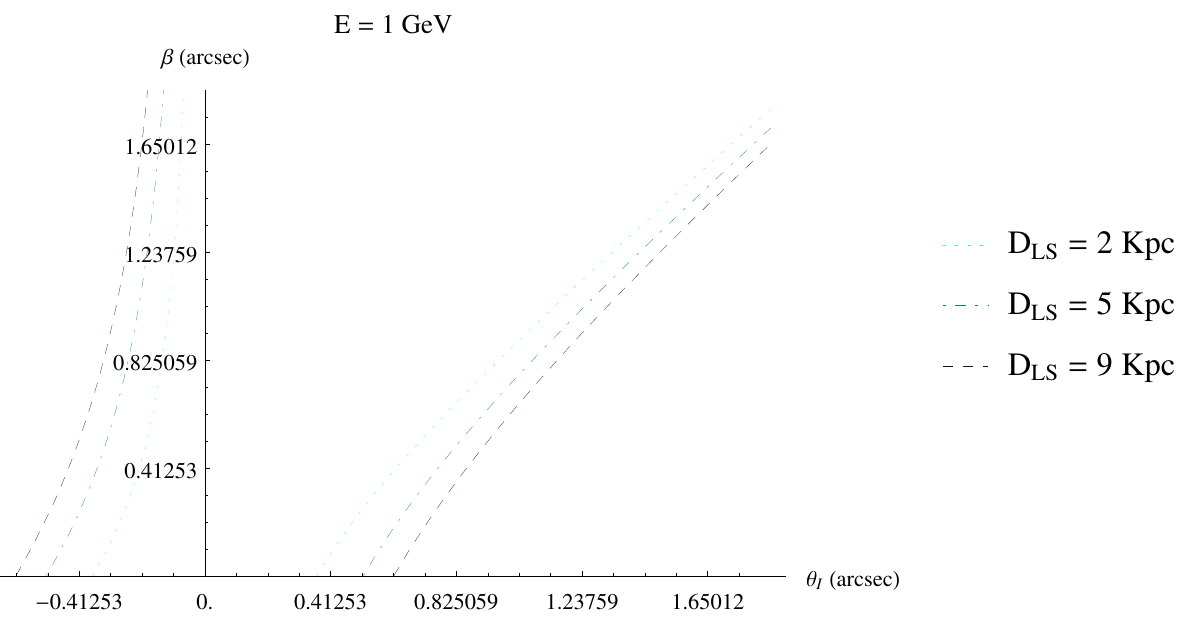}}
\caption{(a): $\beta(\theta_I)$ for the Virbadhra-Ellis lens equation in the neutrino case, for a black hole with M $=10^6\,\text{M}_{\odot}$ and with $D_{OL}$=10 Kpc, $D_{OS}$=19 Kpc. The numerical solution for the classical and the energy-dependent result. (b): $\beta(\theta_I)$ as in (a) but for a 1 GeV neutrino beam.}
\label{VHc1}
\end{figure}

\beq 
\overline{PS}=\overline{PI} -\overline{SI}, 
\eeq
which gives
\bea
\label{lenseq}
D_{OS}\tan\beta=D_{OS}\tan{\theta}_I-D_{LS}(\tan{\theta_I} +\tan(\alpha-{\theta}_I)),
\eea
under the assumption that the point $R$ in Fig. \ref{geo} lies on the vertical plane of the lens. ${\theta}_I$ is the angle at which the image is viewed by the observer and $\beta$ is the unlensed angular position of the source. 
Within this approximation we can use the geometric relation 
\beq
b=D_{OL}\sin{\theta}_I,
\label{bt}
\eeq
which allows to relate the image position ${\theta}_I$ to the angular deflection of the beam $\alpha$. Notice that this approximate relation is justified by the fact that the distances $D_{OL}$ and $D_{OS}$ are very large compared to the radius of closest approach $r_0$. In this limit the two segments $\overline{VH}$ and $\overline{VT}$ are treated as equal.\\
We remind that (\ref{lenseq}) is not the unique lens equation that one can write down, but, differently from Eq. (\ref{thin}), it can be used in the case of strong lensing. It  takes into account the nonlinear contributions to the angular deflection by the introduction of the $\tan(\beta)$ and $\tan(\theta_I)$ terms, which in (\ref{thin}) are not included. We refer to \cite{Bozza:2008ev} for a review of possible lens equations. 

\section{Radiative effects and the geometry of lensing }
Turning to our case study, radiative effects in the lens equations can be introduced by replacing the expression of the angular deflection generated by the source on the source plane, which is a function of the impact parameter 
$b$ $(\alpha=\alpha(b))$ with the new, energy dependent relation $\alpha(b,E)$ whose general form is given by (\ref{genex}). \\
 For simplicity we consider a pointlike source, and a pointlike deflector, as shown in Fig. \ref{geo}.
We recall that for a massless particle the geodesic motion is determined in terms of the energy $E$ and of the angular momentum $L$ at the starting point of the trajectory. The gravitational deflection, however, can be written only as a function of the impact parameter $b$ of the source, with $b=E/L$, which is an important result of the classical approach. For a further clarification of this aspect, which differs from the semiclassical analysis we are interested in, 
we briefly overview the classical case, using the lens geometry as a reference point for our discussion.\\
For a source located on the source plane at an angular opening $\beta$ (in the absence of the deflector), the initial conditions can be expressed in terms of the two components of the initial momentum $ \vec p=(p_r,p_\phi)$ on the plane of the geodesic, or, equivalently, by the pairs $(p_r, E)$ or $(p_\vartheta,E)$, with $E$ the initial energy of the beam. We recall that for a Schwarzschild metric these are defined as 
\beq
p_r = \left(  1- \frac{2 G M }{r} \right)^{\!-1} \dot{r},    
\qquad p_{\vartheta} = -r^2 \dot{\vartheta}, \qquad
p_t =  \left(  1- \frac{2 G M}{r} \right) \dot{t}, 
\qquad p_{\phi} = -  r^2 \sin^2 \vartheta \dot{\phi} \,.
\eeq
We have denoted with $\dot{x}\equiv dx/d s$ the derivative respect to the affine parameter. 
$p_t$ and $p_\phi$ related to the energy and to the angular momentum as 
$p_t=E$ and $p_\phi=-L$, and with the motion taking place on the plane $\vartheta=\pi/2$ $(p_\vartheta=0)$. They are constrained by the mass-shell condition 

\beq
 \left(1-\frac{2 G M }{r} \right) (p^t)^2 - \left( 1-\frac{2 G M}{r} \right)^{\!-1} (p^r)^2 - r^2 (p^\phi)^2=0,
\eeq
with $(p^r=\dot r, p^\phi=\dot \phi,p^t=\dot t)$.

The lens equation, usually written as 
\beq
L(\beta,\theta_I)=0, 
\eeq
can also be written, equivalently, in the form of a constraint between $\beta$ and $b$ using (\ref{bt}). We can use any of the independent variables mentioned above.
For a given initial momentum of the beam, emitted from the plane of the source, the lens equation will then determine the position of the source in such a way that the geodesic motion will reach the observer at its location on the optical axis. In particular, an interesting description emerges if we choose as initial conditions the angular position of the source $(\beta)$ and the value of the impact parameter $b$. These two conditions fix the direction of the trajectory of the beam at its origin on the source plane. In these last variables, the lens equation will then determine one of the two in terms of the other in such a way that outgoing geodesic will reach the observer. \\
The inclusion of an energy dependence in the angle of deflection $\alpha$ renders this picture slightly more complex. For instance, the lens equation will now depend on 3 parameters, which can be chosen to to be $(\beta, \theta_I, E)$ or $(\beta,p_r,p_\phi)$ or any other equivalent combination, with one of the three fixed in terms of the other two by the equation itself. For a monochromatic and spherical source of energy $E$, fixed at a position $\beta$, emitting a beam with a given impact parameter $b$ respect to the deflector, the lens equation may not have a real solution, since the deflector may disperse the beam in such a way that it will never reach the observer. For a fixed spherical source which emits photons or neutrinos of any energy, one can look for solution in the reduced variables $b, E$. Being $b$ related to the primary and secondary images $\theta_{I\pm}$, the beam that reaches the observer will be characterized by a unique energy $E$, assuming that the images are detected at angular positions $\theta_{I\pm} $. \\
The argument above can be repeated by using any triple combination of independent kinematic variables among those mentioned above. \\
Having clarified this point, we now move to a description of the actual implementation of the lens equation is this extended framework. The angular location of the image $\theta_I$ and the impact parameter are related in the geometry of the lens by Eq. (\ref{bt}), and this allows to search for solutions of the lens equation (\ref{lenseq}) in regions characterized by smaller values of the impact parameter ($ 20 <b_h <100$) where the angular deflections are stronger. \\
The key to the derivation of the radiative lens equation are Eqs. (\ref{genex}) and (\ref{bt}). Combining the two relations we obtain 

\begin{align}
\label{genexp}
\alpha(b(\theta_I,E))=&\frac{4 G M}{D_{OL} \sin\theta_I} + \sum_{n\geq 1} \frac{A_{2n}}{\left(D_{OL} \sin\theta_I\right)^{2n}}\nn\\
&+ \sum_{n\geq 1}\left( \frac{2 G M}{D_{OL} \sin\theta_I}\right)^{2n+1}\left(A_{2n+1} + D_1 \ln^n \left(\frac{D_{OL}}{2 G M} \sin\theta_I\right) +\ldots \right),
\end{align}
where the ellipsis refer to the extra logarithmic contributions present in Eq.(\ref{genex}). The expression 
above is known analytically if we manage to solve explicitly the semiclassical equation (\ref{semic}), otherwise it has to be found by a numerical fit. However, it is clear that the ansatz for the fit has, in any case, to coincide with Eqs.~(\ref{genex}) and (\ref{genexp}), due to the typical functional forms of the solutions of Eq. (\ref{semic}). For instance, in the case of a thin lens, the modifications embodied in (\ref{genexp}) can be incorporated into the new equation 

\bea
\label{thin3}
\beta=\theta_I- \alpha(b(\theta_I,E)) \frac{D_{LS}}{D_{OS}}, 
\eea
which is an obvious generalization of (\ref{thin2}), the latter being valid only in the classical GR case.
As we are going to illustrate below, (\ref{thin3}) can be studied numerically for several geometrical configurations, which are obtained by varying the lensing parameters $D_{LS}$ and $D_{OL}$.\\
A similar approach can be followed for the VE or for any other classical lens equation.
The insertion of $\alpha({\theta_I},E)$ given by (\ref{genex}) into (\ref{lenseq}) generates the radiative lens equation 
\beq
\label{lens1}
D_{OS}\tan\beta=D_{OS}\tan{\theta}_I-D_{LS}(\tan{\theta}_I +\tan(\alpha({\theta}_I,E)-{\theta}_I)),
\eeq
which takes into account also the quantum corrections and is now, on the contrary of (\ref{lenseq}), energy dependent. At this point it is clear that all the lens observables, such as magnifications, shears, light curves of microlensing etc. descend rather directly by this general prescription. \\
For instance, we can determine for the Virbadhra-Ellis lens the expression for the magnification using the 
radiative (semiclassical) expression
\bea
\label{mueq}
\mu&=&\frac{\chi_1}{\chi_2} \\
\chi_1&=&D_{OS}\sin\theta_I\,(1+((D_{OL}\tan\theta_I+(D_{OL}-D_{OS})\tan(\alpha(\theta_I,E)-\theta_I))/D_{OS})^2)^{3/2}, \nonumber 
\eeqa
\beqa
\chi_2 &=&(D_{OL}\tan\theta_I+(D_{OL}-D_{OS})\tan(\alpha(\theta_I,E)-\theta_I))\nonumber \\
&&\times (\sec^2\theta_I+(D_{OL}-D_{OS})/D_{OS}(\sec^2\theta_I
+\sec^2(\alpha(\theta_I,E)-\theta_I)\,(\alpha^{\prime}(\theta_I,E)-1)))\nn,
\eea
where $\alpha^{\prime}\equiv \partial \alpha/\partial \theta_I$. As clear from Eqs. (\ref{lens1}) and (\ref{mueq}), both equations are very involved, although they can be investigated very accurately at numerical level. It is also possible to discuss the analytical form of the solutions within the formalism of the $1/b^n$ expansion. In fact, we are entitled to expand all the observables of the fully nonlinear lens in the angular deflection $\alpha$, and work at a certain level of accuracy in the angular parameters. In this work, however, we prefer to proceed with a direct numerical analysis of the full equations, both for the thin  and for the VE lens, leaving the discussion of the explicit solutions to a future work.

\begin{figure}[t]
\centering
\subfigure[]{\includegraphics[scale=.76]{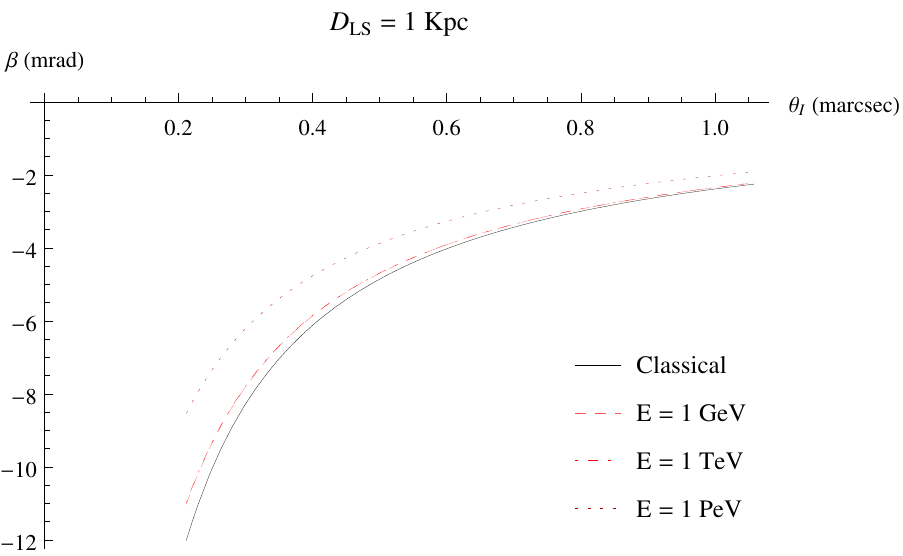}} \hspace{.5cm}
\subfigure[]{\includegraphics[scale=.58]{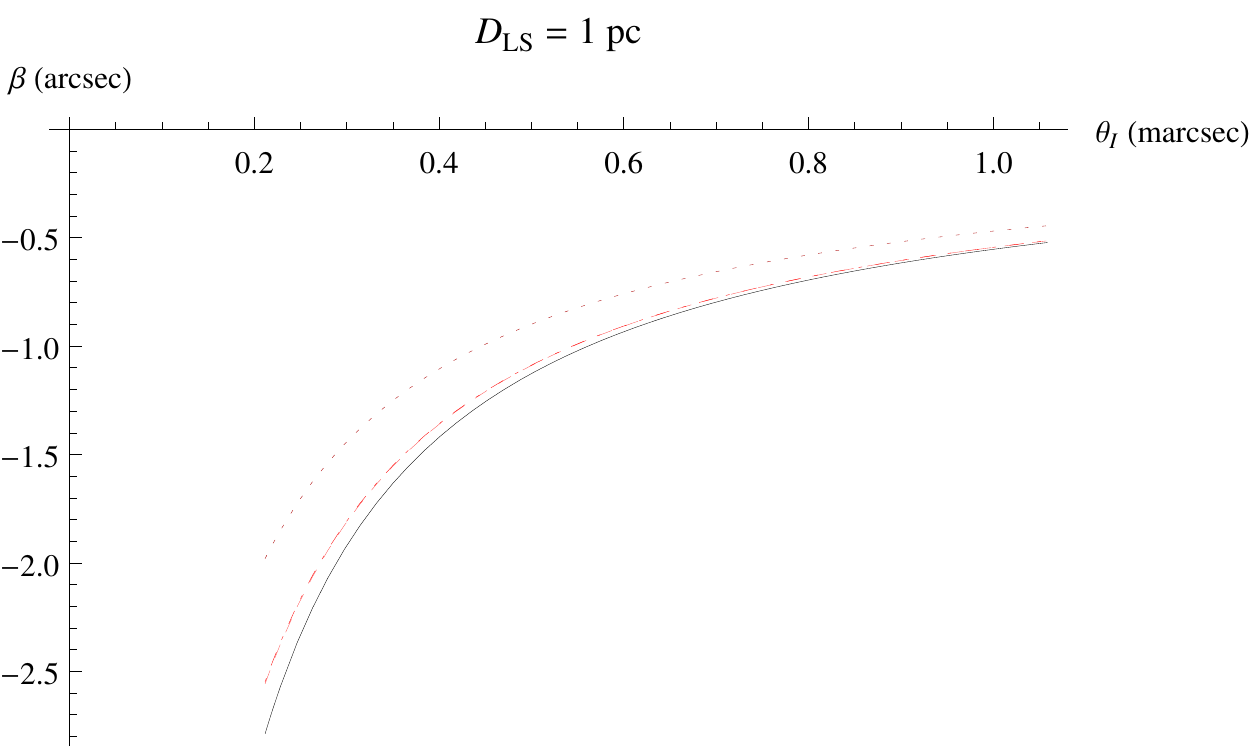}} \hspace{.5cm}
\subfigure[]{\includegraphics[scale=.58]{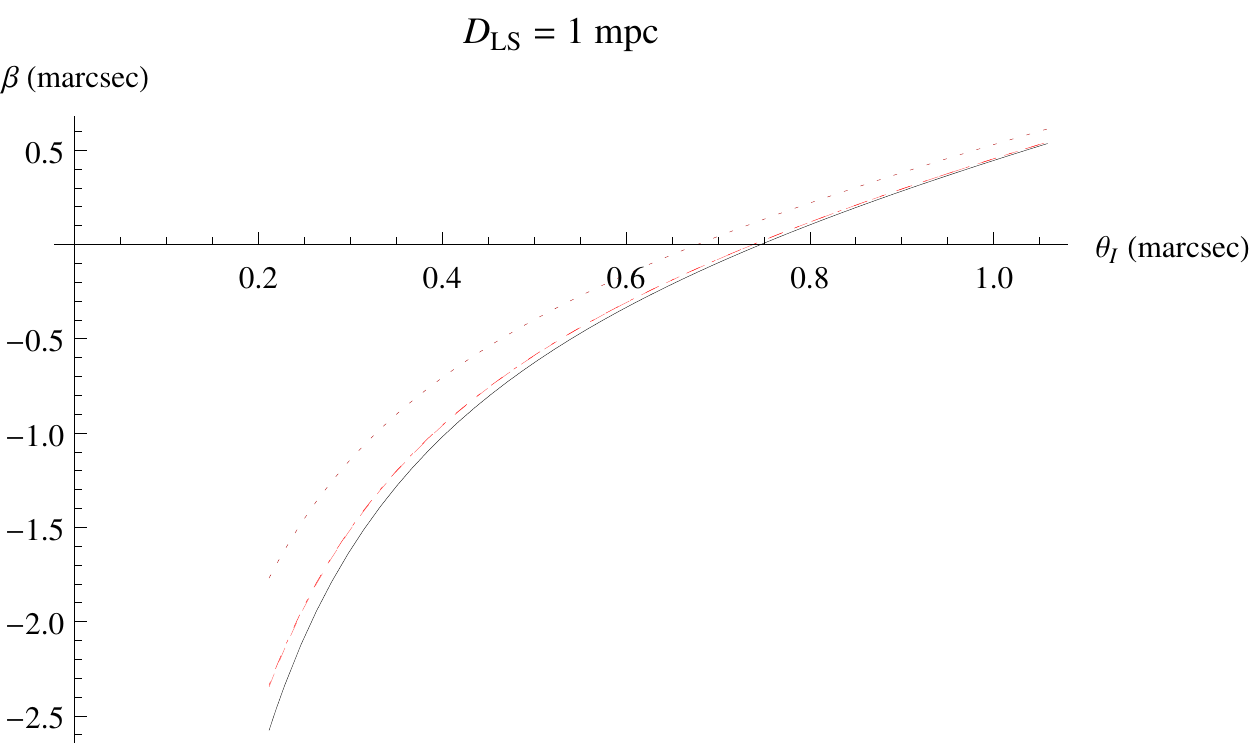}}
\caption{$\beta(\theta_I)$ with $20<b_h<100$ in the neutrino case for a black hole with M = 4.31 $10^6\,\text{M}_\odot$: $D_{LS}$ = 1 kpc (a), $D_{LS}$ = 1 pc (b), $D_{LS}$ = 1 milliparsec (mpc) (c). 
\label{betasagitt1}}
\end{figure}

\begin{figure}[!htb]
\centering
\subfigure[]{\includegraphics[scale=.76]{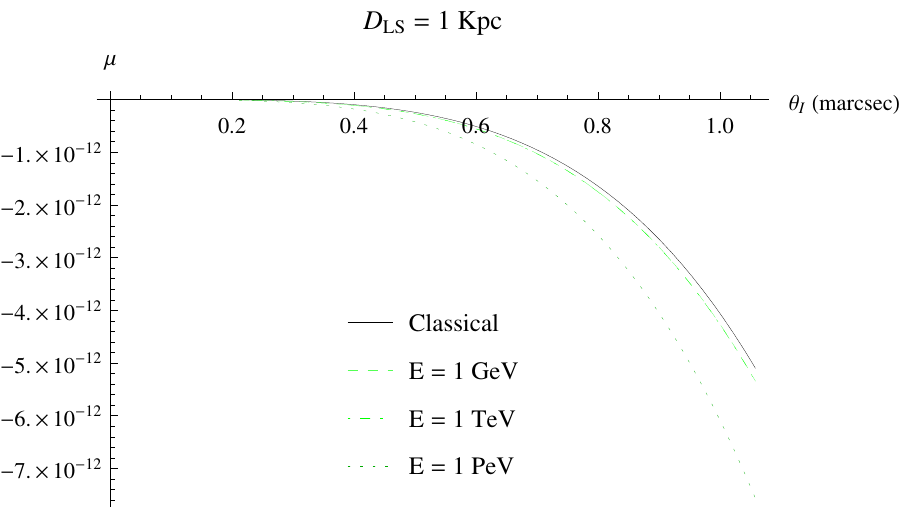}}\hspace{.5cm}
\subfigure[]{\includegraphics[scale=.58]{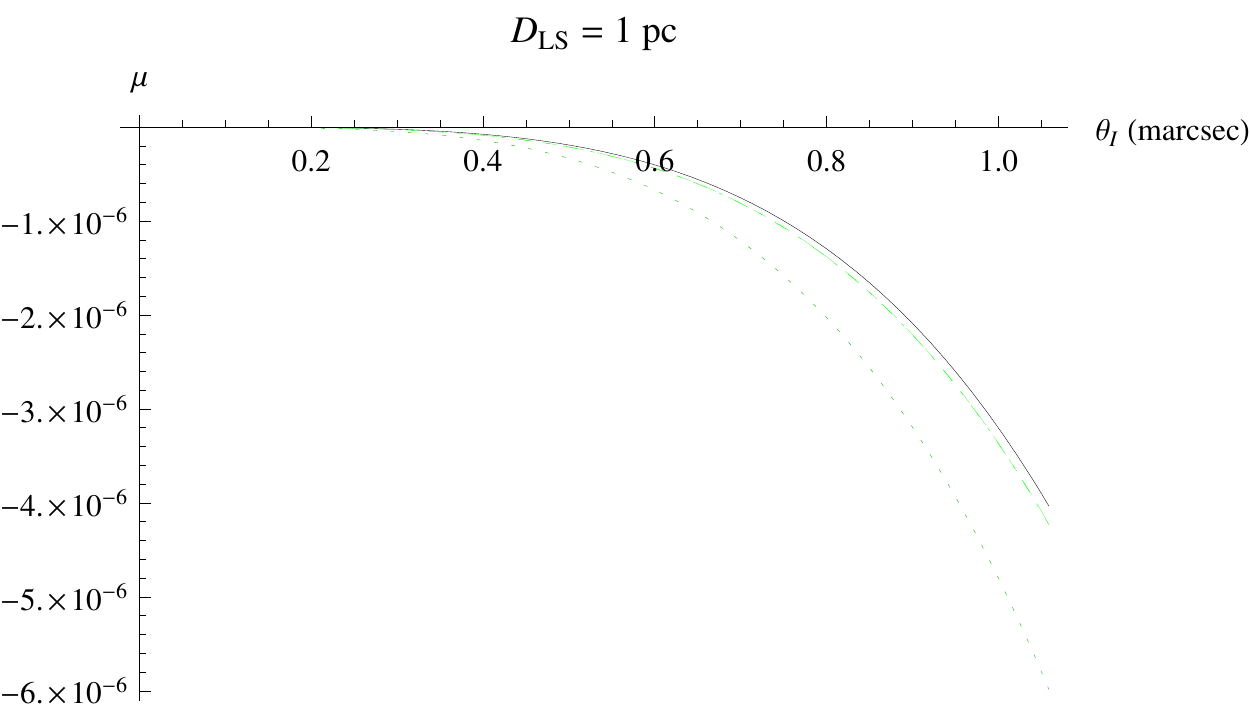}}\hspace{.5cm}
\subfigure[]{\includegraphics[scale=.58]{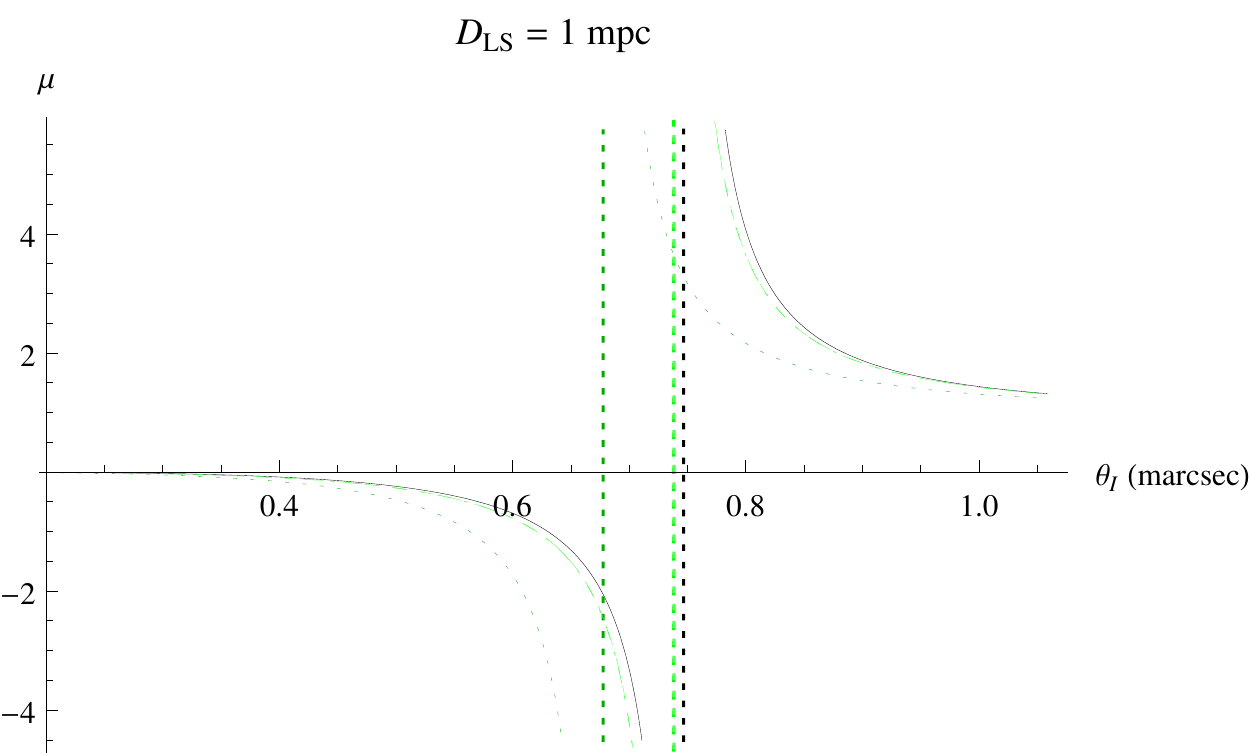}}
\caption{$\mu(\theta_I)$ (Sagittarius) with $20<b_h<100$ in the neutrino case for a black hole with $M = 4.31 \times 10^6\,\text{M}_\odot$: $D_{LS}$ = 1 Kpc (a), $D_{LS}$ = 1 pc (b), $D_{LS}$ = 1 mpc (c). The strong suppression of the magnification parameter for panels (a) and (b) are associated with the secondary image.\label{musagitt1}}
\end{figure}

\begin{figure}[!htb]
\centering
\subfigure[]{\includegraphics[scale=.76]{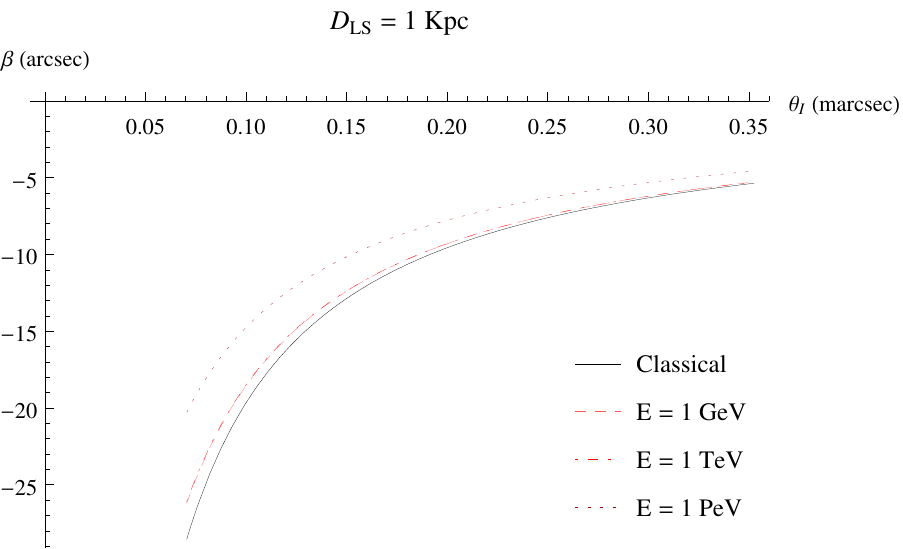}} \hspace{.5cm}
\subfigure[]{\includegraphics[scale=.58]{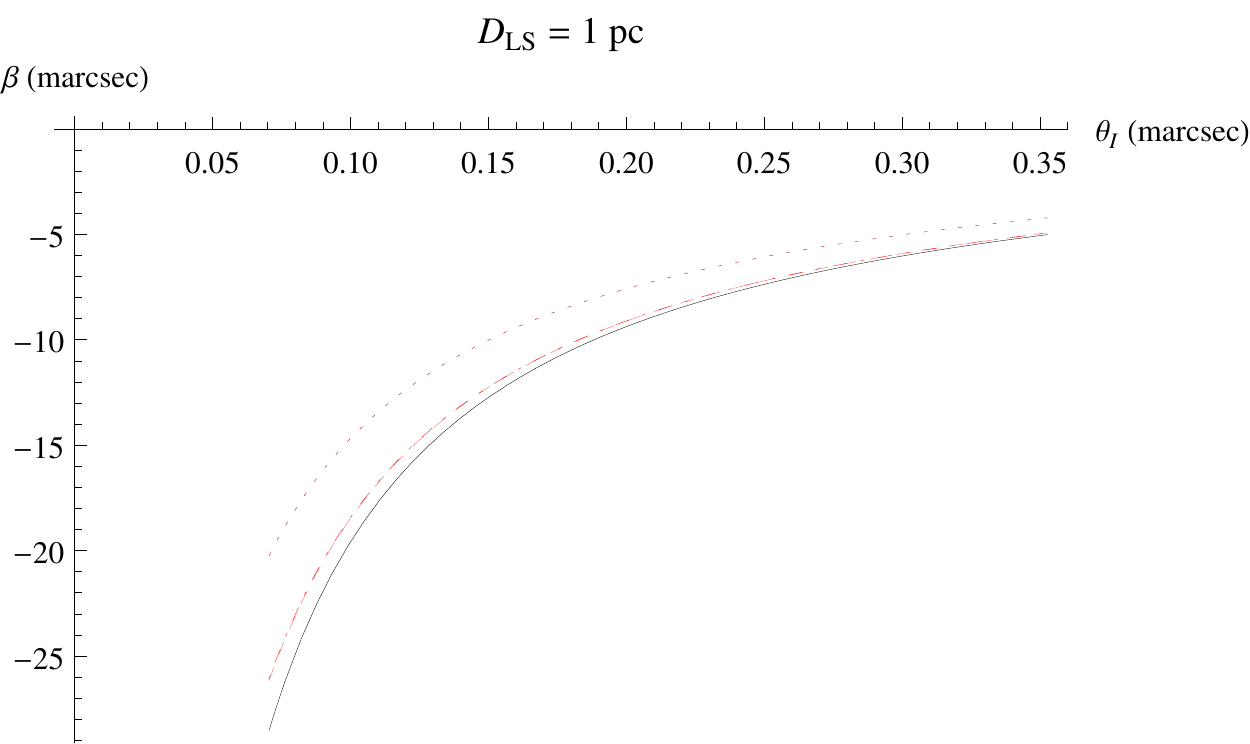}} \hspace{.5cm}
\subfigure[]{\includegraphics[scale=.58]{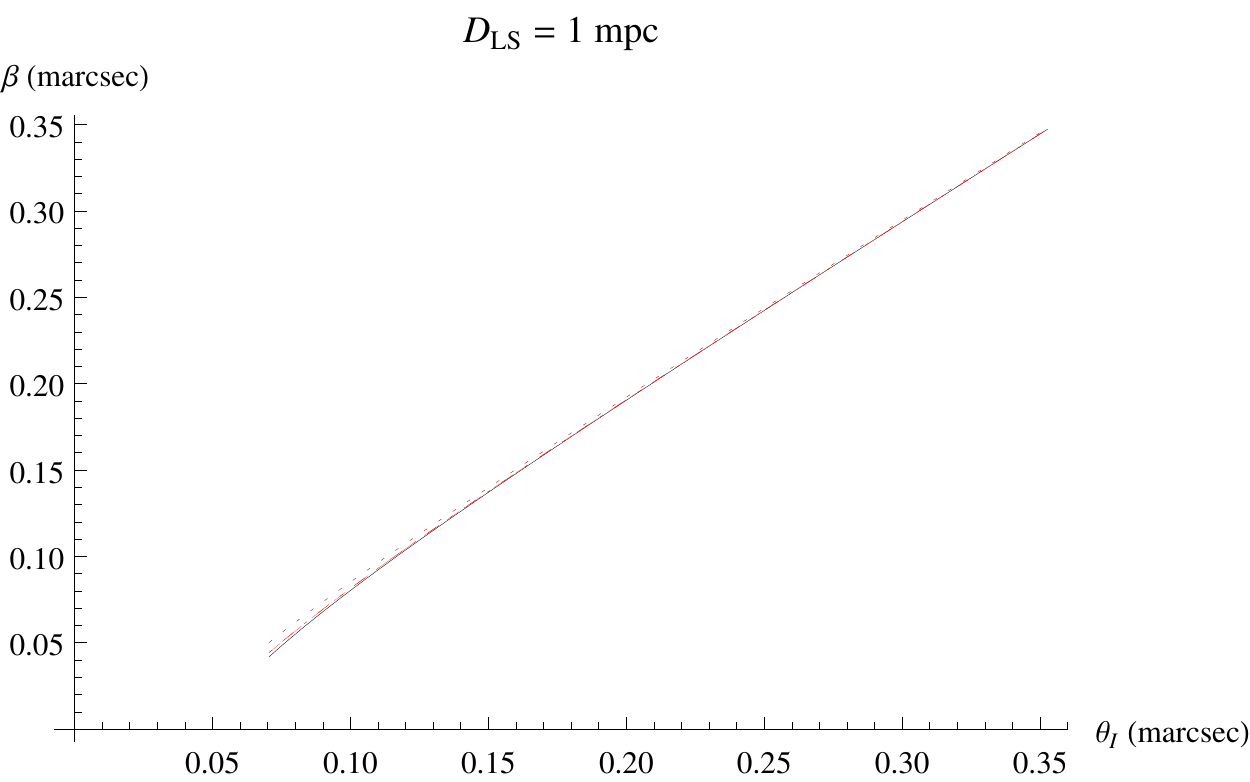}}
\caption{$\beta(\theta_I)$ with $20<b_h<100$ in the neutrino case for a black hole with M = 1.40 $10^8\,\text{M}_\odot$: $D_{LS}$ = 1 Kpc (a), $D_{LS}$ = 1 pc (b), $D_{LS}$ = 1 mpc (c) (Andromeda galactic center). 
}
\label{Andro1}
\end{figure}

\begin{figure}[!htb]
\centering
\subfigure[]{\includegraphics[scale=.76]{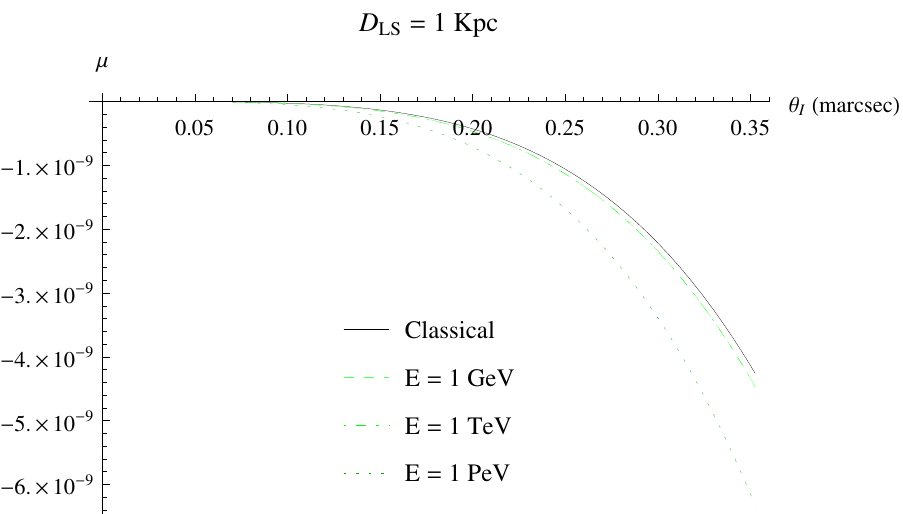}}\hspace{.5cm}
\subfigure[]{\includegraphics[scale=.58]{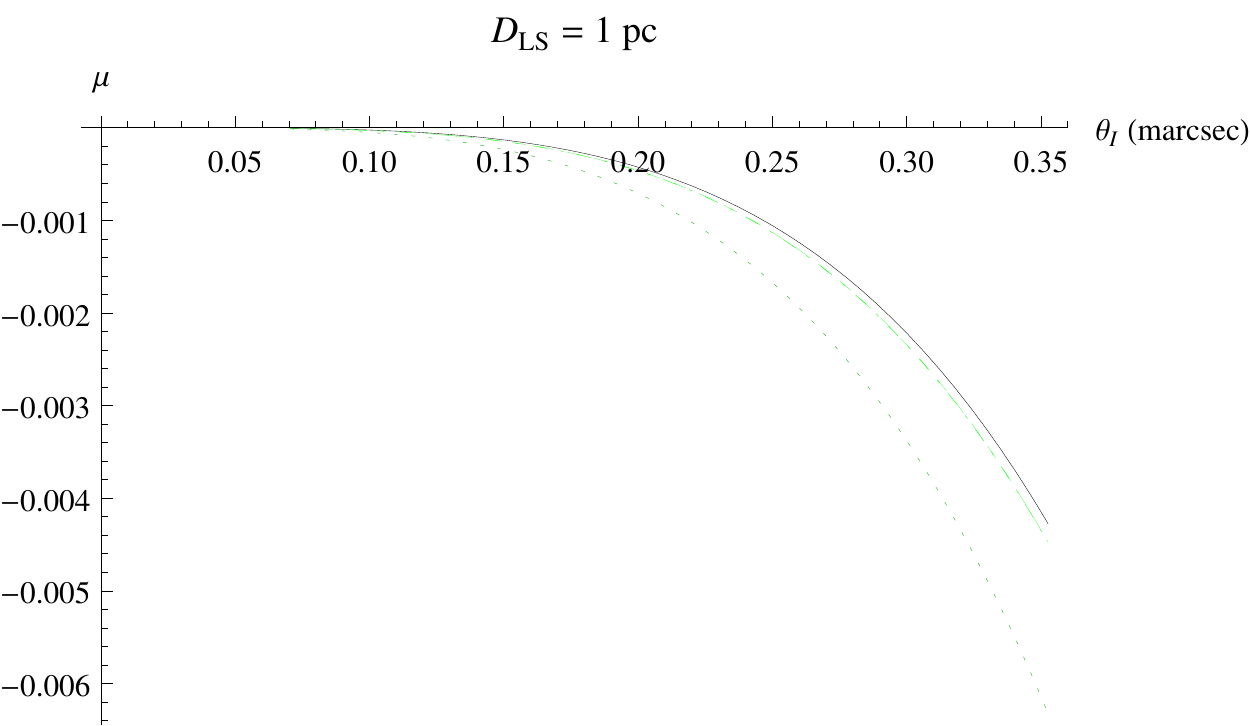}}\hspace{.5cm}
\subfigure[]{\includegraphics[scale=.58]{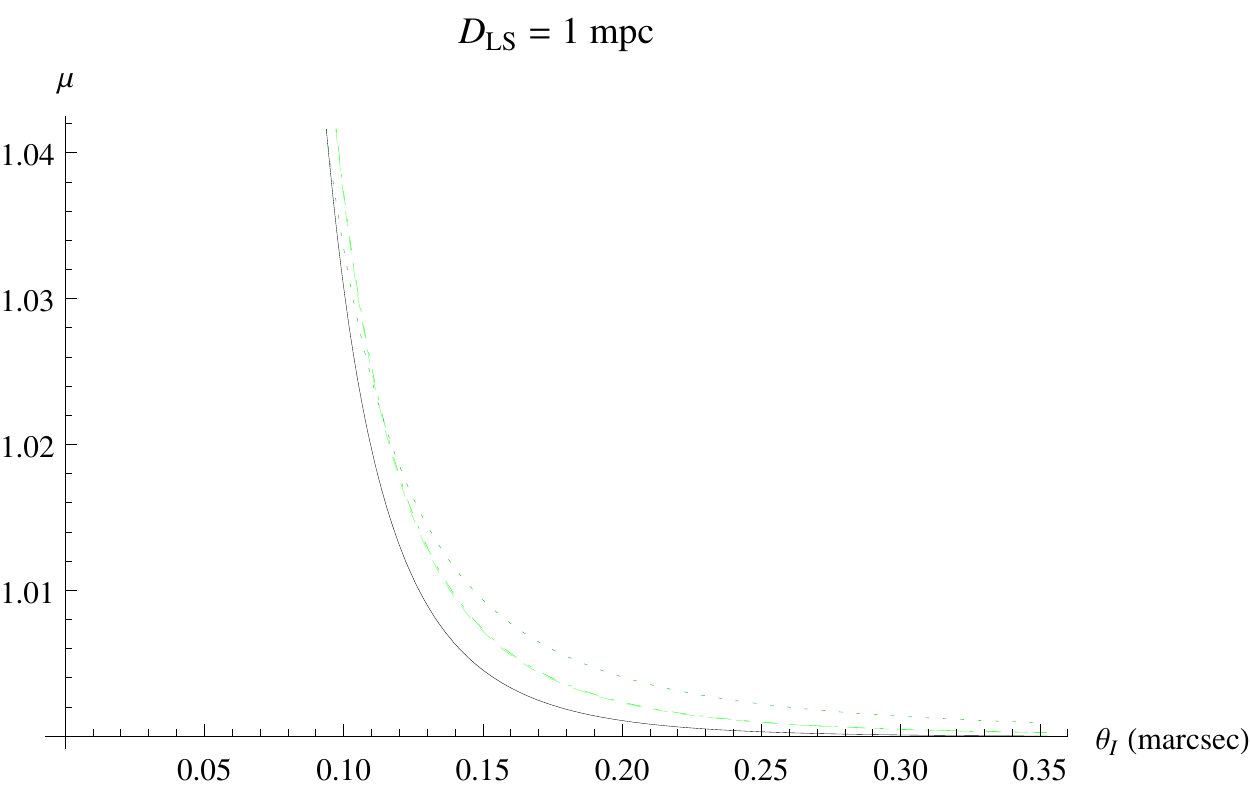}}
\caption{$\mu(\theta_I)$  with $20<b_h<100$ in the neutrino case for a black hole with M = 1.40 $10^8\,\text{M}_\odot$: $D_{LS}$ = 1 Kpc (a), $D_{LS}$ = 1 pc (b), $D_{LS}$ = 1 mpc (c) (Andromeda galactic center). The strong suppression of the magnification parameter, in this case, is associated to the secondary image for this geometric configuration.}
\label{Andro2}
\end{figure}

\begin{figure}[!htb]
\centering
\subfigure[]{\includegraphics[scale=.76]{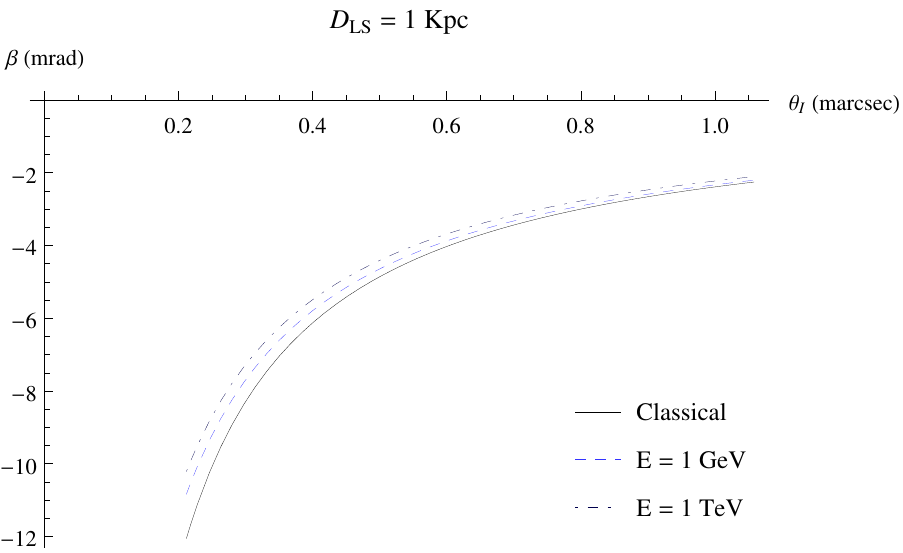}} \hspace{.5cm}
\subfigure[]{\includegraphics[scale=.58]{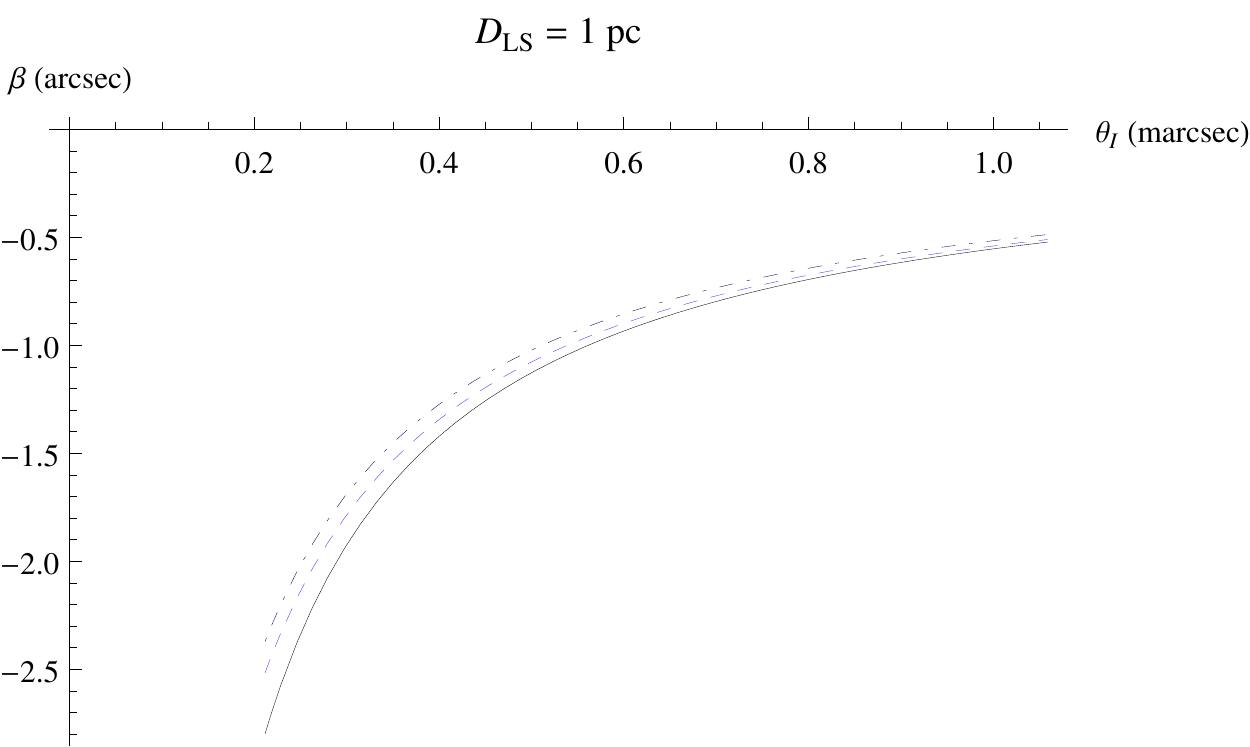}} \hspace{.5cm}
\subfigure[]{\includegraphics[scale=.58]{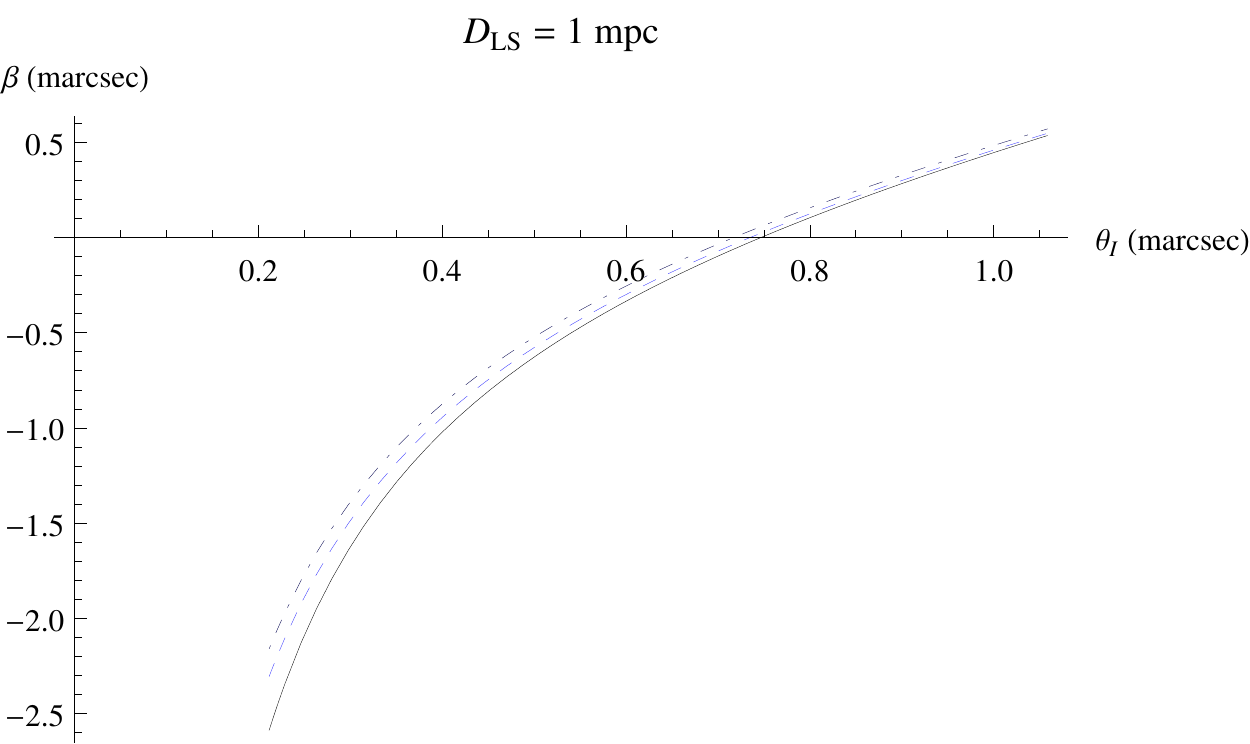}}
\caption{$\beta(\theta_I)$ with $20<b_h<100$ in the photon case for a black hole with M = 4.31 $10^6\,\text{M}_\odot$: $D_{LS}$ = 1 Kpc (a), $D_{LS}$ = 1 pc (b), $D_{LS}$ = 1 mpc (c) (Sagittarius). 
\label{betasagitt}}
\end{figure}

\begin{figure}[!htb]
\centering
\subfigure[]{\includegraphics[scale=.76]{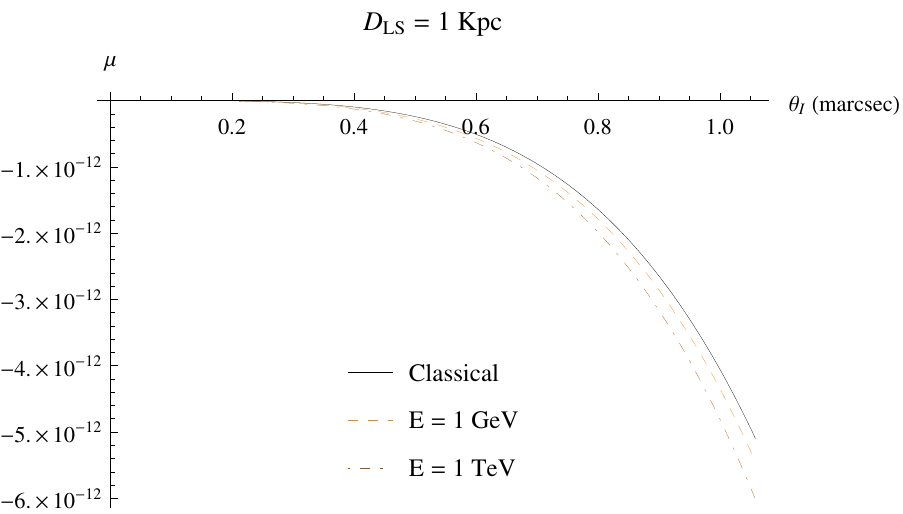}}\hspace{.5cm}
\subfigure[]{\includegraphics[scale=.58]{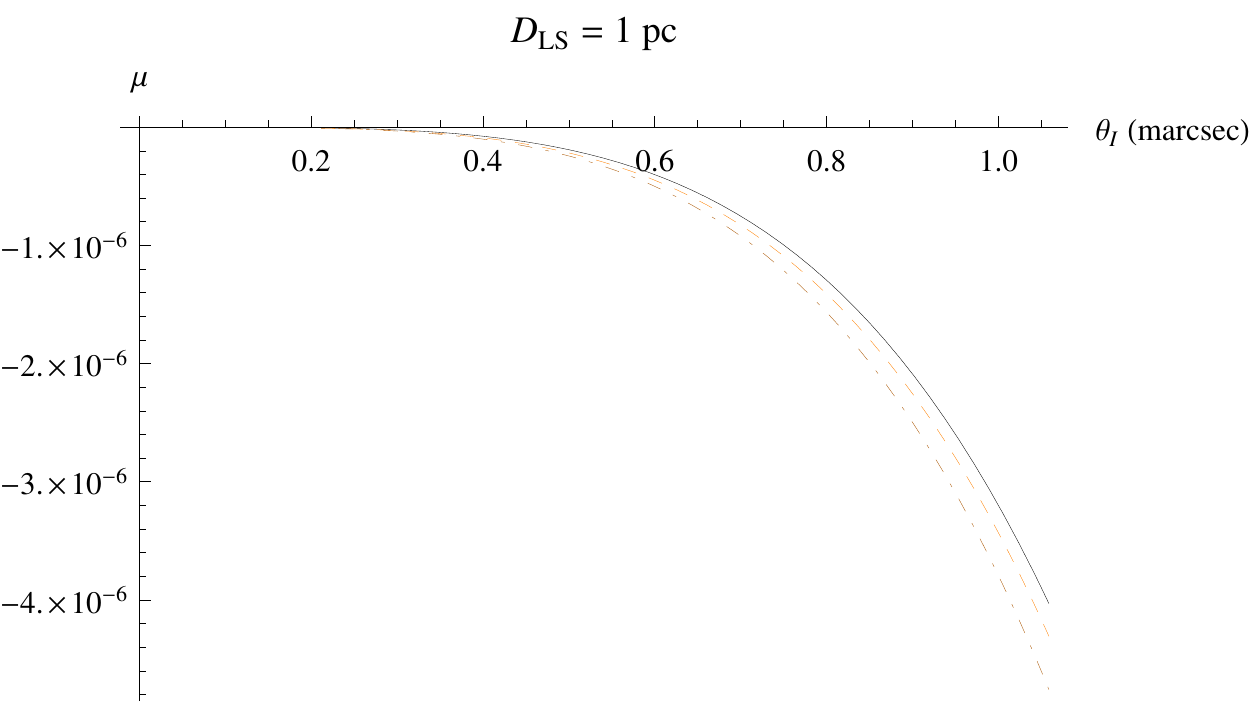}}\hspace{.5cm}
\subfigure[]{\includegraphics[scale=.58]{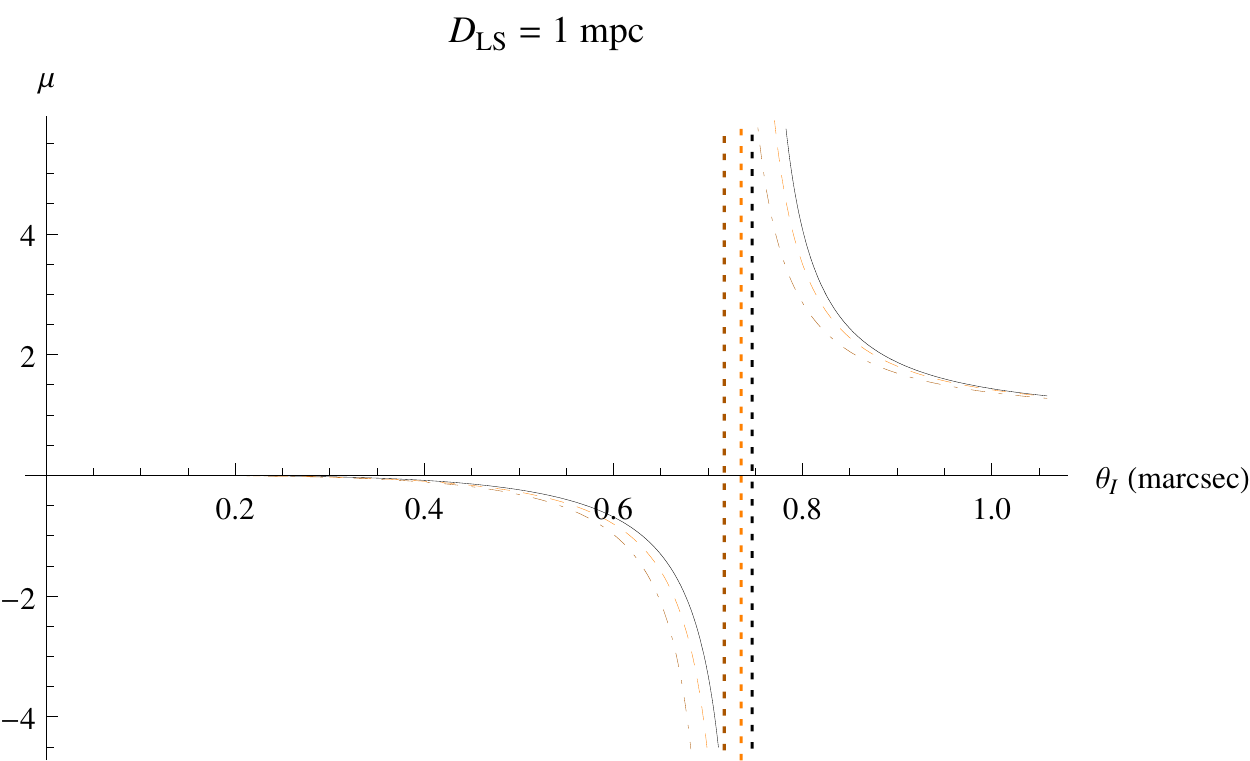}}
\caption{$\mu(\theta_I)$ sagittarius with $20<b_h<100$ in the photon case for a black hole with M = 4.31 $10^6\,\text{M}_\odot$: $D_{LS}$ = 1 Kpc (a), $D_{LS}$ = 1 pc (b), $D_{LS}$ = 1 mpc (c) (Sagittarius).\label{musagitt}}
\end{figure}

\begin{figure}[!htb]
\centering
\subfigure[]{\includegraphics[scale=.76]{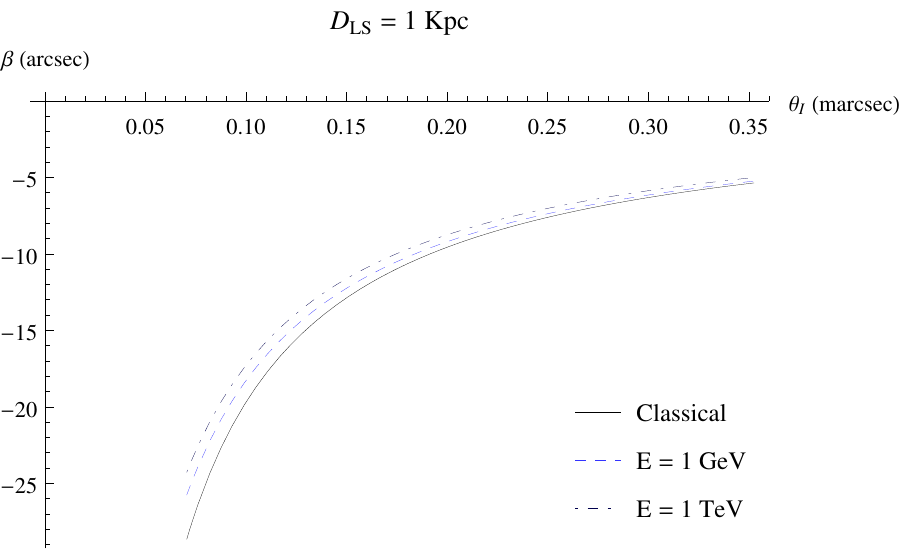}} \hspace{.5cm}
\subfigure[]{\includegraphics[scale=.58]{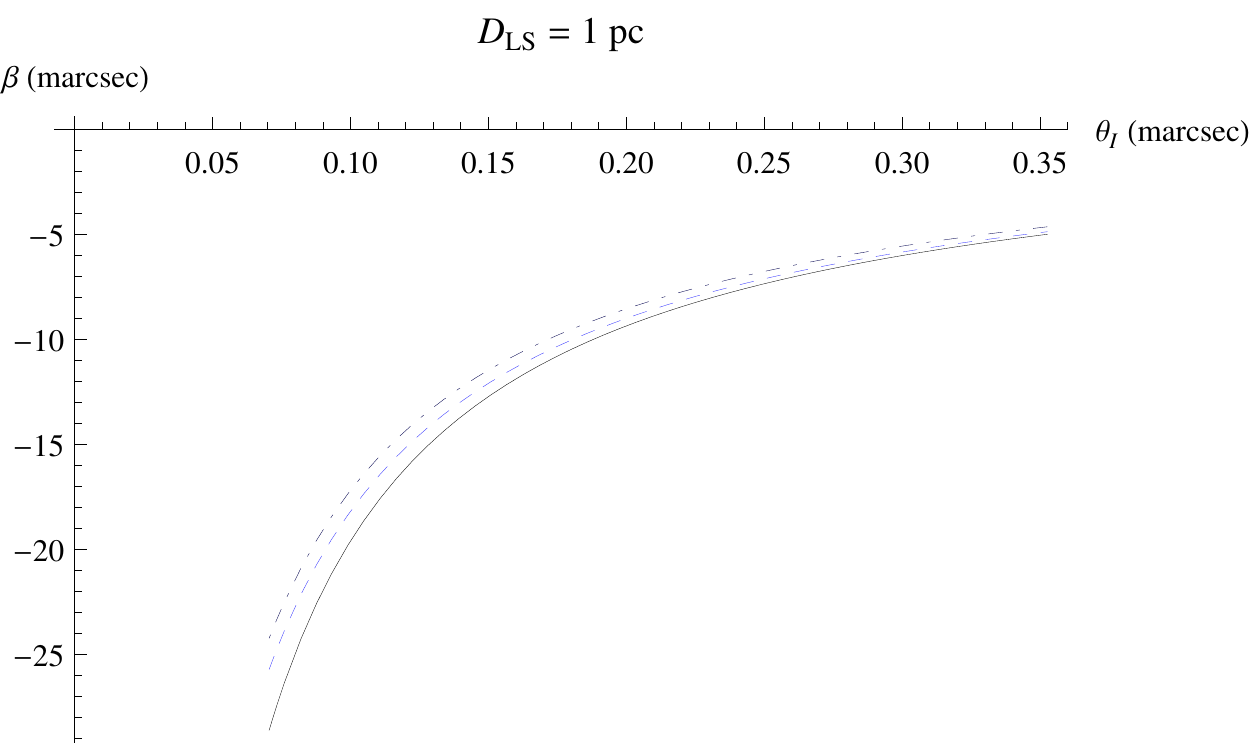}} \hspace{.5cm}
\subfigure[]{\includegraphics[scale=.58]{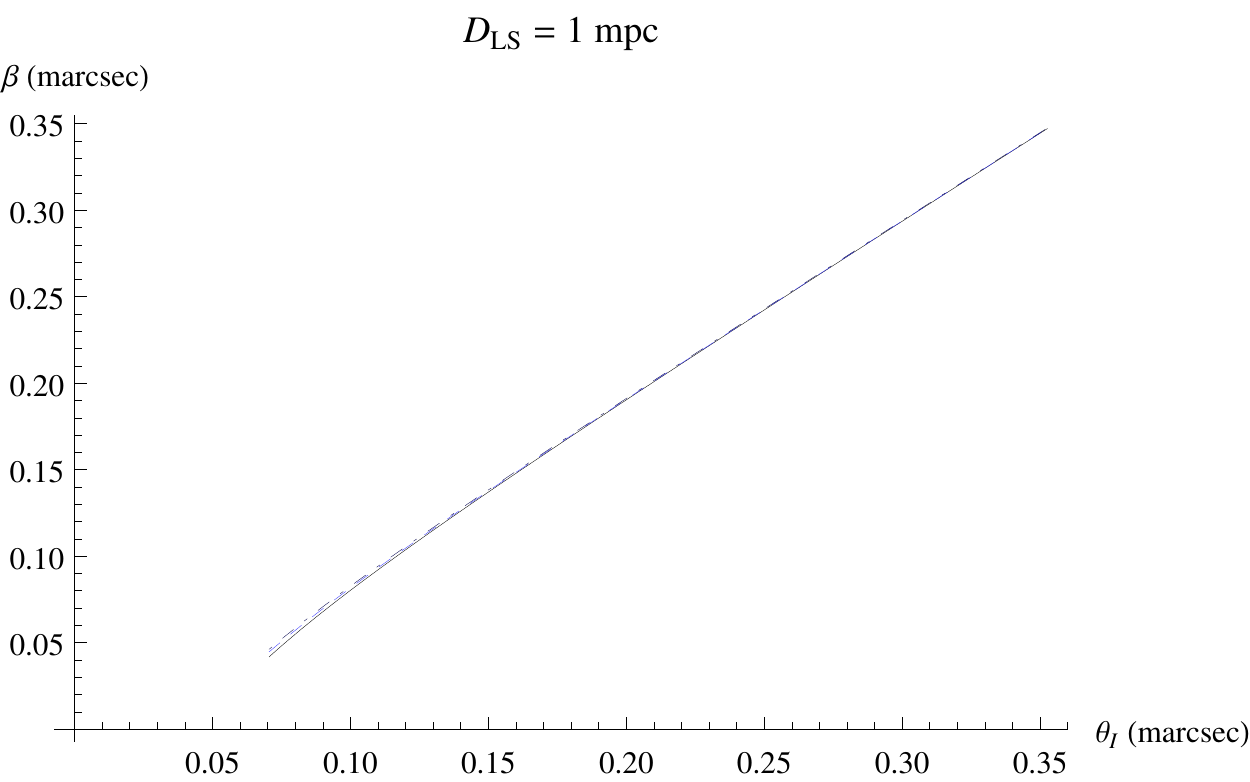}}
\caption{$\beta(\theta_I)$ with $20<b_h<100$ in the photon case for a black hole with M = 1.40 $10^8\,\text{M}_\odot$: $D_{LS}$ = 1 Kpc (a), $D_{LS}$ = 1 pc (b), $D_{LS}$ = 1 mpc (c) (Andromeda). 
\label{betaAndr}}
\end{figure}

\begin{figure}[!htb]
\centering
\subfigure[]{\includegraphics[scale=.76]{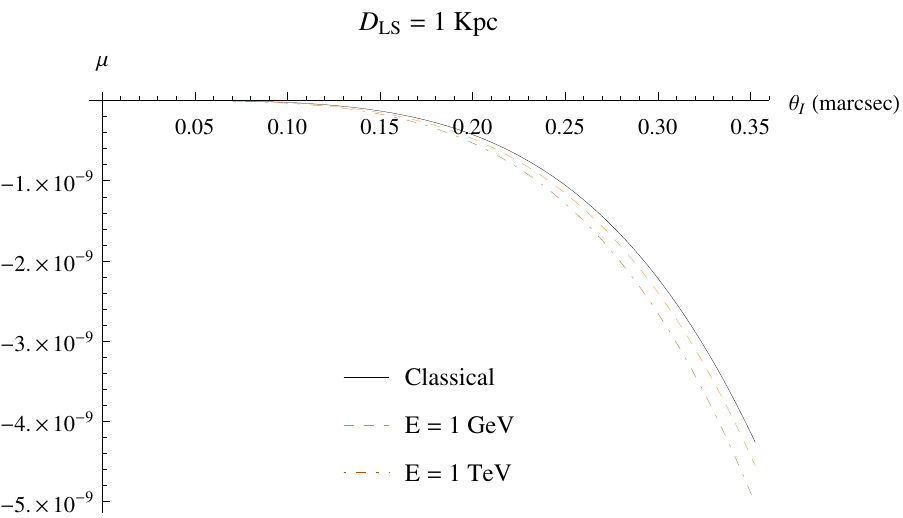}}\hspace{.5cm}
\subfigure[]{\includegraphics[scale=.58]{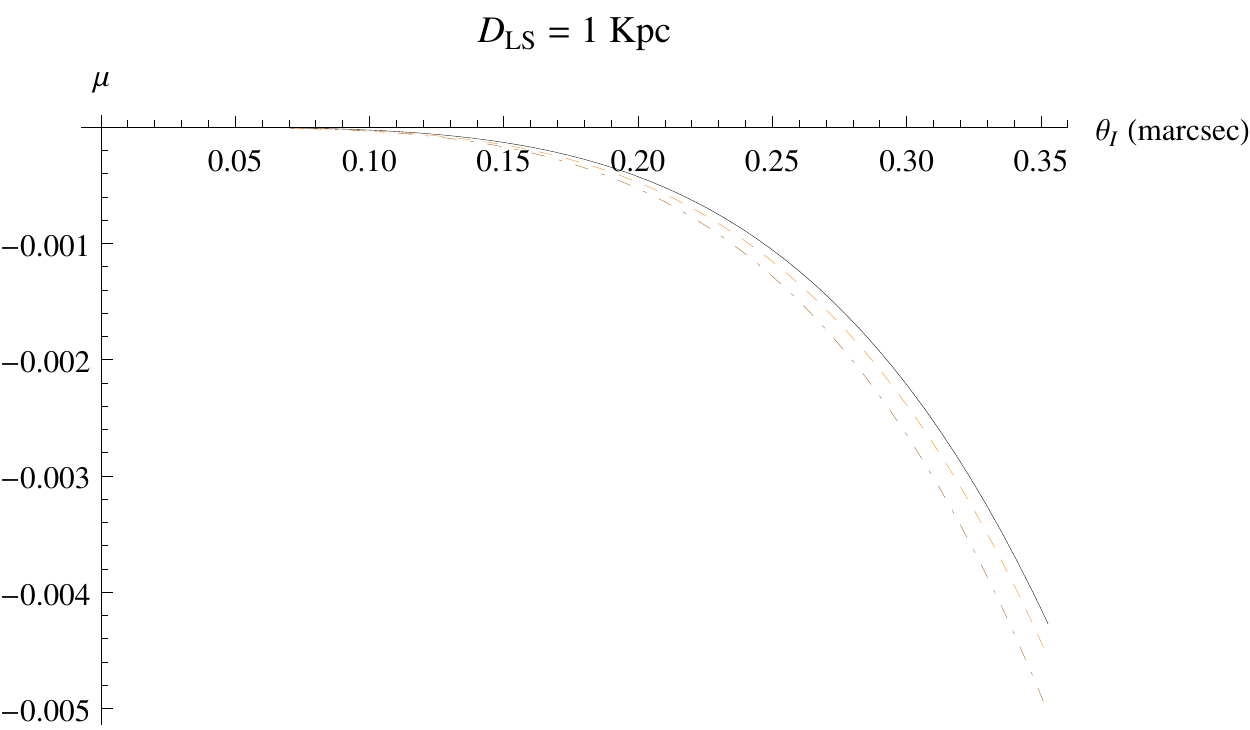}}\hspace{.5cm}
\subfigure[]{\includegraphics[scale=.58]{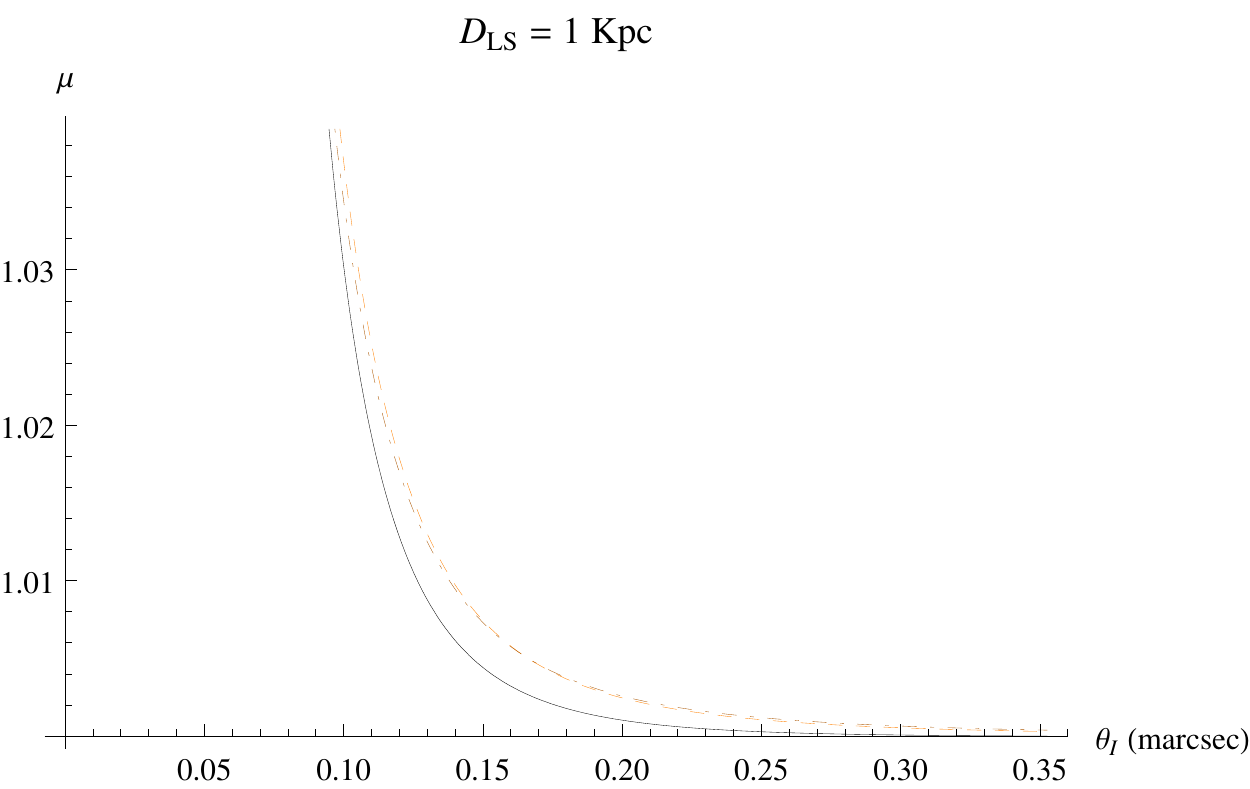}}
\caption{$\mu(\theta_I)$ with $20<b_h<100$ in the photon case for a black hole with M = 1.40 $10^8\,\text{M}_\odot$: $D_{LS}$ = 1 Kpc (a), $D_{LS}$ = 1 pc (b), $D_{LS}$ = 10 mpc (c). The strong suppression of the magnification parameter, in this case, is associated to the secondary image for this geometric configuration 
(Andromeda).\label{muAndr}}
\end{figure}

\subsection{Numerical analysis for neutrino and photon lensing}
The analysis that we present in this section, especially in the neutrino case, is of exploratory nature.  It has the aim to test the consistency of the theoretical approach presented in the previous sections, rather than being an explicit proposal for the detection of such effects.  \\
We start investigating the behaviour of the solutions for the VE lens equation, which are shown in Figs.~\ref{VHc1}, by plotting the angular position of the source $(\beta)$ as a function of the location of the image $\theta_I$, for neutrino beams. In panel (a) we have chosen distances between lens/source and observer of galactic size. The branch of the solution with $\theta_I>0$ describes a primary image, while for $\theta_I <0$ the points $(\beta,\theta_I)$ on the curve describe configurations corresponding to secondary images. One can test the energy dependence of the solution by varying the energy of the original beam in the lens equation (\ref{lens1}), which, for this specific geometry, is clearly unnoticeable. In general, in fact, specific geometries $(D_{OL},D_{LS})$ select impact parameters $b_h$ in the beam path which are quite large. In this case the impact parameter turns out to be pretty large ($b_h\sim10^5$), corresponding to very weak deflections, and causes a superposition between the two curves, the one describing the classical GR solution of (\ref{lenseq}), and the semiclassical one, obtained by solving (\ref{lens1}). The curves intersect the $\theta_I$ (image) axis in two opposite points $(\theta_I=\pm \theta_E)$, giving rise to the Einstein ring, which are obtained for $\beta=0$, i.e. for a complete alignment of the lens/source/observer along the optical axis. In this figure the negative $\beta$ range is symmetric and hence it is not shown. It can be obtained by a parity flip of the two positive branches with $\beta\to-\beta$ and $\theta_I\to - \theta_I$.
\\
In Fig.~\ref{VHc1} (b) we investigate the dependence of the classical solution for three values of $D_{LS}$. The solution curves so generated exhibit variations which are clearly far more significant that any radiative correction which might affect the lensing geometry. Notice that for a given solution of the lens equation $\theta_I$, as we increase the distance $D_{LS}$, for a given angular position of the image, the source  moves towards the optical axis. This behaviour, in the secondary image, appears to be reversed: in this case larger values of $D_{LS}$ require larger angular values of  $\beta$, for a given angular position $\theta_I$ of the image. \\
Solutions with smaller values of the impact parameters are those which are more favourable from the point of view of the semiclassical treatment, since in these cases the deflections are larger and induce larger 
gradients into the lens equation (\ref{lens1}).
For this reason we investigate two lensing configurations corresponding to the case of the supermassive black holes located at the center of our galaxy (Sagittarius A$^*$, 8 kpc) and of the nearby galaxy Andromeda ({$D_{OL}\sim 780$ kpc), and vary the distance $D_{LS}$ between the source and the lens. \\ 
We show in Fig.~\ref{betasagitt1} the solutions for the Sagittarius configuration. In panels (a) and (b) the solutions that we identify correspond to secondary images obtained for very small values of $\theta_I$, of the order of a milliarcsecond. These are the only configurations which guarantee close encounters between the cosmic ray beam and the black hole, with $20<b_h<100$. As the distance $D_{LS}$ gets reduced, 
the equation has solutions with values of $\theta_I$ which define primary images, being $\beta >0$ (panel $\textrm{c}$). At the same time, as shown in panel (\textrm{c}), the angular position of the source moves from $\beta<0$ to $\beta>0$. A primary image is shown to form when $D_{LS}$ is 1 milliparsec (mpc). \\ 
Simple considerations show that such lensing configurations are not unreasonable. For instance, for the supermassive black hole that we consider (with M=$4.31\times 10^6 M_\odot$), this distance is of the order of $5\times 10^3 r_s$, with $r_s$ denoting its Schwarzschild radius ($\sim 6 \times 10^6$ km). Being the center of our galaxy rather densely populated by massive compact sources, one could envisage a distribution of these covering a large array of possible distances from the center of the black hole. For instance, it has been found that stars may orbit the supermassive galactic black hole with orbital periods even of the order of $T\sim 11.4$ ys, corresponding to orbital distances $(\text{R})$  from its center as close as few astronomical units (AU),  $R=T^{2/3}\sim 5$ AU, i.e. $R\sim 125$ $r_s$. While such distances 
may correspond to realistic lensing configurations, the lensing resolutions of these specific events, which is of the order of a milliarcsecond, remains a challenging aspect of these studies. This is due to the strong quasi-alignment required between source, lens and observer along the optical axis, which might be difficult to measure.  

The energy dependence of the lensing configuration, extracted from Eq.~(\ref{lens1}), is illustrated by plotting the solutions for several values of the initial energy of the neutrino beam, corresponding to 1 GeV, 1 TeV and 1 PeV respectively. A comparison with the classical GR solution is included. Differences among the various predictions for the position of the source can be as large as $15 \%$.\\
The analysis for the magnification of this lensing configuration is presented in Fig.~\ref{musagitt1}. Panels (a) and (b)  show the strong suppression of the secondary images identified in Fig.~\ref{betasagitt1} (a) and (b). The  magnification of the secondary and primary images of Fig. \ref{betasagitt1} (\textrm{c}) is shown in Fig.~\ref{musagitt1} (\textrm{c}). The $\mu<0$ and $\mu>0$ regions, in this figure, are separated by asymptotes for $\theta_I=\theta_E$, the Einstein angle of the lens. These regions in $\mu$ correspond to the secondary ($\beta<0$) and ($\beta>0$) branches of the panel $(\textrm{c})$ in Fig.~\ref{musagitt1} and therefore refer to secondary and primary pictures respectively. \\ The dependence on the energy 
of the incoming neutrino beam appears in the form of 3 displaced curves and respective asymptotes. The three vertical asymptotes therefore characterize the dependence of the Einstein radius on the energy. \\
The analysis is repeated in Figs.~\ref{Andro1} and \ref{Andro2} for lensing events detected on Earth from the galactic center of the Andromeda galaxy. In this case as before, we vary the distance between the lens and the 
source, with $D_{LS}=1$ kpc, 1 pc and 1 milliparsec in panels (a), (b) and (\textrm{c}) respectively. The first two plots correspond to secondary images while panel (\textrm{c}) describes a primary image solution of the lens equation. The corresponding plots for the magnifications, in the three cases, are shown in Fig.~\ref{Andro2}, evidencing its strong suppression for the two secondary images (panels (a) and (b)), and the enhancement for the primary image in Fig.~\ref{Andro1} (\textrm{c}), shown in panel (\textrm{c}). Also in this case we illustrate the dependence of the image solutions on the energy of the incoming neutrino beam. \\
The patterns found for the neutrino lenses remain valid also for the photons, as one can easily figure out by a cursory look at Figs. \ref{betasagitt} and {\ref{musagitt} for the supermassive black hole in Sagittarius. 
In panels (a) and (b) of Fig. \ref{betasagitt} the secondary images found as solution of the lens equation are suppressed in magnitude, as shown in Fig. \ref{musagitt} (a) and (b), while the solution in Fig. \ref{betasagitt} (\textrm{c}), corresponding to one primary and one secondary image, is associated with the magnification given in Fig. \ref{musagitt} ({\textrm{c}). In these two sets of figures the energy dependence of the result is quite small. \\
Finally, the numerical result for the photon lensing case, generated by the supermassive black hole in the Andromeda galactic center is discussed in Figs. \ref{betaAndr} and \ref{muAndr}. While the two secondary images found in \ref{betaAndr} (a) and (b) give suppressed magnifications, the solution in panel (\textrm{c}) corresponds to a primary image. The corresponding magnification, shown in Fig. \ref{muAndr} has, obviously, a single branch, with an Einstein radius located at approximately $7\times 10^{-5}$ arcsec.

\section{Post-Newtonian corrections: the case of primordial black holes}
We have seen in the previous sections that the $b_h(\alpha)$ expression for the deflection 
does not suffer from any apparent divergence (from the gravity or external field side) due to well-defined structure of the Newtonian cross section. The expression given in (\ref{leading}), in fact, is similar to the ordinary Rutherford scattering encountered in electrodynamics.\\
 The dependence of the resulting cross section on the scale $G M/c^2$, the Schwarzschild radius, manifests as an overall dimensionful constant. 
Therefore, the inclusion of the electroweak corrections - and the logarithmic dependence on the energy of the terms in the expansion that follows - do not appear in combination with the macroscopic scale $r_s$. This allows, in principle, an extension of the perturbative computation up to any order in the electroweak coupling constant $\alpha_w$. It is also clear that this result is expected to be valid for any renormalizable field theoretical model, when combined with an external static gravitational field of Coulomb type, as in the case of the Newtonian limit of GR.\\
From now on, we will be using the notation nPN to indicate the (post-Newtonian) order in the potential at which we expand the Schwarzschild metric. For instance, contributions of a certain nPN order involve corrections in the external field proportional to $\Phi^{n+1}$, with 0PN denoting the ordinary (lowest order) Newtonian (i.e. zeroth post-Newtonian) contributions proportional to $\Phi$, as given in Eq. (\ref{h2}).
The inclusion of the higher order corrections in the external potential modifies this simple picture due 1) to the need of introducing a cutoff regulator in the computation of the Fourier transform of the higher powers of the Newtonian potential and 2) to the presence of the Schwarzschild radius $r_s$ in the actual expansion. These features emerge already at the first post-Newtonian order (1PN) for an uncharged black hole and at order 0PN for the Reissner-Nordstrom (RN) metric (charged black hole). \\
Both points 1) and 2) are, in a way, expected, since the microscopic expression for the transition matrix element given by (\ref{volume}), in fact, cannot be extrapolated to the case of a macroscopic source, with the presence of a macroscopic scale such as the black hole horizon. This seems to indicate that the 
success of the Newtonian approximation is essentially due to the rescaling of $r_s$ found in the expression of the cross section, which is a feature of this specific order, and is therefore limited to a $1/r$ potential. It is then natural to ask if there is any other realistic case in which the post-Newtonian corrections can be included in an analysis of this type. Obviously, the answer is affirmative, as far as we require that $r_s$ is microscopic and that the energy of the beam, which is an independent variable of a scattering event, is at most of the order of $1/r_s$. Under these conditions, we are then allowed to extend our analysis through higher orders in $\Phi$, with scatterings in which the dimensionless parameter $r_s q$ with $q$ the impact parameter, is at most of $\mathcal{O}(1)$. 
This specific situation is encountered in the case of primordial black holes, where $r_s$ can be microscopic. 
We are going to illustrate this point in some detail, since it becomes relevant in the case of primordial black holes. 
\subsection{Post Newtonian contributions in classical GR} 
To illustrate this point we extend the expansion of the Schwarzschild metric at order 0PN given in (\ref{SCH3}). A similar expansion will be performed on the RN metric. \\
For this purpose, it is convenient to perform a 
change of coordinates on the Schwarzschild metric 
\bea
\label{ds}
ds^2=\left(1-\frac{2\,G\,M}{r}\right)dt^2-\left(1-\frac{2\,G\,M}{r}\right)^{-1}dr^2-r^2d\Omega
\eea
in such a way that this takes an isotropic form. The radial change of coordinates is given by 
\bea
r=\rho\left(1+\frac{G\,M}{2\rho}\right)^2
\eea
which allows to rewrite (\ref{ds}) as 
\bea
ds^2=A(\rho)dt^2-B(\rho)(d\rho^2+\rho^2\,d\Omega)
\eea
with
\bea
\label{ABsch}
A(\rho)=\frac{(1-G\,M/2\rho)^2}{(1+G\,M/2\rho)^2}\qquad\qquad B(\rho)=(1+G\,M/2\rho)^4 \,.
\eea
Post-Newtonian (weak field) corrections can be obtained by an expansion of $A$ and $B$ taking $M/\rho\ll1$. Up to third order in $\Phi$ this is given by
\bea
&&A(\rho)=1+2\,\Phi+2\,\Phi^2+\frac{3}{2}\,\Phi^3\\
&&B(\rho)=1-2\,\Phi+\frac{3}{2}\,\Phi^2-\frac{1}{2}\,\Phi^3.
\eea
In the RN spacetime for a charged black hole the analysis runs similar. The interest in this metric is due to the fact that the lowest order potential, in this case, involves charge-dependent $1/r^2$ contributions which, for an uncharged black hole, appear at first post-Newtonian order (1PN). The metric, in this case, is given by the expression
\bea
\label{RN}
ds^2=\left(1-\frac{2\,G\,M}{r}+\frac{G\,Q^2}{r^2}\right)dt^2-\left(1-\frac{2\,G\,M}{r}+\frac{G\,Q^2}{r^2}\right)^{-1}dr^2-r^2d\Omega,
\eea
with $Q$ denoting the overall charge of the black hole. It has two concentric horizons which become degenerate in the maximally charged case. The two horizons are the solution of the equation
\bea
\left(1-\frac{2\,G\,M}{r}+\frac{G\,Q^2}{r^2}\right)=0
\eeqa
with solutions $r=G\,M\pm\sqrt{G^2M^2-G\,Q^2}$.
The RN black hole has a maximum allowed charge $Q=M\sqrt G$, in order to avoid a naked singularity.
In this case, the radial change of variables which brings the metric into a symmetric form is given by
\bea
r=\rho\left(1+\frac{G\,M+\sqrt G\,Q}{\rho}\right)\left(1+\frac{G\,M-\sqrt G\,Q}{\rho}\right),
\eea
so that the RN spacetime in isotropic coordinates is
\begin{align}
\label{RNiso}
ds^2=&\frac{\left(1-\frac{G^2\,M^2-G\,Q^2}{4\rho^2}\right)^2}{\left(1+\frac{G\,M+\sqrt G\,Q}{2\rho}\right)^2\left(1+\frac{G\,M-\sqrt G\,Q}{2\rho}\right)^2}dt^2\nn\\
&-\left(1+\frac{G\,M+\sqrt G\,Q}{2\rho}\right)^2\left(1+\frac{G\,M-\sqrt G\,Q}{2\rho}\right)^2(d\rho^2+\rho^2\,d\Omega).
\end{align}
 We just recall that for a massless particle in this metric background the angle of deflection and the impact parameter are given by the expressions
\bea
&&\alpha(r_0)=2\,\int^\infty_{r_0}\frac{dr}{r\sqrt{\frac{r^2}{r_0^2}\left(1-\frac{2\,GM}{r_0}+\frac{GQ^2}{r_0^2}\right)-\left(1-\frac{2\,GM}{r}+\frac{GQ^2}{r^2}\right)}}-\pi\\
&&b(r_0)=\frac{r_0}{\sqrt{1-\frac{2\,GM}{r_0}+\frac{GQ^2}{r_0^2}}}
\eea
where $r_0$ is the closest distance of approach. It's convenient to normalize $r$, $r_0$ and $Q$ to the Schwarzshild radius $r_s=2\, G M$ and introduce the variables 
\bea
x=\frac{r}{2\, G M}\qquad x_0=\frac{r_0}{2\, G M}\qquad q=\frac{Q}{2\, G M}.
\eea
With this redefinitions the deflection can be expressed in the form \cite{Eiroa:2002mk}
\bea
\alpha(x_0)=G(x_0)\,\mathrm{F}(\phi_0, \lambda)-\pi
\label{elli}
\eea
with
\bea
G(x_0)=\frac{4\,x_0}{\sqrt{1-\frac{1}{x_0}+\frac{q^2}{x_0^2}}}\frac{1}{\sqrt{(r_1-r_3)(r_2-r_4)}}
\eea
and with
\bea
\mathrm{F}(\phi_0,\lambda)=\int_0^{\phi_0}(1-\lambda\,\sin^2\phi)^{-1/2}d\phi
\eea
being an elliptic integral of the first kind with arguments
\bea
&&\phi_0=\arcsin \sqrt{\frac{r_2-r_4}{r_1-r_4}}\\
&&\lambda=\frac{(r_1-r_4)(r_2-r_3)}{(r_1-r_3)(r_2-r_4)}.
\eea
The $r_i$ are the roots of the fourth order polynomial
\bea
P(x)=x^4+\frac{x_0^2}{1-\frac{1}{x_0}+\frac{q^2}{x_0^2}}\,(x-x^2-q^2)
\eea
ordered so that $r_1>r_2>r_3>r_4$. The comparison between Schwarzschild and RN deflection angle is shown in Figure \ref{schRN}. The plots 
describe the behaviour of the angular deflection as a function of the impact parameter $b_h$ for a RN and Schwarzschild metric in the region with 
$10 < b_h < 50$ (top left) and $4 < b_h < 10$ (top right) for the maximally charged case. The differences tend to be very pronounced as we approach the horizon of the Schwarzschild metric.\\
As pointed out in \cite{Amore:2006pi} in the Schwarzschild case, the $1/b$ expansion for the deflection angle does not reproduce the photon sphere singularity of the Schwarzschild metric, which is achieved using the exact GR expression in terms of elliptic function given in (\ref{elli}), but it represents nevertheless an improvement respect to the $0PN$ order.
Expanding the RN metric in $M/\rho\ll1$ up to the third order, the $2PN$ approximation gives
\bea
&&A(\rho)=1-\frac{2\,G\,M}{\rho}+\frac{2\,G^2\,M^2+G\,Q^2}{\rho^2}-\frac{3\,G^3\,M^3+5\,G^2\,M\,Q^2}{2\,\rho^3}\\
&&B(\rho)=1+\frac{2\,G\,M}{\rho}+\frac{3\,G^2\,M^2-G\,Q^2}{2\,\rho^2}+\frac{G^3\,M^3-G^2\,M\,Q^2}{2\,\rho^3}.
\eea
Inserting this expansion into the deflection integral, we can account in a systematic way of the $1/b$ corrections in the angle of deflection $\alpha$  
\bea
\label{alphaRN}
\alpha(b)=4\,\frac{G\,M}{b}+\left(5-\frac{G\,Q^2}{M^2}\right)\frac{3\pi}{4}\,\frac{G^2M^2}{b^2}+\left(\frac{128}{3}-16\frac{G\,Q^2}{M^2}\right)\,\frac{G^3M^3}{b^3}.
\eea
 The deflection (\ref{alphaRN}) in the maximally charged case is given by the expression
\bea
\alpha^{\textit{m.c.}}=4\,\frac{GM}{b}+3\pi\,\frac{(GM)^2}{b^2}+\frac{80}{3}\,\frac{(GM)^3}{b^3}.
\eea
In the next subsection we are going to illustrate how the inclusion of these expansions at nPN order affects the computation of the quantum corrections to the angular deflection. The corrections are embodied in a geometric form factor whose expression is entirely controlled by the $1/b$ expansion.

\begin{figure}[t]
\centering
\subfigure[]{\includegraphics[scale=.7]{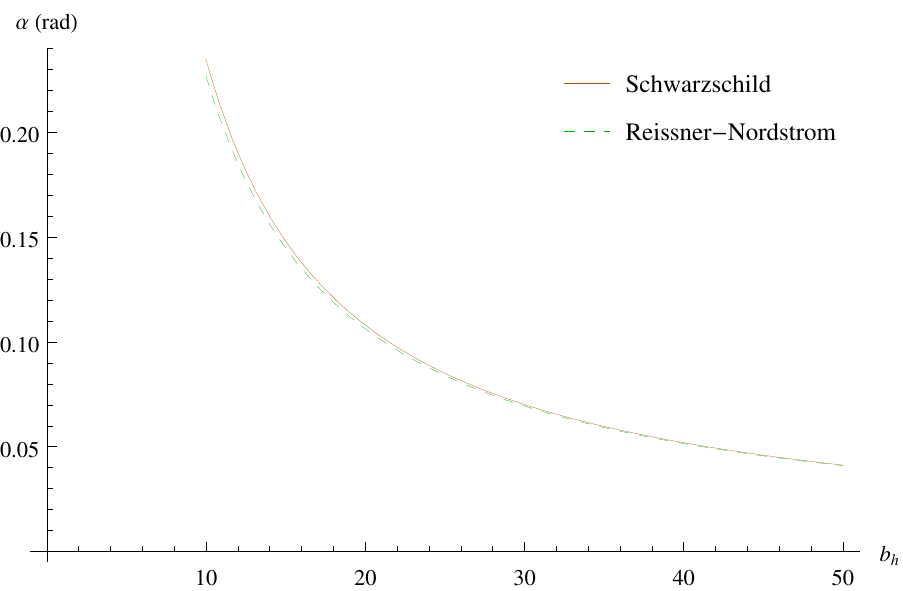}}\hspace{.5cm}
\subfigure[]{\includegraphics[scale=.7]{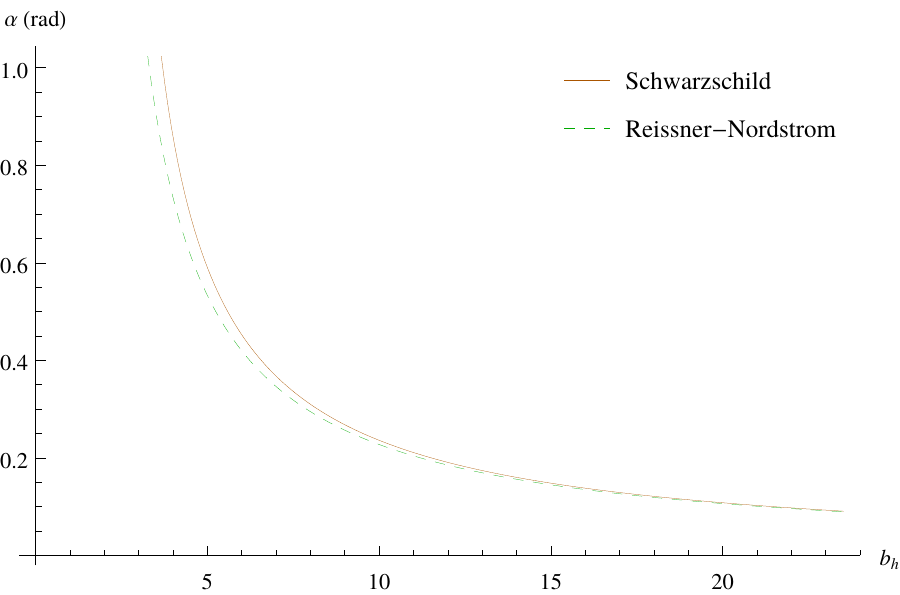}}
\caption{Comparison of the deflection angle for the Schwarzshild case and the maximally charged Reissner-Nordstrom case in the near-horizon (a), in the very-near-horizon (b).\label{schRN}}
\end{figure}
\subsection{Quantum effects at 2nd PN  order} 
The inclusion of the PN corrections to the external background requires a recalculation of the cross section, with the inclusion of the additional 
terms in the fluctuation of the metric in momentum space. 
As usual we consider a static source, so that the metric is written as
\begin{equation}
h_{\mu\nu}(q)=2\pi\delta(q_0) h_{\mu\nu}(\vec{q}\,).
\end{equation}
At leading order in the external field $\Phi$ both the timelike and the spacelike components are equal ( $h_{00}\equiv h_{ii}$), while at higher orders they are expressed in terms of two form factors $h_0$ and $h_1$
 \begin{equation}
h_{\mu\nu}(\vec{q}\,)=h_0(\vec{q}\,) \delta_{0\mu}\delta_{0\nu}+h_1(\vec{q}\,)\bigl( \eta_{\mu\nu}-\delta_{0\mu}\delta_{0\nu} \bigr),
\end{equation}
which at higher order in the weak external field are given by 
\begin{equation}
\label{eq:ffh0h1}
\begin{split}
h_0(\vec{q}\,)&=-\frac{2}{\kappa} \int d^3\vec{x}\, \biggl[  \frac{\Phi}{c^2}  + \biggl( \frac{\Phi}{c^2} \biggl)^{\!\!2} + \frac{3}{4} \biggl( \frac{\Phi}{c^2} \biggl)^{\!\!3\,}   \biggr] \text{e}^{i \vec{q} \cdot \vec{x}}  \\
h_1(\vec{q}\,)&= -\frac{2}{\kappa} \int d^3\vec{x}\,  \biggl[  -\frac{\Phi}{c^2}  +\frac{3}{4} \biggl( \frac{\Phi}{c^2} \biggl)^{\!\!2} - \frac{1}{4} \biggl( \frac{\Phi}{c^2} \biggl)^{\!\!3\,}  \biggr] \text{e}^{i \vec{q} \cdot \vec{x}}, 
\end{split}
\end{equation}
where we have explicitly reinstated the dependence on the speed of light. Below we will conform to our 
previous notations in natural units, with $c=1$.
\begin{itemize} 
\item{\bf \large Neutrinos}
\end{itemize}
The computation, at this stage, follows rather closely the approach of the previous sections, giving for 
 the averaged squared matrix element in the neutrino case 
\begin{equation}
\left|iS_{fi}\right|^2=\frac{\kappa^2}{16 V^2 E_1 E_2} 2\pi \delta(q_0) \,\mathcal{T} \,\frac{1}{2} \, \text{tr}\biggl[ \slashed{p}_2 V^{\mu\nu}(p_1,p_2) \slashed{p}_1 V^{\rho\sigma}(p_1,p_2) \biggr] h_{\mu\nu}(\vec{q}\,) h_{\rho\sigma}(\vec{q}\,),
\end{equation} 
with $\mathcal{T}$ being the time of the transition,
and the differential cross section 
\begin{equation}
d \sigma=\frac{d \mathcal{W}}{\mathcal{T}}=\frac{\left|S_{fi}\right|^2}{j_i \mathcal{T}}dn_f.
\end{equation}
We have denoted with $d n_f$ the density of final states in the transition amplitude, and with $j_i$ the incoming flux density. 
After integration over the final states, and using $|\vec{p}_1|=|\vec{p}_2|$, we obtain the expression
\begin{equation}
\left.\frac{d\sigma}{d\Omega}\right|_{\text{2PN}}^{(0)}=\frac{\kappa^2}{16 \pi^2} E^4 \cos^2\frac{\theta}{2} F_g(q)^2,
\end{equation}
where we have introduced the gravitational form factor of the external source
\beq
\label{fg}
F_g(q)\equiv \Bigl( h_0(\vec{q}\,)-h_1(\vec{q}\,) \Bigr).
\eeq
Notice the complete analogy between the corrections coming from a distributed source charge, for a potential scattering in quantum mechanics, and the gravity case. 
In the evaluation of $F_g$ in momentum space we are forced to introduce a cutoff $\Lambda$, being the Fourier transforms of the cubic contributions in $\Phi$ divergent. The singularity is generated by the integration around the region of $r\sim 0$ in the Fourier transform of the potential. The relevant integrals in this case are given by 
\begin{equation}
\begin{split}
I_n&=\int d^3\vec{x}\,\, \frac{1}{{|\vec {x}|}^n} \text{e}^{i\vec{q}\cdot \vec{x}}\\
\end{split}
\end{equation}
with 
\beq
I_1=\frac{4\pi}{\vec{q}\,^2}, \qquad  I_2 = \frac{2\pi^2}{|\vec{q}|}.  
\eeq
and with $I_3$ requiring a regularization with an ultraviolet cutoff  in space ($\Lambda$) 
\beq
I_3 =  \frac{4\pi}{|\vec{q}\,| }  \int_\Lambda^{\infty} dr\,\, \frac{\sin(|\vec{q}\,| r)}{r^2}.\\
\eeq
The choice of $\Lambda$ is dictated by simple physical considerations. Given the fact that consistency of the expansion requires that $r_s q \lesssim \mathcal{O}(1)$, it is clear the appropriate choice in the regulator is given by the condition that this coincides with the Scwarzschild radius, i.e. $\Lambda\sim r_s$.  Expressed in terms of the cutoff, we obtain for the geometric form factors the expressions 
\begin{equation}
\begin{split}
h_0(\vec{q}\,)&=-\frac{2}{\kappa} \biggl[ - \frac{4 \pi}{|\vec{q}\,|^2}GM  + \frac{2\pi^2}{|\vec{q}\,|}(GM)^2 - \frac{3}{4} \frac{4\pi}{|\vec{q}\,|}\biggl( \frac{\sin(\Lambda |\vec{q}\,|)}{\Lambda} - |\vec{q}\,| \text{Ci}(\Lambda |\vec{q}\,| )\biggl)(GM)^{3}   \biggr]  \\
h_1(\vec{q}\,)&= -\frac{2}{\kappa} \biggl[    \frac{4 \pi}{|\vec{q}\,|^2}GM  +\frac{3}{4} \frac{2\pi^2}{|\vec{q}\,|}(GM)^2 + \frac{1}{4} \frac{4\pi}{|\vec{q}\,|}\biggl( \frac{\sin(\Lambda |\vec{q}\,|)}{\Lambda} - |\vec{q}\,| \text{Ci}(\Lambda |\vec{q}\,|) \biggl)(GM)^{3}  \biggr] , \\
\end{split}
\end{equation}
where we have indicated with $\text{Ci}$ the cosine integral function 
\begin{equation}
\text{Ci}(x)= \int_{\infty}^x dt\, \frac{\cos t}{t}.
\end{equation}
From the previous equations we obtain the cross section
\begin{equation}
\left.\frac{d\sigma}{d\Omega}\right|_{\text{2PN}}^{(0)}=\frac{1}{4 \pi^2}E^4 \cos^2\frac{\theta}{2} \biggl[ \frac{8 \pi}{|\vec q\,|^2}GM  - \frac{\pi^2}{2|\vec{q}\,|}(GM)^2 + \frac{4\pi}{|\vec{q}\,|} \biggl( \frac{\sin(\Lambda |\vec{q}\,|)}{\Lambda} - |\vec{q}\,| \mathrm{Ci}(\Lambda |\vec{q}\,|) \biggl)(GM)^{3}   \biggr]^2,
\end{equation}
which is valid at Born level and includes the weak field corrections up to the third order in $\Phi$. In the expression of the cross sections, we use the subscript nPN, with $n=0,1,2$ to indicate a n-{th} order expansion of the metric in the gravitational potential, while the superscripts ((0), (1) and so on) label the perturbative order in $\alpha_w$. The leading order cross section at order 2PN, for instance, takes the form 

\begin{equation}
\left.\frac{d\sigma}{d\Omega}\right|_{\text{2PN}}^{(0)}=\left.\frac{d\sigma}{d\Omega}\right|_{\text{0PN}}^{(0)}\,\mathcal{PN}_2(E,\theta) ,
\end{equation}
with
\begin{align}
\mathcal{PN}_2(E, \theta)\equiv&\biggl[1-\frac{\pi}{8}(GM)\,E\sin\frac{\theta}{2}+\frac{1}{2}(GM)^2\,E \sin\frac{\theta}{2}\biggl( \frac{1}{\Lambda}\sin\left(2\,\Lambda\,E\sin\frac{\theta}{2}\right)\nn\\
&- 2E\sin\frac{\theta}{2} \mathrm{Ci}\biggl(2\,\Lambda\,E\sin\frac{\theta}{2}\biggr)\biggr)\biggr]^2,
\label{PN}
\end{align}
where we have factorized the tree level result ${d\sigma}/{d\Omega}|_{\text{0PN}}^{(0)}$ given in (\ref{leading}). The post-Newtonian form factor $\mathcal{PN}_2(E, \theta)$ induces an energy dependence of the cross section which is unrelated to the electroweak corrections. 
The analysis, in fact, can be extended at one loop in the electroweak theory. In this case, a lengthy computation gives the 2PN result 
\begin{equation} 
 \frac{d\sigma}{d\Omega}\biggr|_{\text{2PN}}^{\text{(1)}} =\frac{d\sigma}{d\Omega }\biggr|_{\text{0PN}}^{(0)}  \left[ 1 + \frac{4 G_F}{16 \pi^2\sqrt{2}}  \left( f_W^1(E,\theta) + f_Z^1(E,\theta) - \frac{1}{4} \Sigma^L_W - \frac{1}{4} \Sigma^L_Z  \right)  \right]\,\mathcal{PN}_2(E, \theta),
 \label{third}
\end{equation}
where we have inserted the one loop expression given in \eqref{sigmaOL}.\\
We can obtain an explicit solution of the corresponding semiclassical equation at order 1PN. Using the expression of the $\mathcal{PN}$ function at this order
\bea
\mathcal{PN}_1(E, \theta)&\equiv&\biggl[1- \frac{\pi}{8}(GM)\,E\sin\frac{\theta}{2}\biggr]^2
\eea
 on the right hand side of (\ref{PN}) in order to generate the 1PN cross section at Born level, and solving the 
 corresponding semiclassical equation (\ref{semic}) we obtain 
\begin{align}
\label{nbpn}
\left.b^2\right|_{\text{1PN}}^{(0)}&=4\,(GM)^2\Big(-1+\csc^2\frac{\alpha}{2}+2\ln\sin\frac{\alpha}{2}\Big) + E\, (GM)^3\pi\Big(4+(\cos\alpha-3)\csc\frac{\alpha}{2}\Big)\nn\\
&-\frac{1}{32}E^2(GM)^4\pi^2\Big(1+\cos\alpha+4\ln\sin\frac{\alpha}{2}\Big).
\end{align}
At this point, we can invert Eq.~(\ref{nbpn}) for $\alpha(b)$ obtaining
\begin{align}
\left.\alpha\right|_{\text{1PN}}^{(0)}&=\frac{2}{b_h}- \frac{\pi}{2}\frac{1}{b^2_h}E\,(GM)-\frac{1}{b^3_h}\Big(\ln b_h\,\Big(2-\frac{\pi^2}{32}\,E^2(GM)^2\Big)+\frac{2}{3}- E\,(GM)\,\pi\nn\\
&-\frac{3\pi^2}{64}\,E^2(GM)^2\Big) + \mathcal O(b_h^4)
\end{align}
for the tree level post Newtonian one.\\
For the Reissner-Nordstrom geometry the situation is similar. The post-Newtonian form factor is then given by
\begin{align}
\left.\mathcal{PN}(E, \theta)\right|^{RN}=&\left[1- \frac{\pi}{8}(GM)\big(1+3\frac{Q^2}{G\,M^2}\big)\,E\sin\frac{\theta}{2}\right.\nn\\
&\left.+(GM)^2\big(1+\frac{Q^2}{G\,M^2}\big)\,E \sin\frac{\theta}{2}\biggl( \frac{1}{\Lambda}\sin\left(2\,\Lambda\,E\sin\frac{\theta}{2}\right) - 2E\sin\frac{\theta}{2} \mathrm{Ci}\biggl(2\,\Lambda\,E\sin\frac{\theta}{2}\biggr)\biggr)\right]^2,
\label{PNrn}
\end{align}
and the impact parameter in the 1PN approximation is
\begin{align}
\label{nbpnRN}
\left.b^2\right|_{\text{1PN}}^{(0)\,RN}&=4\,(GM)^2\Big(-1+\csc^2\frac{\theta}{2}+2\ln\sin\frac{\theta}{2}\Big)+E\, (GM)^3\big(1+3\frac{Q^2}{G\,M^2}\big)\pi\Big(4+(\cos\theta-3)\csc\frac{\theta}{2}\Big)\nn\\
&-\frac{1}{32}E^2(GM)^4\big(1+3\frac{Q^2}{G\,M^2}\big)^2\pi^2\Big(1+\cos\theta+4\ln\sin\frac{\theta}{2}\Big)
\end{align}
The inversion formula in this case is
\begin{align}
\left.\alpha\right|_{\text{1PN}}^{(0)\,RN}=&\frac{2}{b_h}-\frac{\pi}{2}\frac{1}{b^2_h}E\,(GM)\big(1+3\frac{Q^2}{G\,M^2}\big)-\frac{1}{b^3_h}\Big(\ln b_h\,\Big(2-\frac{\pi^2}{32}\,E^2(GM)^2\big(1+3\frac{Q^2}{G\,M^2}\big)^2\Big)\nn\\
&+\frac{2}{3}-E\,(GM)\big(1+3\frac{Q^2}{G\,M^2}\big)\,\pi-\frac{3\pi^2}{64}\,E^2(GM)^2\big(1+3\frac{Q^2}{G\,M^2}\big)^2\Big)+ \mathcal O(b_h^4).
\end{align}
\begin{itemize} 
\item{\bf \large Photons} 
\end{itemize}
We can extend the analysis presented above for neutrinos to the photon case. Here the cross section takes the form

\begin{equation}
\left.\frac{d\sigma}{d\Omega}\right|_{\gamma,\text{2PN}}^{(0)}=\frac{\kappa^2}{16\pi^2} E^4 \cos^4\frac{\theta}{2}  F_g(q)^2 
\end{equation}
and, as in the neutrino case, we have
\begin{equation}
\left.\frac{d\sigma}{d\Omega}\right|_{\gamma,\text{2PN}}^{(0)}=\left.\frac{d\sigma}{d\Omega}\right|_{\gamma,\text{0PN}}^{(0)}\,\mathcal{PN}_2(E, \theta)
\end{equation}
where we inserted the tree level cross section for the photon
\begin{equation}
\left.\frac{d\sigma}{d\Omega}\right|_{\gamma,\text{0PN}}^{(0)}= (GM)^2 \cot^4 \frac{\theta}{2} \,.
\end{equation}
 In the 0PN Newtonian limit, this cross section has been computed in \cite{Coriano:2014gia}, and takes the form
\bea
\label{phDCS}
\left.\frac{d \sigma}{d \Omega}\right|_{\gamma,\text{0PN}}^{\text{(1)}}=\left.\frac{d\sigma}{d\Omega}\right|_{\gamma,\text{0PN}}^{(0)}\left\{1+2\left[ \sum_{{f_k}} N^c_{f_k} F^3_{f_k}(E, \theta, m_{f_k}, Q_{f_k})+F^3_W(E, \theta)\right]\right\}
\eea
where 
\begin{align}
F^3_{f_k}(E, \theta)=&\frac{1}{36}\frac{\alpha_w}{\pi}\frac{Q^2_{f_k}}{E^2}(35\,E^2-39\,m^2_{f_k}\csc^2\theta/2)\nonumber\\
&-\frac{1}{12}\frac{\alpha_w}{\pi}\frac{Q^2_{f_k}}{E^2}(4\,E^2-5\,m^2_{f_k}\csc^2\theta/2)\sqrt{1+m^2_{f_k}\frac{\csc^2\theta/2}{E^2}}\log\left(\frac{1+\sqrt{1+m^2_{f_k}\frac{\csc^2\theta/2}{E^2}}}{-1+\sqrt{1+m^2_{f_k}\frac{\csc^2\theta/2}{E^2}}}\right)\nonumber\\
&+\frac{1}{16}\frac{\alpha_w}{\pi}\frac{m^2_{f_k}Q^2_{f_k}}{E^4}\csc^4\theta/2\left(E^2\cos\theta-E^2+m^2_{f_k}\right)\log^{2}\left(\frac{1+\sqrt{1+m^2_{f_k}\frac{\csc^2\theta/2}{E^2}}}{-1+\sqrt{1+m^2_{f_k}\frac{\csc^2\theta/2}{E^2}}}\right)
\end{align}
and
\begin{align}
F^3_W(E, \theta)=&-\frac{1}{24}\frac{\alpha_w}{\pi}\frac{1}{E^2}(125\,E^2-39\,m^2_W\csc^2\theta/2)\nonumber\\
&+\frac{1}{8}\frac{\alpha_w}{\pi}\frac{1}{E^2}(14\,E^2-5\,m^2_W\csc^2\theta/2)\sqrt{1+m^2_W\frac{\csc^2\theta/2}{E^2}}\log\left(\frac{1+\sqrt{1+m^2_W\frac{\csc^2\theta/2}{E^2}}}{-1+\sqrt{1+m^2_W\frac{\csc^2\theta/2}{E^2}}}\right)\nonumber\\
&-\frac{1}{32}\frac{\alpha_w}{\pi}\frac{1}{E^4}\left(16\,E^4-16\,E^2\,m^2_W\csc^2\theta/2+3\,m_W^4\csc^4\theta/2\right)\log^2\left(\frac{1+\sqrt{1+m^2_W\frac{\csc^2\theta/2}{E^2}}}{-1+\sqrt{1+m^2_W\frac{\csc^2\theta/2}{E^2}}}\right)
\end{align}
are the relevant electroweak form factors entering in the computation. In the previous equations the sum $f_k$ is over all Standard Model fermions, with $m_{fk}$ and $Q_{f_k}$ their masses and charges. $N^c_{f_k}$ is 1 for leptons and 3 for quarks.
Proceeding similarly to the neutrino case, the one loop cross section in the 2PN approximation takes the form
\begin{equation}
\left.\frac{d \sigma}{d \Omega}\right|_{\gamma,\text{2PN}}^{\text{(1)}}=\left.\frac{d \sigma}{d \Omega}\right|_{\gamma,\text{0PN}}^{\text{(1)}}\,\mathcal{PN}_2(E, \theta),  
\label{twop}
\end{equation}
with $\mathcal{PN}_2$ given by (\ref{PN}), which can be inserted again in (\ref{semic}) and investigated numerically. Solving at order 1PN the analogous of (\ref{twop}), the solution of (\ref{semic}) gives
\begin{align}
\label{pbpn}
\left.b^2\right|_{\gamma,\text{1PN}}^{(0)}&=2\,(GM)^2\Big(-1+2\,\csc^2\frac{\alpha}{2}+\cos\alpha+8\ln\sin\frac{\alpha}{2}\Big)-\frac{2}{3}E\,(GM)^3\pi\Big(1+3\,\csc\frac{\alpha}{2}\Big)\Big(\cos\frac{\alpha}{4}-\sin\frac{\alpha}{4}\Big)^6\nn\\
&-\frac{1}{256}E^2(GM)^4\pi^2\Big(11+12\cos\alpha+\cos2\alpha+32\ln\sin\frac{\alpha}{2}\Big).
\end{align}
In the photon case the inversion formulae at orders 0PN and 1PN are given by
\bea
\left.\alpha\right|_{\gamma, \text{0PN}}^{(0)}=\frac{2}{b_h}-\frac{1}{b_h^3}\Big(4\,\ln b_h-\frac{1}{3}\Big)-\frac{1}{b_h^5}\Big(12\,\ln^2 b_h+10\,\ln b_h+\frac{17}{20}\Big) + \mathcal O(b_h^7)
\eea
and
\begin{align}
\left.\alpha\right|_{\gamma, \text{1PN}}^{(0)}=&\frac{2}{b_h}-\frac{1}{b_h^2}\frac{\pi}{2}E\,(GM)-\frac{1}{b_h^3}\Big(\ln b_h\big(4-\frac{1}{16}\,\pi^2E^2(GM)^2\big)-\frac{1}{64}\,\pi^2E^2(GM)^2\nn\\
&-\frac{4}{3}\,\pi\,E\,(GM)-\frac{1}{3}\Big) + \mathcal O(b_h^4)
\end{align}
respectively.
\subsection{Range of applicability}
The structure of the one-loop 2PN result for neutrinos and photons shows the complete factorization between the quantum corrections and the background-dependent contributions. While the former 
are process dependent, the latter are general. Obviously, this result is not unexpected, and follows rather closely other typical similar cases in potential scattering in quantum mechanics. An example is the case of an electron scattering off a finite charge distribution characterized by a geometrical size $R$, where the finite size corrections are all contained in a geometric form factor. \\
We recall that for a Coloumb interaction of the form $V( r )=e^2/r$, the cross section is given in terms of the pointlike $( p )$ amplitude   
\beq
f(\theta)_{\textrm{p}} =- 2 \frac{m e^2}{\vec{q}^{\,2}}
\label{point}
\eeq
with $\vec{q}= \vec{k} - \vec{k}'$ and $|\vec{q}|=2 |\vec{k}|\sin\theta/2$
being the momentum transfer of the initial (final) momentum of the electron $\vec{k}$ ($\vec{k}'$) and charge $e$. The scattering angle is measured with respect to the z-direction of the incoming electron. The charge of the static source has also been normalized to $e$. The corresponding cross section is given by 
\beq
\frac{d\sigma}{d\Omega}_\textrm{p}=|f(\theta)_{\textrm{p}}|^2=  \frac{(2 m) ^2 e^4}{16 k^4 \sin^4\theta/2},
\eeq
and the modification induced by the size of the charge distribution $(\rho(x))$ is contained in 
\beq
F( \vec{q})=\int d \vec{x}\rho( \vec{x}) e^{i  \vec{q}\cdot  \vec{x}} 
\eeq
with 
\beq
\frac{d\sigma}{d\Omega}=\frac{d\sigma}{d\Omega}_\textrm{p}|F( q)|^2 .
\eeq
For a uniform charge density, for instance, the geometrical form factor $F(\vec{q})$, which is the transform of the charge distribution, introduces a dimensionless variable $ q R$ in the cross section which is absent in the point-like (Coulomb) case, of the form 
\beq
F(q)=3 \frac{\sin(q\, R) -  q\, R \cos(q\, R)}{(q\, R)^3}.
\label{coulomb}
 \eeq
The validity of the expression above is for $q\, R\lesssim 1$, and the presence of the geometrical form factor is responsible for the fluctuations measured in the cross section as a result of the finite extension of the charged region. \\
In the analysis of the nPN corrections in gravity, the situation is clearly analogous, with the size of the horizon taking the role of the classical charge radius $R$. For ordinary (macroscopic) horizons 
(e.g. of a km size) $r_s\sim G M$ invalidates the perturbative expansion due to the 
appearance of the $G M E$ parameter in the expression of the post Newtonian factor 
$\mathcal{P N}(E,\theta)$, which is small only if $E\sim 1/GM$, a choice which is not relevant for our analysis, since it applies to particle beams whose energy is in the very far infrared. \\
\begin{figure}[t]
\centering
\subfigure[]{\includegraphics[scale=.7]{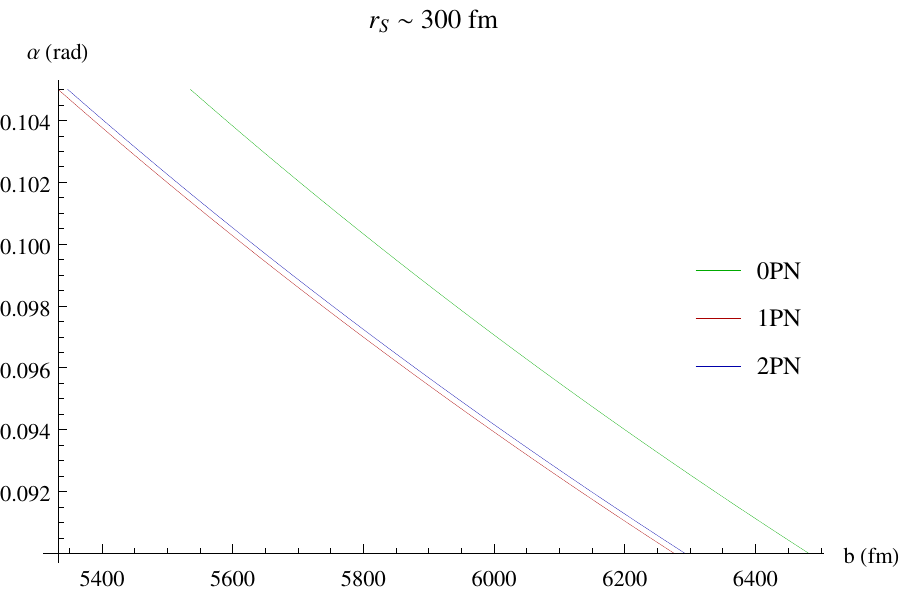}}\hspace{.5cm}
\subfigure[]{\includegraphics[scale=.7]{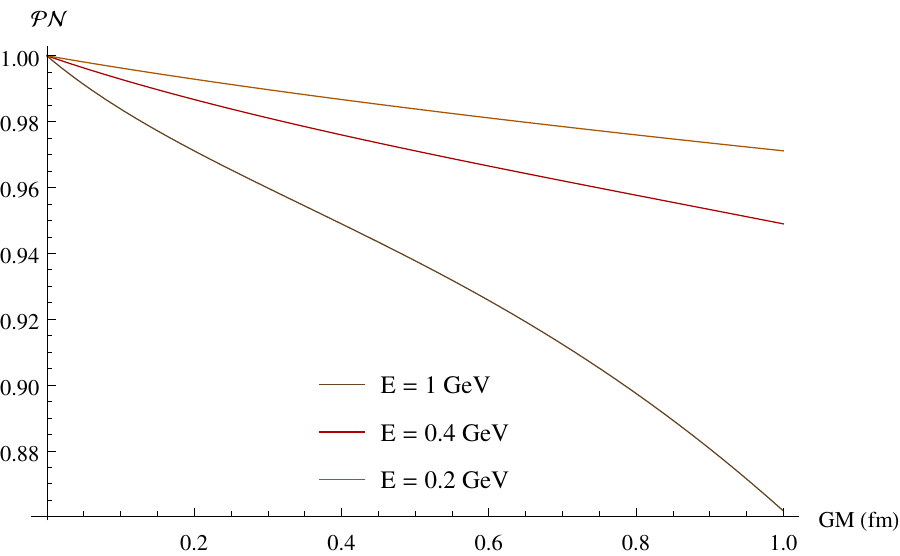}}\hspace{.5cm}
\subfigure[]{\includegraphics[scale=.7]{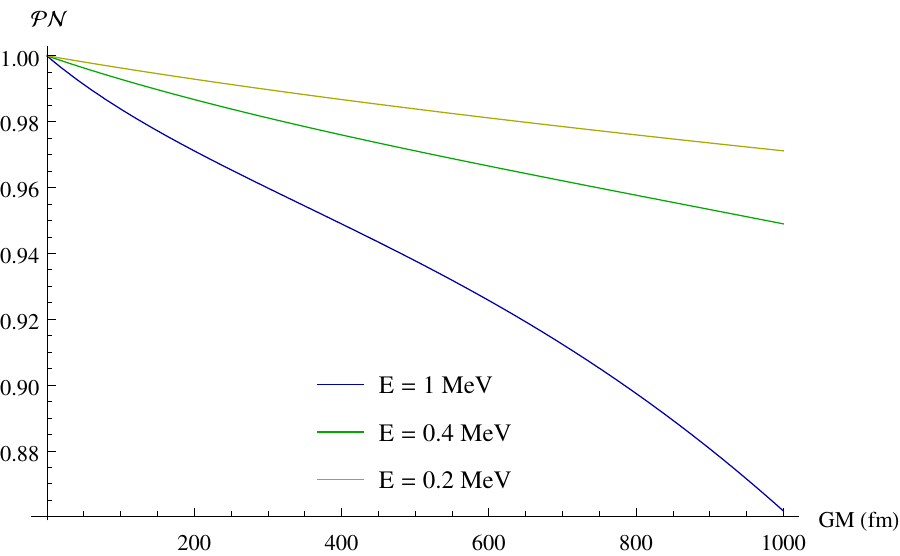}}\hspace{.5cm}
\caption{Comparison of nPN approximations for $\alpha(b)$ in the photon case with $\text{M}_{PBH}=10^{-16}\,\text{M}_\odot$ and for $E G M \approx 1$ (a). In (b) and (c) we show the $\mathcal{PN}$ function for different energies.\label{pnPBH}}
\end{figure}
By imposing that the cutoff $\Lambda$ coincides with the Schwarzschild radius 
($\Lambda\equiv r_s$), one can immediately realize that the post-Newtonian expansion gets organized only in terms of this parameter ($G M E$). In the regions of strong deflections, which are those that concern our analysis, we can reasonably assume that $y\equiv \sin\theta/2\sim \mathcal{O}(1)$, if we use the GR prediction to estimate the bending angle. This allows to discuss the convergence of the PN expansion only in terms of the energy $E$ of the incoming beam and of the size of the horizon. The analogous of the charge oscillations given by (\ref{coulomb}), in the gravitational case, are then uniquely related to the post-Newtonian function 
$\mathcal{PN}$, and hence to the size of the parameter $ \Lambda E\sim 
r_s E$ which defines its expansion in powers of the gravitational potential. Assuming a small value of $x\equiv G M E$, we can indeed rewrite (\ref{PN}) via a small-x expansion, obtaining
 \beq
\mathcal{PN}(x,y)= 1 -\frac{\pi}{8} x y + x^2 y^2\left( 1 -\gamma_E - \log x - \log 4 y\right).
\eeq
This expression can be used to investigate the range of applicability of these corrections in terms of the two factors appearing in $x$, the energy of the incoming beam and the size of the horizon of the gravitational source. The requirement that such a parameter be small defines a unique range of applicability of such corrections in the quantum case. \\
One possible application of the formalism which renders the PN corrections to the gravitational scattering quite sizable is in the context of primordial black holes \cite{Carr:1974nx}, which have found a renewed interest in the current literature \cite{Carr:2014eya,Khlopov:2008qy}. \\ 
We just mention that primordial black holes (PBHs) have been considered a candidate component of dark matter since the 70's, and conjectured to have formed in the early universe by the gravitational collapse of large density fluctuations, with their abundances and sizes tightly constrained by various theoretical arguments. These range from Hawking radiation, which causes their decay to occur at a faster rate compared to a macroscopic black hole (of solar mass); bounds from their expected microlensing events;  their influence on the CMB, just to mention a few \cite{Clesse:2015wea}. For instance, the mechanism of thermal emission by Hawking radiation sets a significant lower bound on their mass ($\sim5\times 10^{14} g$), in order for them to survive up to the present age of the universe. This bound  satisfies also other constraints, such as those coming from the possible interference of their decay with the formation of light elements at the nucleosynthesis time.  
With the launch of the FERMI gamma ray space telescope \cite{FERMI}, the interest in this kind of component has found new widespread interest. The unprecedented sensitivity of its detector in the measurement of interferometric patterns generated by high energy cosmic rays (femtolensing events), such as Gamma Ray Bursts \cite{Gould1992}, has allowed to consider new bounds on their abundances \cite{Barnacka:2012bm}. The hypothesis of having PBHs as a dominant component of the dark matter of the universe provides remarkable constrains on their allowed mass values, except for a 
mass range $10^{18} \textrm{kg} < M_{\textrm{PBH}} <10^{23} \textrm{kg}$, where it has been argued that they can still account for the majority of it. In other mass ranges several analyses indicate that the PBH fraction of dark matter cannot exceed $1\% $ of the total \cite{Clesse:2015wea}. \\
PN corrections turn out to be significant for PBH in this mass range, due to the large variation induced on the  $\mathcal{P N}$ function by the 1PN and 2PN terms. These may play a considerable role in a PBH mediated lensing event. We illustrate this behaviour by showing plots of the post Newtonian behaviour of the relevant expressions for lensing. In Fig. \ref{pnPBH} (a) we plot the angular deflection as a function of the impact parameter for the Newtonian 0PN, and relative post Newtonian corrections. We have considered a primordial black hole with a mass of $10^{-16}$ $M_{\odot}$, which carries a microscopic Schwarzschild radius (300 fm) and chosen $E=1/(GM)=0.6$ MeV for the incoming photon beam. The impact of the corrections on the gravitational cross section are quite large, as one can easily figure out from panel (b), where we plot the factor $\mathcal{P N}$ as a function of the Schwarzschild radius for these compact massive objects, for $b_h\sim$ 1 fm. For a more massive primordial black hole, with $200 < b_h <1000$, the pattern is quite similar, as shown in panel (\textrm{c}). In both cases the post-Newtonian corrections appear to be significant, of the order of 15-20 $\%$ and could be included in a more accurate analysis of lensing for these types of dark matter candidate solutions. 

\section{Conclusions and perspectives} 
We have presented a discussion of neutrino lensing at 1-loop in the electroweak theory.
In our approach the gravitational field is a static background, and the propagating matter fields are obtained by embedding the Standard Model Lagrangian on a curved spacetime, as discussed in previous works \cite{Coriano:2013iba,Coriano:2011zk}. As in a previous study \cite{Coriano:2014gia}, also in our current case the field theoretical corrections to the gravitational deflection are in close agreement with the predictions of general relativity. The agreement holds both asymptotically, for very large distances from the center of the black hole, of the order of $10^6$ horizon sizes $(b_h)$, but also quite close to the photon sphere $(\sim 20\, b_h)$. In this respect, the similarity of the results for photons and neutrinos indicates the consistency of the semiclassical approach that we have implemented. 
As noticed in \cite{Accioly1}, the inclusion of the quantum effects causes the appearance of an energy dependent dispersion of a particle beam, which implies a violation of the classical equivalence principle. 

Various types of lens equations have been formulated in the past using classical GR, and we have illustrated the modifications induced on their expressions by the 
inclusion of the suppressed $1/b^n$ corrections in the impact parameter to the angular deflections.
We have then developed a formalism which allows to include the semiclassical results, due to the radiative effects in the propagation of a photon or a neutrino, in a typical lensing event.
We have considered both the case of a thin lens, which is quadratic in the deflection angle, and the fully nonlinear case, taking as an example the Virbhadra-Ellis lens equation. Radiative and post-Newtonian effects induce 
a dependence of the angle of deflection with the appearance of extra $1/b^n$ suppressed contributions and of extra logarithms of the impact parameter, that we have studied numerically for 
some realistic geometric configurations. In general, radiative effects are significant only for configurations of the source/lens/observer which involve small impact parameters in the deflection $(b_h\sim 20)$, and require angular resolutions in the region of few milliarcseconds. 
Our results are valid for a Schwarzschild metric, considered both in the Newtonian and in the post-Newtonian approximation, but they can be extended to other metrics as well. \\
We have also discussed the consistency of the post-Newtonian approach. We have shown that such corrections can be consistently taken into account in the case of microscopic horizon sizes, such as primordial black holes. These corrections have been shown to factorize and be accounted for by a post-Newtonian function. Our analysis can be extended in several directions, from the case of Kerr-Newman metrics to the study of microlensing and Shapiro delays, and to dynamical gravity. We hope to return on some of these topics in a future work.

\vspace{1cm}

\centerline{\bf Acknowledgements}
We thank Kostas Skenderis, Pietro Colangelo, Emidio Gabrielli and Subir Sarkar for discussions and exchanges. The work of Claudio Corian\`o is supported by a {\em The Leverhulme Trust Visiting Professorship} at the University of Southampton at the STAG Research Centre and Mathematical Sciences. He thanks Kostas Skenderis, Marika Taylor and the members of the Centre for the kind hospitality provided. 

\appendix
\section{Fluctuations to first order in the Newtonian potential} 
The solution for the static massive source is obtained from the linearized equation
\beq
\square\left(h_{\mu\nu}-\frac{1}{2}\eta_{\mu\nu} h\right)=-{\kappa}T^{ext}_{\mu\nu} \,,
\eeq
where $h\equiv h^{\mu\nu}\eta_{\mu\nu}$, and can be rewritten as
\beq
\square h_{\mu\nu}={\kappa}\, S_{\mu\nu}, \qquad \mbox{with} \qquad S_{\mu\nu}=
- \left( T^{ext}_{\mu\nu}-\frac{1}{2}\eta_{\mu\nu}T^{ext}\right) \,.
\label{oneq}
\eeq
The external field $h_{\mu\nu}$ is obtained by convoluting the static source with the retarded propagator 
\beq
G_R(x,y)= \frac{1}{4 \pi}\, \frac{\delta(x_0-y_0-|\vec{x}-\vec{y}|)}{|\vec{x}-\vec{y}|},
\eeq
normalized as
\beq
\square G_R(x,y)=\delta^4(x-y). 
\eeq
The solution of Eq.(\ref{oneq}) takes the form 
\beq
h^{ext}_{\mu\nu }(x)={\kappa}\, \int d^4 y\, G_R(x,y)\, S_{\mu\nu}(y)
\eeq
with the EMT of the external localized source, defining $S_{\mu\nu}$, given by
\beq
T^{ext}_{\mu\nu}= \frac{P_{\mu}P_{\nu}}{P_0}\delta^3(\vec{x}) \,.
\eeq
For a compact source of mass $M$ at rest at the origin, with $P_{\mu}=(M,\vec{0}$), we have
\beq
T^{ext}_{\mu\nu}=M\delta^0_\mu\delta^0_\nu\delta^3(\vec{x})
\eeq
which gives
\beq
S_{\mu\nu}=\frac{M}{2}\bar{S}_{\mu\nu} \qquad 
\bar{S}_{\mu\nu}\equiv \eta_{\mu\nu}-2 \delta^0_{\mu}\delta^0_{\nu}
\eeq
and 
\bea
h_{\mu\nu}(x)
&=& \frac{2 G M}{\kappa |\vec{x}|}\bar{S}_{\mu\nu},
\label{hh}
\eea
where the field generated by a local (point-like, $L$) mass distribution has the typical $1/r$  $(r\equiv |\vec{x}|)$ behaviour.
The fluctuations are normalized in such a way that $h_{\mu\nu}$ has mass dimension 1, as an ordinary bosonic field, with 
$\kappa$ of mass dimension ${-1}$. 

\section{ $1/b^n$ corrections to lensing for discrete and continuous mass distributions in GR } 
\label{continuous}
The method of extracting the $1/b^n$ corrections to the angular deflection can be extended to the case of a countinuos mass distribution in the lens plane, as shown in Fig.~\ref{lenspicture}.
For this purpose we can consider the case of a distributed lens with a surface density
\bea
\Sigma(\vec \xi)=\int \rho(\vec \xi, z)\,dz.
\eea 
In the Newtonian approximation we have the usual relation
\bea
\label{alphaN}
\hat{\vec \alpha}(\vec \xi)=4\,G\,M\,\int d^2\xi' \frac{\vec \xi - \vec \xi'}{|\vec \xi - \vec \xi'|^2}\,\Sigma(\vec \xi')
\eea
or, after the rescaling $\vec \xi=\xi_0\vec x$,
\bea
\hat{\vec \alpha}(\vec x)=4\,G\,M\,\int d^2 x' \xi_0 \frac{\vec x - \vec x'}{|\vec x - \vec x'|^2}\,\Sigma(\xi_0 \vec x') \,.
\eea 
The deflection angle is
\bea
\vec \alpha(\vec x)=4\,G\,M\,\frac{D_{LS}\,D_{OL}}{D_{OS}}\,\nabla_x\int d^2 x' \log |\vec x - \vec x'| \Sigma(\xi_0 \vec x)
\eea
and is the gradient of the lensing potential
\bea
\label{lpN}
\vec \alpha(\vec x)=\nabla_x \mathbf{\Psi}(\vec x)\qquad\qquad\mathbf{\Psi}(\vec x)\equiv\frac{1}{\pi}\,\int d^2 x' \log |\vec x - \vec x'|\,\frac{\Sigma(\xi_0 \vec x)}{\Sigma_{cr}}
\eea
where we have introduced the critical surface density
\bea
\Sigma_{cr}=\frac{1}{4\pi\,G\,M}\frac{D_{OS}}{D_{LS}\,D_{OL}}.
\eea
It is possible to introduce the corrections to this behaviour by extending to the case of a continuous distribution the result given in Eq. (\ref{exp}).
Inserting the generalized $1/b$ expansion given in (\ref{exp}), we derive a generalized version of  (\ref{alphaN}) that remains valid 
also in the case of strong lensing and which is given by  
\bea
\label{alphaPPN}
\hat{\vec \alpha}(\vec \xi)=\int d^2\xi' (\vec \xi-\vec \xi')\left(4\,GM\,\frac{\Sigma(\vec \xi')}{|\vec \xi-\vec \xi'|^2}+\frac{15\pi}{4}\,(GM)^2\,\frac{\Sigma^2(\vec \xi')}{|\vec \xi-\vec \xi'|^3}+\frac{128}{3}\,(GM)^3\,\frac{\Sigma^3(\vec \xi')}{|\vec \xi-\vec \xi'|^4}\right)\nn\\.
\eea
This allows to define a generalized deflection potential
\bea
\vec \alpha(\vec x)=\nabla_x\mathbf{\Psi}(\vec x),
\eea
with
\bea
\mathbf{\Psi}(\vec x)=\frac{1}{\pi\,\Sigma_{cr}}\int d^2 x'\left(\log |\vec x - \vec x'|\,\Sigma(\xi_0 \vec x)-\frac{15\pi}{16}\frac{1}{|\vec x-\vec x'|}\frac{G\,M}{\xi_0}\Sigma^2(\xi_0 \vec x)-\frac{128}{24}\frac{1}{|\vec x-\vec x'|^2}\frac{G^2M^2}{\xi_0^2}\Sigma^3(\xi_0 \vec x)\right).\nn\\
\eea

\section{Full cross section}
\label{fcs}
We give the explicit expression of the neutrino differential cross section in terms of $E$ and $\theta$ used for the calculation of the form factors
\begin{align}
\frac{d \sigma}{d \Omega}=&\,G^2M^2\frac{\cos^2\theta/2}{\sin^4\theta/2}\Biggl\{ 1 + \frac{2 G_F}{16\sqrt{2}\pi^2} \frac{1}{6 E^2 \sin^2{\theta/2}} \Biggl[ 6\Bigl(m_f^2-4(2 m_W^2+m_Z^2)\Bigr) E^2\sin^2(\theta/2) \nn
\end{align}
\vspace{-.7 cm}
\begin{align}
& -4\Bigl(m_f^4+m_f^2m_Z^2-2m_W^4-m_Z^2\Bigr)+  \Bigl( m_f^2+ 2m_W^2 \Bigr)\mathcal{A}_0\bigl(m_f^2\bigr) \nn
\end{align}
\vspace{-.7 cm}
\begin{align}
&+\Bigl( 24 E^2\sin^2(\theta/2)- 7m_Z^2 \Bigr)  \mathcal{A}_0\bigl(m_Z^2\bigr) - \Bigl( m_f^2+ 2m_W^2 \Bigr) \mathcal{A}_0\bigl(m_W^2\bigr)\nn
\end{align}
\vspace{-.7 cm}
\begin{align}
&+  \Bigl( -34 m_Z^2 E^2\sin^2(\theta/2) -\frac{3 m_Z^6}{E^2\sin^2(\theta/2)} +21 m_Z^4\Bigr)\mathcal{B}_0\bigl(-4E^2\sin^2(\theta/2),0,0\bigr) \nn
\end{align}
\vspace{-.7 cm}
\begin{align}
&+ \Bigl( 16 m_Z^2 E^2\sin^2(\theta/2)+\frac{3m_Z^6}{E^2\sin^2(\theta/2)} -14m_Z^4\Bigr)  \mathcal{B}_0\bigl(-4E^2\sin^2(\theta/2),m_Z^2,m_Z^2\bigr) \nn
\end{align}
\vspace{-.7 cm}
\begin{align}
&+\Bigl( 4 (m_f^2+8m_W^2)E^2\sin^2(\theta/2) +\frac{3}{E^2\sin^2(\theta/2)} (m_f^6-3m_f^2m_W^4+2m_W^6) \nn
\end{align}
\vspace{-.7 cm}
\begin{align}
&-4(m_f^4-m_f^2m_W^2+7m_W^4) \Bigr) \mathcal{B}_0\bigl(-4E^2\sin^2(\theta/2),m_W^2,m_W^2\bigr) \nn
\end{align}
\vspace{-.7 cm}
\begin{align}
&+ \Bigl( 2 (m_f^2-34m_W^2)E^2\sin^2(\theta/2)  -\frac{3}{E^2\sin^2(\theta/2)}(m_f^6-3m_f^2m_W^4+2m_W^6)\nn
\end{align}
\vspace{-.7 cm}
\begin{align}
& + 5m_f^4-23m_f^2m_W^2+42m_W^4 \Bigr)  \mathcal{B}_0\bigl(-4E^2\sin^2(\theta/2),m_f^2,m_f^2\bigr) \nn
\end{align}
\vspace{-.7 cm}
\begin{align}
&+\Bigl( 48m_W^2 +\frac{6}{E^2\sin^2(\theta/2)} \bigl(  m_f^4 -2m_W^4 + m_f^2m_W^2 \bigr) \Bigr) \mathcal{B}_0\bigl(0,m_f^2,m_W^2\bigr) \nn
\end{align}
\vspace{-.7 cm}
\begin{align}
&+\Bigl( 12m_Z^6 - \frac{3m_Z^8}{E^2\sin^2(\theta/2)}\,  \Bigr)\mathcal{C}_0\bigl( -4E^2\sin^2(\theta/2),0,m_Z^2,m_Z^2 \bigr)\nn
\end{align}
\vspace{-.7 cm}
\begin{align}
& +\Bigl( 48 m_Z^2 E^4\sin^4(\theta/2) -72m_Z^4E^2\sin^2(\theta/2)\nn
\end{align}
\vspace{-.7 cm}
\begin{align}
&-\frac{3m_Z^8}{E^2\sin^2(\theta/2)} +27m_Z^6 \Bigr) \mathcal{C}_0\bigl( -4E^2\sin^2(\theta/2),m_Z^2,0,0 \bigr)\nn
\end{align}
\vspace{-.7 cm}
\begin{align}
&+\Bigl( \, \frac{1}{4E^2\sin^2(\theta/2)}  \Bigl(m_f^8-2(m_W^2-E^2\sin^2(\theta/2))(m_W^3-4m_WE^2\sin^2(\theta/2))^2 \nn
\end{align}
\vspace{-.7 cm}
\begin{align}
& +m_f^4(-4E^2\sin^2(\theta/2)+8m_W^2E^2\sin^2(\theta/2)-3m_W^4)  -m_f^6(E^2\sin^2(\theta/2)+m_W^2) \nn
\end{align}
\vspace{-.7 cm}
\begin{align}
&+ m_f^2 (24m_W^2E^4\sin^4(\theta/2)-25m_W^2E^2\sin^2(\theta/2)+5m_W^6) \Bigr)  \Bigr) \mathcal{C}_0\bigl( -4E^2\sin^2(\theta/2),m_W^2,m_f^2,m_f^2 \bigr) \nn
\end{align}
\vspace{-.7 cm}
\begin{align}
&+\Bigl( 24m_W^6 -\frac{6m_W^8}{E^2\sin^2(\theta/2)}-30m_W^4m_f^2+\frac{15m_W^6m_f^2}{E^2\sin^2(\theta/2)} -\frac{9m_W^4m_f^4}{E^2\sin^2(\theta/2)}-\frac{3m_W^2m_f^6}{E^2\sin^2(\theta/2)}\nn
\end{align}
\vspace{-.7 cm}
\begin{align}
& +\frac{3m_f^8}{E^2\sin^2(\theta/2)}+24m_W^2m_f^4+24m_f^4E^2\sin^2(\theta/2)-18m_f^6 \Bigr) \mathcal{C}_0\bigl( -4E^2\sin^2(\theta/2),m_f^2,m_W^2,m_W^2 \bigr) \Biggr] \nn
\end{align}
\vspace{-.7 cm}
\begin{align}
& - \frac{ G_F}{16\sqrt{2}\pi^2} \Bigl( \Sigma_{Z}^{L}+\Sigma_{W}^{L} \Bigr) \Biggr\} \,.
\end{align}

\section{Coefficients of the semiclassical expansion}
\label{expansion}
The coefficients present in the expansion in Eq. (\ref{OLb}) are given by 
\bea
C_1(E)&=&\frac{1}{24\sqrt 2}\frac{G_F}{\pi^2}\frac{E^2}{(m_f^2-m_W^2)^4}\left(9\,m_f^8+62\,m_f^6m_W^2+125\,m_f^4m_W^4-94\,m_f^2m_W^6+22\,m_W^8\right.\nn\\
&&+\left.(20\,m_f^6m_W^2-30\,m_f^4m_W^4+4\,m_f^2m_W^6)\ln(m_f^2/m_W^2)\right)\nn \\
C_2(E)&=&\frac{1}{600\sqrt 2}\frac{G_F}{\pi^2}\frac{E^4}{m_Z^2(m_f^2-m_W^2)^6}\left(20\,m_Z^2\left(2\,m_f^{10}-84\,m_f^8m_W^2+36\,m_f^6 m_W^4+76 m_f^4m_W^6\right.\right.\nn\\
&&\left.\left.+21\,m_f^2m_W^8-6\,m_W^{10}\right)\ln(m_f^2/m_W^2)+(m_f^2-m_W^2)\left(69\,m_f^{10}-69\,m_W^{10}+150\,m_W^8m_Z^2\right.\right.\nn\\
&&\left.\left.-5\,m_f^8(69\,m_W^2-55\,m_Z^2)+15\,m_f^6(46\,m_W^4+135\,m_W^2m_Z^2)-15\,m_f^4(46\,m_W^6+165\,m_W^4m_Z^2)\right.\right.\nn\\
&&\left.\left.+5\,m_f^2(69\,m_W^8-175\,m_W^6m_Z^2)\right)\right)\nn \\
D_2(E)&=&\frac{1}{160\sqrt 2}\frac{G_F}{\pi^2}\frac{E^4}{m_Z^2}\nn \\
E_2(E)&=&\frac{1}{160\sqrt 2}\frac{G_F}{\pi^2}\frac{E^4}{m_Z^2}\left(-11+6\,\ln 4+\ln(E^2/m^2_Z)\right)\nn \\
F_2(E)&=&\frac{1}{160\sqrt 2}\frac{G_F}{\pi^2}\frac{E^4}{m_Z^2}\left(-3+\ln 16+2\,\ln(E^2/m^2_Z)\right)\nn \\
G_2(E)&=&\frac{1}{160\sqrt 2}\frac{G_F}{\pi^2}\frac{E^4}{m_Z^2}\left(-1+\,\ln 16+2\,\ln(E^2/m^2_Z)\right).
\eea
 We find 
 
\begin{align}
\label{lowb}
b_h(E,\theta)&=\frac{2}{\theta}+\frac{1}{12}\big\{-2+6\,C_1(E)+3\,C_2(E)-3\,E_2(E)-12\,F_2(E)-3\,G_2(E)\big\}\,\theta\nn\\
&-6\,\big\{1+2\,C_1(E)\big\}\,\theta\ln\theta+\frac{1}{2880}\big\{-4\,(17+75\,C_2(E)+15\,E_2(E)\nn\\
&-300\,F_2(E)-345\,G_2(E)+30\,(1+96\,D_2(E))\ln 2)+15\big(-3\,C^2_2(E)+6\,C_2(E)(E_2(E)\nn\\
&+4\,F_2(E)+G_2(E))-12\,C^2_1(E)(\ln 4-1)^2-3\,(E_2(E)+4\,F_2(E)+G_2(E)+\ln 4)^2\nn\\
&-4\,C_1(E)\big(3\,C_2(E)-3(-2+E_2(E)+4\,F_2(E)+G_2(E)+4\ln^2 2)\nn\\
&+(12\,F_2(E)-1)\ln 4\big)+2(C_2(E)+2\,C_1(E)C_2(E)-2\,C_1(E)(E_2(E)+G_2(E)))\ln 64\big)\big\}\theta^3\nn\\
&+\frac{1}{48}\big\{2-3\,C_2(E)+192\,D_2(E)+3\,E_2(E)+12\,F_2(E)+3\,G_2(E)+\ln 64\nn\\
&+2\,C_1(E)(-1-3\,C_2(E)+3\,E_2(E)+12\,F_2(E)+3\,G_2(E)\nn\\
&+6\,C_1(E)(\ln 4-1)+\ln 4096\big\}\,\theta^3\ln\theta-\frac{1}{16}\big\{1+2\,C_1(E)^2\big\}\,\theta^3\ln^2\theta \,.
\end{align}

\section{Scalar integrals}
\label{scalarint}
In this appendix we collect the definitions of the scalar integrals appearing in the computation of the matrix element. One-, two- and three- point functions are denoted respectively as $\mathcal A_0$, $\mathcal B_0$ and $\mathcal C_0$ with
\bea
\mathcal A_0(m_0^2) &=& \frac{1}{i \pi^2} \int d^n l \frac{1}{l^2 - m_0^2} \,, \nn \\
\mathcal B_0(p_1^2, m_0^2, m_1^2) &=& \frac{1}{i \pi^2} \int d^n l \frac{1}{(l^2 -m_0^2)((l+p_1)^2 -m_1^2)} \,,  \\ 
\mathcal C_0(p_1^2, (p_1-p_2)^2, p_2^2,m_0^2, m_1^2, m_2^2) &=&  \frac{1}{i \pi^2} \int d^n l \frac{1}{(l^2 -m_0^2)((l+p_1)^2 -m_1^2)((l+p_2)^2 -m_2^2)} \,. \nn
\eea
Because the kinematic invariants on the external states of our computation are fixed, $q^2 = (p_1 - p_2)^2$, $p_1^2 = p_2^2 = 0$, we have defined the shorter notation for the three-point scalar integrals
\bea
\mathcal C_0(m_0^2, m_1^2, m_2^2) = \mathcal C_0(m^2, q^2, m^2, m_0^2, m_1^2, m_2^2) \,,
\eea
with the first three variables omitted.

\section{Deflection integral}
\label{deflection}
The deflection integral for the GR solution can be recast in the form 
\begin{equation}
\label{alpha1}
\alpha(x_0)= 2\, x_0^{3/2} \int_{0}^{1/x_0} \frac{dt}{ \sqrt{x_0^3 t^3 - x_0^3 t^2 + x_0 -1}}  - \pi,
\end{equation}
with $x_0$ being the point of closest approach between the deflector and the beam. It is an elliptic integral of first kind. In the indefinite form, the general expression of these types of integrals is given by
\begin{equation}
\begin{split}
\label{io}
I_0&= \int \frac{dz}{\sqrt{R(z)}}  \\
\end{split}
\end{equation}
with the polynomial at the denominator 
\begin{equation}
R(z)= c_0z^n + c_1z^{n-1}+ \dots+ c_n,
\end{equation}
being of degree $n$. For $n=3$, which is the GR case, the Weierstrass form of (\ref{io}) is obtained 
by introducing the roots $(e_1,e_2,e_3)$ and given by
\beq
\label{eq:rditi}
I_0=\int \frac{dt}{\sqrt{4(t-e_1)(t-e_2)(t-e_3)}}.
\eeq
By the transformation 
\begin{equation}
\label{tr}
t=e_3 - \frac{e_1-e_3}{z^2}
\end{equation}
it can be brought to the Legendre form 
\begin{equation}
I_0=\int \frac{C d\zeta}{\sqrt{(1-\zeta^2)(1-k^2\zeta^2)}},
\end{equation}
where $k$ is its modulus. In the case of a finite integration, the form that is needed is
\begin{equation}
F(x;k)=\int_0^x \frac{dz}{\sqrt{(1-z^2)(1-k^2z^2)}},
\end{equation}
also re-expressed as
\begin{equation}
\label{eq:ellipticsin}
F(\phi;k)=\int_0^{\phi} \frac{d\vartheta}{\sqrt{1-k^2\sin^2\vartheta}}
\end{equation}
by a simple change of the integration variable.
For (\ref{alpha1}) the corresponding roots are 
\begin{equation}
\begin{split}
e_1&= \frac{1}{2 x_0} \left( x_0 - 1 - \sqrt{x_0^2+2 x_0 - 3} \right)   \\
e_2&= \frac{1}{2 x_0} \left( x_0 - 1 + \sqrt{x_0^2+2 x_0 - 3} \right)   \\
e_3&= \frac{1}{x_0}
\end{split}
\end{equation}
and the transformation (\ref{tr}) is given by
\begin{equation}
t= \frac{1}{x_0} - \frac{x_0-3-\sqrt{x_0^2+2x_0-3}}{ 2 x_0 z^2}
\end{equation}
which takes to the Legendre form 
\begin{equation}
I_0=-\sqrt{\frac{8 x_0}{x_0-3-\sqrt{x_0^2+2x_0-3}}} \int \frac{dz}{\sqrt{(1-z^2)(1-k^2(x_0) z^2)}},
\end{equation}
with
\begin{equation}
k^2(x_0)=\frac{x_0-3+\sqrt{x_0^2+2x_0-3}}{x_0-3-\sqrt{x_0^2+2x_0-3}}.
\end{equation}
Keeping into account the finite integration region, (\ref{alpha1}) becomes 
\begin{equation}
\label{eq:legdef} 
\alpha(x_0)= -4 \Sigma(x_0) \int_{\Delta(x_0)}^{\infty} \frac{dz}{\sqrt{(1-z^2)(1-k^2(x_0) z^2)}},
\end{equation}
with
\begin{equation}
\Delta(x_0)= \sqrt{\frac{x_0-3-\sqrt{x_0^2+2x_0-3}}{2 x_0} }.
\end{equation}
The boundaries of integration in (\ref{eq:legdef}) can be expressed in the Jacobi form by the substitution $z\to \zeta=1/z$, obtaining 
\begin{equation}
\label{eq:int3}
\int_{\Delta(x_0)}^{\infty} \frac{dz}{\sqrt{(1-z^2)(1-k^2(x_0) z^2)}} = \frac{1}{k(x_0)}\int_{0}^{\tau(x_0)}  \frac{d\zeta}{\sqrt{(1-\zeta^2)(1-\lambda^2(x_0) \zeta^2)}},
\end{equation}
with $\tau(x_0)=(\Delta(x0))^{-1}$ e $\lambda^2(x_0)=(k^2(x_0))^{-1}$. After reabsorbing a factor in front of the integral in the definition of $\Sigma(x_0)$, we obtain
\begin{equation}
\alpha(x_0)= -4 \Sigma(x_0) F(\phi(x_0);\lambda(x_0)) -\pi,
\end{equation} 
with
\begin{subequations}
\begin{align}
&\Sigma(x_0)=\sqrt{\frac{x_0^2\left( x_0-3-\sqrt{x_0^2+2x_0-3}  \right)}{6-4x_0}}\\
&\phi(x_0)=\arcsin(\tau(x_0))  \\  
&\tau(x_0)=\sqrt{\frac{2}{x_0-3-\sqrt{x_0^2+2x_0-3}}} \\
&\lambda^2(x_0)=\frac{x_0-3-\sqrt{x_0^2+2x_0-3}}{x_0-3+\sqrt{x_0^2+2x_0-3}}\, .
\end{align} 
\end{subequations}

\section{Feynman Rules} 
\label{rules}
We collect here all the Feynman rules involving an external gravitational field that have been used in this work. All the momenta are incoming

\begin{itemize}
\item{ graviton - gauge boson - gauge boson vertex}
\\ \\
\begin{minipage}{95pt}
\includegraphics[scale=0.6]{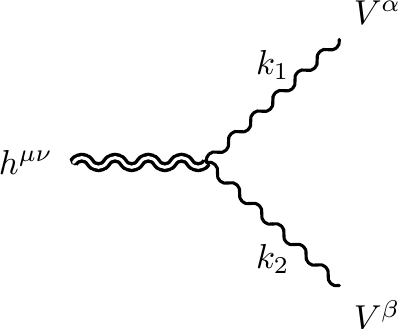}
\end{minipage}
\begin{minipage}{70pt}
\bea
= - i \frac{\kappa}{2} \bigg\{ \left( k_1 \cdot k_2  + M_V^2 \right) C^{\mu\nu\alpha\beta}
+ D^{\mu\nu\alpha\beta}(k_1,k_2) + \frac{1}{\xi}E^{\mu\nu\alpha\beta}(k_1,k_2) \bigg\}
\nn
\eea
\end{minipage}
\bea
\label{FRhVV}
\eea
where $V$ stands for the vector gauge bosons $g$, $\gamma$, $Z$ and $W$.
\item{graviton - fermion - fermion vertex}
\\ \\
\begin{minipage}{95pt}
\includegraphics[scale=0.6]{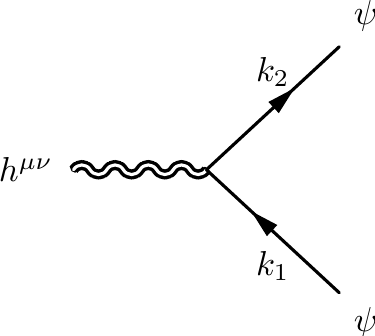}
\end{minipage}
\begin{minipage}{70pt}
\bea
=- i \frac{\kappa}{8} \bigg\{ \gamma^\mu \, (k_1 + k_2)^\nu + \gamma^\nu \,(k_1 + k_2)^\mu - 2 \, \eta^{\mu\nu} \left( \ksl_1 + \ksl_2 - 2 m_f \right)\bigg\}
\nn
\eea
\end{minipage}
\bea
\label{FRhFF}
\eea
\item{graviton - scalar - scalar vertex}
\\ \\
\begin{minipage}{95pt}
\includegraphics[scale=0.6]{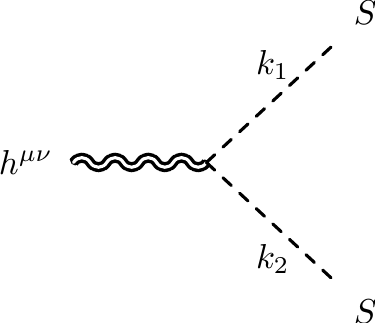}
\end{minipage}
\begin{minipage}{70pt}
\bea
&=&  i \frac{\kappa}{2} \bigg\{ k_{1\, \rho} \, k_{2 \, \sigma} \, C^{\mu\nu\rho\sigma}  - M_S^2 \, \eta^{\mu\nu} \bigg\} \nn \\
&-&  i \frac{\kappa}{2}  2 \chi  \bigg\{ (k_1+k_2)^{\mu}(k_1+k_2)^{\nu} - \eta^{\mu\nu} (k_1+k_2)^2 \bigg\} \nn
\eea
\end{minipage}
\bea
\label{FRhSS}
\eea
where $S$ stands for the Higgs $H$ and the Goldstones $\phi$ and  $\phi^{\pm}$. The first line is the contribution coming from the minimal energy-momentum tensor while the second is due to the improvement term.
\item{graviton - scalar - fermion - fermion vertex}
\\ \\
\begin{minipage}{95pt}
\includegraphics[scale=0.6]{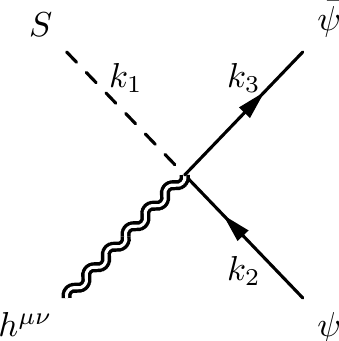}
\end{minipage}
\begin{minipage}{70pt}
\bea
=  \frac{\kappa}{2} \left( C^L_{S\bar \psi \psi} \, P_L + C^R_{S\bar \psi \psi} \, P_R \right) \, \eta^{\mu\nu}
\nn
\eea
\end{minipage}
\bea
\label{FRhSFF}
\eea
where  the coefficients are defined as
\bea
&& C^L_{h \bar \psi \psi}  = C^R_{h \bar \psi \psi} = - i \frac{e}{2 s_W} \frac{m}{m_W}  \,, \qquad
C^L_{\phi \bar \psi \psi}  = - C^R_{\phi \bar \psi \psi} =  i\frac{e}{2 s_W} \frac{m}{m_W} 2 I_3 \,, \nn \\
&& C^L_{\phi^+ \bar \psi \psi} = i \frac{e}{\sqrt{2} s_W} \frac{m_{\bar \psi}}{m_W} V_{\bar \psi \psi} \,, \qquad  C^R_{\phi^+ \bar \psi \psi} = - i \frac{e}{\sqrt{2} s_W} \frac{m_{\psi}}{m_W} V_{\bar \psi \psi} \,, \nn \\
&& C^L_{\phi^- \bar \psi \psi} = - i \frac{e}{\sqrt{2} s_W} \frac{m_{\bar \psi}}{m_W} V^*_{\bar \psi \psi} \,, \qquad  C^R_{\phi^- \bar \psi \psi} =  i \frac{e}{\sqrt{2} s_W} \frac{m_{\psi}}{m_W} V^*_{\bar \psi \psi} \,.
\eea
\item{graviton - gauge boson - fermion - fermion vertex}
\\ \\
\begin{minipage}{95pt}
\includegraphics[scale=0.6]{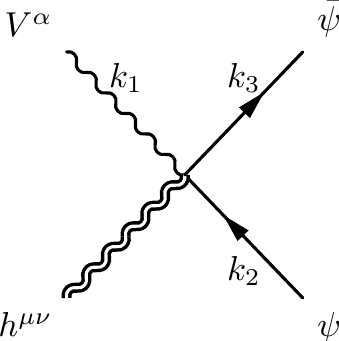}
\end{minipage}
\begin{minipage}{70pt}
\bea
= - \frac{\kappa}{2} \left( C^L_{V \bar\psi \psi} \, P_L + C^R_{V \bar\psi \psi} \, P_R\right) C^{\mu\nu\alpha\beta} \gamma_\beta
\nn
\eea
\end{minipage}
\bea
\label{FRhAFF}
\eea
with
\bea
&& C^L_{g \bar\psi \psi} = C^R_{g \bar\psi \psi} = i g_s T^a \,, \qquad  C^L_{\gamma \bar\psi \psi} = C^R_{\gamma \bar\psi \psi} = i e Q \,, \nn \\
&& C^L_{Z \bar\psi \psi} = i \frac{e}{2 s_W c_W} (v + a) \,, \quad C^R_{Z \bar\psi \psi} = i \frac{e}{2 s_W c_W} (v - a) \,, \nn \\
&& C^L_{W^+ \bar\psi \psi} = i \frac{e}{\sqrt{2} s_W} V_{\bar\psi \psi}  \,, \quad C^L_{W^- \bar\psi \psi} = i \frac{e}{\sqrt{2} s_W} V^*_{\bar\psi \psi}  \,, \quad C^R_{W^\pm \bar\psi \psi} = 0 \,,
\eea
and $v = I_3 - 2 s_W^2 Q$, $a = I_3$.
\end{itemize}
The tensor structures $C$, $D$ and $E$ which appear in the Feynman rules defined above are given by
\bea
&& C_{\mu\nu\rho\sigma} = \eta_{\mu\rho}\, \eta_{\nu\sigma} +\eta_{\mu\sigma} \, \eta_{\nu\rho} -\eta_{\mu\nu} \, \eta_{\rho\sigma} \,, \nn \\
&& D_{\mu\nu\rho\sigma} (k_1, k_2) = \eta_{\mu\nu} \, k_{1 \, \sigma}\, k_{2 \, \rho} - \biggl[\eta^{\mu\sigma} k_1^{\nu} k_2^{\rho} + \eta_{\mu\rho} \, k_{1 \, \sigma} \, k_{2 \, \nu}
  - \eta_{\rho\sigma} \, k_{1 \, \mu} \, k_{2 \, \nu}  + (\mu\leftrightarrow\nu)\biggr] \,,  \\
&& E_{\mu\nu\rho\sigma} (k_1, k_2) = \eta_{\mu\nu} \, (k_{1 \, \rho} \, k_{1 \, \sigma} +k_{2 \, \rho} \, k_{2 \, \sigma} +k_{1 \, \rho} \, k_{2 \, \sigma})
-\biggl[\eta_{\nu\sigma} \, k_{1 \, \mu} \, k_{1 \, \rho} +\eta_{\nu\rho} \, k_{2 \, \mu} \, k_{2 \, \sigma} +(\mu\leftrightarrow\nu)\biggr] \,. \nn
\eea

\bibliographystyle{h-physrev5}

 \end{document}